\input harvmac
\input epsf
\epsfverbosetrue
\def\p{\partial}
\def\ap{\alpha'}
\def\half{{1\over 2}}
\Title{\vbox{\rightline{}\rightline{}}}
{\vbox{\centerline{Introduction to M Theory }}}
\vskip10pt

\baselineskip=12pt
\centerline{Miao Li}
\medskip
\centerline{\sl Enrico Fermi Inst. and Dept. of Physics}
\centerline{\sl University of Chicago}
\centerline{\sl 5640 S. Ellis Ave., Chicago IL 60637, USA}
\baselineskip=16pt
\vskip2cm
\noindent
This is an introduction to some recent developments in string theory
and M theory. We try to concentrate on the main physical aspects, and often
leave more technical details to the original literature. \foot{Lectures 
delivered at the duality workshop at CCAST, China, Sept. 1998.}

\Date{Nov. 1998}
\nref\mtheory{There are a number of reviews, see for example, J.H.
Schwarz, ``Lectures on Superstring and M Theory Dualities'', 
hep-th/9607201; M.R. Douglas, ``Superstring Dualities, Dirichlet Branes and 
the Small Scale Structure of Space'', hep-th/9610041; P. Townsend,
``Four Lectures on M-theory'', hep-th/9612121; ``M-theory from its   
superalgebra'', hep-th/9712004; N. A. Obers and B. Pioline, 
``U-duality and M-Theory'', hep-th/9809039.}
\nref\gauge{A. Hanany and E. Witten, ``Type IIB Superstrings, BPS Monopoles, 
And Three-Dimensional Gauge Dynamics'', hep-th/9611230; E. Witten, 
`` Solutions Of Four-Dimensional Field Theories Via M Theory'', 
hep-th/9703166; A. Giveon and D. Kutasov, ``Brane Dynamics and Gauge Theory'',
hep-th/9802067.}
\nref\bhrev{J. Maldacena, ``Black Holes in String Theory'', hep-th/9607235;
A. Peet, `` The Bekenstein Formula and String Theory (N-brane Theory)'',
hep-th/9712253.}
\nref\gsw{M.B. Green, J.H. Schwarz and E. Witten, ``Superstring theory'',
Cambridge Press, 1987.}
\nref\gpr{A. Giveon, M. Porrati and E. Rabinovici, ``Target Space Duality 
in String Theory'', hep-th/9401139.}
\nref\nahm{W. Nahm, in ``monopoles in quantum field theory'', eds. N.S.
Craigie, P. Goddard and W. Nahm, World Scientific, 1982.}`
\nref\om{ C. Montonen and D. Olive, ``Magnetic monopoles as gauge 
particles?'',  Phys. Lett.72B (1977) 117.}
\nref\uduality{C. Hull and P. Townsend, ``Unity of Superstring Dualities'',
hep-th/9410167; E. Witten, ``String Theory Dynamics In Various Dimensions'',
hep-th/9503124.}
\nref\dbrane{J. Dai, R. Leigh and J. Polchinski, ``New connections between
string theories'', Mod. Phys. Lett. A4 (1989) 2073; J. Polchinski,
`` Dirichlet-Branes and Ramond-Ramond Charge'', hep-th/9510017;
``TASI Lectures on D-Branes'',  hep-th/9611050;
C. P. Bachas, ``Lectures on D-branes'', hep-th/9806199.}
\nref\ssy{J. Scherk and J.H. Schwarz, ``Dual models for nonhadrons'',
Nucl. Phys. B81 (1974) 118; T. Yoneya, ``Connection of dual models to
electrodynamics and gravidynamics'', Prog. Theor. Phys. 51 (1974) 1907.}
\nref\matri{T. Banks, W. Fischler, S. Shenker and L. Susskind,
``M theory as a matrix model: A conjecture ", hep-th/9610043; 
L. Susskind, ``Another conjecture about m(atrix) theory'', 
hep-th/9704080. }
\nref\witten{E. Witten, ``Bound states of strings and p-branes",    
hep-th/9510135.}
\nref\maxsugra{W. Nahm, ``Supersymmetries and their representations",
Nucl. Phys. B135 (1978) 149; E. Cremmer, B. Julia and J. Scherk, 
``Supergravity theory in eleven dimensions", Phys. Lett. 76B (1978)
409.}
\nref\tdual{M. Dine, P Huet and N. Seiberg,``Large and small radius
in string theory", Nucl. Phys. B322 (1989) 301.}
\nref\pgin{P. Ginsparg, ``on toroidal compactification of heterotic 
superstrings", Phys. Rev. D35 (1987) 648.}
\nref\hw{P. Horava and E. Witten, ``Heterotic and Type I String Dynamics 
from Eleven Dimensions'', hep-th/9510209; `` Eleven-Dimensional Supergravity 
on a Manifold with Boundary'', hep-th/9603142.}
\nref\sen{A. Sen, ``String string duality conjecture in six dimensions
and charged solitonic strings'', hep-th/9504027; J. Harvey and 
A. Strominger, ``The heterotic string is a soliton", hep-th/9504047.}
\nref\gross{D. Gross and P, Mende, ``String theory beyond the Planck
scale", Nucl. Phys. B303 (1988) 407; D. Amati, M. Ciafaloni and
G. Veneziano,  ``Superstring collisions at Planck energies",  Phys. 
Lett. B197 (1987) 81.}
\nref\gso{F. Gliozzi, J. Scherk and D. Olive, ``Supergravity and the 
spinor dual model", Phys. Lett. B65 (1976) 282.}
\nref\fms{D. Friedan, E. Martinec and S. Shenker, ``Conformal invariance,
supersymmetry and
string theory", Nucl. Phys. B271 (1986) 93.}
\nref\ghmr{D. Gross, J. Harvey, E. Martinec and R. Rohm, ``Heterotic 
string theory. 1 and 2", Nucl. Phys. B256 (1985) 253; Nucl. Phys.
B267 (1986) 75.}
\nref\gsm{M. Green and J.H. Schwarz, ``Infinity cancellations in 
$SO(32)$ superstring theory", Phys. Lett. B151 (1985) 21.}
\nref\towns{P. Townsend, ``The eleven dimensional supermembrane revisited", 
Phys. Lett. B350 (1995) 184.}
\nref\schwarz{J.H. Schwarz, ``Covariant field equations of chiral N=2 D=
10 supergravity'', 
Nucl. Phys. B226 (1983) 269.}
\nref\jhs{J.H. Schwarz, ``An $SL(2,Z)$ multiplet of type IIB 
superstrings", hep-th/9508143.}
\nref\bhr{E. Bergshoeff, C.M. Hull and T. Ortin,
``Duality in the type--II superstring effective action'',
hep-th/9504081.}
\nref\pajs{P. Aspinwall, ``Some relationships between dualities in 
string theory", hep-th/9508154; J.H. Schwarz, ``The power of M
theory", hep-th/9510086.}
\nref\ptown{P. Townsend, ``Three lectures on supermembranes ", 1988.}
\nref\dkl{M. Duff, R. Khuri and J.X. Lu, ``String solitons", 
hep-th/9412184.}
\nref\andy{A. Strominger, ``Open p-branes", hep-th/9512059;
P. Townsend, ``D-branes from M-branes", hep-th/9512062.}
\nref\aspinwall{P. Aspinwall, ``K3 surfaces and string duality",
hep-th/9611137.}
\nref\ss{J.H. Schwarz and A. Sen, ``Duality symmetries of 4D 
heterotic strings", hep-th/9305185. }
\nref\sam{N. Seiberg, ``Observations on the moduli space of 
superconformal field theories", Nucl. Phys. B303 (1988) 286; 
P. Aspinwall and D. Morrison,
``String theory on K3 surfaces", hep-th/9404151.}
\nref\sgms{A. Strominger, ``Massless black holes and conifolds in 
string theory", hep-th/9504090; B. Greene, D. Morrison and
A. Strominger, ``Black hole condensation and the unification of string
vacua", hep-th/9504145.}
\nref\lowd{J. Schwarz, ``Classical symmetries of some 
two-dimensional models", hep-th/9503078; A. Sen, ``Duality symmetry 
group of two dimensional heterotic string theory", hep-th/9503057; 
T. Banks and L. Susskind, ``The number of states of two dimensional
critical string theory",  hep-th/9511193. }
\nref\jpw{J. Polchinski and E. Witten, ``Evidence for heterotic-type I
string duality", hep-th/9510169; C. Vafa, ``Evidence for F-theory",  
hep-th/9602022. }
\nref\rl{R. Leigh, ``Dirac-Born-Infeld action from Dirichlet sigma
model", Mod. Phys. Lett. A4 (1989) 2073.}
\nref\mld{M. Li, ``Boundary states of D-branes and dy-strings", 
hep-th/9510161; M.R. Douglas, ``Branes within branes",  
hep-th/9512077.}
\nref\ewit{E. Witten, ``Small instantons in string theory", 
hep-th/9511030.}
\nref\marev{T. Banks, ``Matrix theory", hep-th/9710231; Biggati and 
L. Susskind, ``Review of matrix theory", hep-th/9712072; W. Taylor, 
``Lectures on D-branes, gauge theory and matrices", hep-th/9801182.}
\nref\wt{W. Taylor, ``D-brane field theory on compact spaces", 
hep-th/9611042.}
\nref\mtst{L. Motl, ``Proposals on nonperturbative superstring
interactions", hep-th/9701025; R. Dijkgraaf, E. Verlinde and 
H. Verlinde,  ``Matrix string theory", hep-th/9703030.}
\nref\bs{T. Banks and N. Seiberg, ``Strings from matrices", 
hep-th/9702187.}
\nref\grts{O. Ganor, S. Ramgoolam and W. Taylor, ``Branes, fluxes and 
duality in matrix theory", hep-th/9611202;
L. Susskind, ``T duality in matrix theory and S duality in field 
theory", hep-th/9611164.}
\nref\tfour{M. Rozali, ``Matrix theory and U-duality in seven dimensions",
hep-th/9702136; M. Berkooz, Rozali and N. Seiberg,
``Matrix description of M theory on $T^4$ and $T^5$", hep-th/9704089.}
\nref\seiberg{N. Seiberg, ``Matrix description of M theory on
$T^5$ and $T^5/Z_2$", hep-th/9705221.}
\nref\dan{U. Danielsson, G. Ferretti and B. Sundborg, ``D-particle
dynamics and bound states", hep-th/9603081; D. Kabat and P. Pouliot,  
``A comment on zero-brane quantum mechanics", hep-th/9603127.}
\nref\dkps{M.R. Douglas, D. Kabat, P. Pouliot and S. Shenker,
``D-branes and short distances in string theory", hep-th/9608024.}
\nref\nonre{S. Paban, S. Sethi and M. Stern, ``Constraints constraints 
from extended supersymmetry in quantum mechanics'', hep-th/9805018.}
\nref\jy{A. Jevicki and T. Yoneya, ``Space-time uncertainty principle and 
conformal symmetry in D-particle dynamics'', hep-th/9805069.}
\nref\sens{A. Sen, ``D0-branes on $T^n$ and matrix theory", 
hep-th/9709220 ; N. Seiberg, ``Why matrix theory is
correct", hep-th/9710009. }
\nref\beken{J. Bekenstein,  Lett. Nuovo. Cim. 4 (1972) 737; 
S. Hawking, Comm. Math. Phys. 43 (1975) 199.}  
\nref\dbh{A. Strominger and C. Vafa, ``Microscopic origin of the
Bekenstein-Hawking entropy",  hep-th/9601029; 
C. Callan and J. Maldacena, ``D-brane approach to black hole quantum
mechanics", hep-th/9602043.}
\nref\radiation{A. Dhar, G. Mandel and S. Wadia, ``Absorption vs 
decay of black holes in string theory and T-symmetry", 
hep-th/9605234; S. Das and S. Mathur, ``Comparing decay rates for 
black holes and D-branes", hep-th/9606185; J. Maldacena and 
A. Strominger, ``Black hole greybody factors
and D-brane spectroscopy", hep-th/9609026.}
\nref\mbh{J. Maldacena,  ``Statistical entropy of near extremal 
five-branes", hep-th/9605016; M. Li and E. Martinec, ``Matrix
black holes", hep-th/9703211; ``On the entropy of matrix black holes",
hep-th/9704134; R. Dijkgraaf, E. 
Verlinde and H. Verlinde,``5D black
holes and matrix strings", hep-th/9704018.}
\nref\msbh{T. Banks, W. Fischler, I. Klebanov and L. Susskind,
``Schwarzschild black holes from matrix theory I and II", 
hep-th/9709091, hep-th/9711005; I. Klebanov and L. Susskind, ``
Schwarzschild black holes in various dimensions from matrix theory", 
hep-th/9709108; G. Horowitz and E. Martinec, ``Comments on black 
holes in matrix theory", hep-th/9710217.}
\nref\lnmbh{M. Li, ``Matrix Schwarzschild black holes in large N
limit", hep-th/9710226; M. Li and E. Martinec, ``Probing matrix 
black holes", hep-th/9801070.}
\nref\mald{J. M. Maldacena, ``The large N limit of superconformal field 
theories and supergravity'',  hep-th/9711200.}
\nref\gkpw{S. S. Gubser, I. R. Klebanov and A. M. Polyakov, ``Gauge theory 
correlators from non-critical string theory'', hep-th/9802109; 
E. Witten, `` Anti de Sitter space and holography'', hep-th/9802150.}
\nref\hwit{E. Witten, ``Anti-de Sitter space, thermal phase transition, 
and confinement in gauge theories'', hep-th/9803131.}
\nref\pres{I. R. Klebanov, hep-th/9702076; S. Gubser, I. R. Klebanov
and A. A. Tseytlin, ``String theory and classical absorption by threebranes'',
hep-th/9703040; M. R. Douglas, J. Polchinski and A. Strominger, 
``Probing five-dimensional black holes with D-branes  hep-th/9703031.}
\nref\correl{W. Mueck and K. S. Viswanathan, 
``Conformal field theory correlators from classical scalar field theory on
$AdS_{d+1}$'',   hep-th/9804035;
D. Z. Freedman, S. D. Mathur, A. Matusis and L. Rastelli,
``Correlation functions in the CFT(d)/AdS(d+1) correspondence'',
hep-th/9804058;
H. Liu and A. A. Tseytlin, ``$D=4$ super Yang Mills, 
$D=5$ gauged supergravity and $D=4$ conformal supergravity'', hep-th/9804083;
G. Chalmers, H. Nastase, K. Schalm and R. Siebelink, 
``R-current correlators in $N=4$ super Yang-Mills theory from Anti-de 
Sitter'', hep-th/9805105;
S. Lee, S. Minwalla, M. Rangamani and N. Seiberg,
``Three-point functions of chiral operators in D=4, $N=4$ SYM at large N'',
hep-th/9806074.}
\nref\wils{J. M. Maldacena, ``Wilson loops in large N field theories'',
hep-th/9803002; S.-J. Rey and J. Yee, ``Macroscopic strings as heavy 
quarks of large N gauge theory and anti-de Sitter supergravity'',
hep-th/9803001.}
\nref\jky{A. Jevicki, Y. Kazama and T. Yoneya, 
``Quantum metamorphosis of conformal transformation in D3-Brane Yang-Mills
theory'',  hep-th/9808039.}
\nref\gkp{S. S. Gubser, I. R. Klebanov and A. Peet, ``Entropy and 
temperature of black 3-branes   hep-th/9602135.} 
\nref\susswit{L. Susskind and E. Witten, ``The holographic bound in 
anti-de Sitter space'', hep-th/9805114.}
\nref\reyith{A. Brandhuber, N. Itzhaki, J. Sonnenschein and S. Yankielowicz,
``Wilson loops in the large N limit at finite temperature'',
hep-th/9803137; S.-J. Rey, S. Theisen and J.-T. Yee, 
``Wilson-Polyakov loop at finite temperature in large N gauge theory and 
anti-de Sitter supergravity'',   hep-th/9804135.}
\nref\hawp{S. Hawking and D. Page, ``Thermodynamics of black holes in
anti-de Sitter space'', Comm. Math. Phys. 87 (1983) 577.}
\nref\phase{M. Li, ``Evidence for large N phase transition in $N=4$
super Yang-Mills theory at finite temperature hep-th/9807196; 
Y.-H. Gao and M. Li, ``Large N strong/weak coupling phase transition and 
the correspondence principle   hep-th/9810053.}
\nref\joeg{G. Horowitz and J. Polchinski, ``A correspondence principle for 
black holes and strings'', hep-th/9612146.}
\nref\liyo{M. Li and T. Yoneya, ``Short-distance space-time structure and 
black holes in string theory : a short review of the present status'',
hep-th/9806240.}

\newsec{Introduction}

In the past four years, a series of exciting developments in
the area of supersymmetric field theories and string theory has completely
changed the landscape of these subjects. Duality has been the central
theme of these developments. By now, it is a common belief that different
string theories all have the same origin, although this unique
theory still remains somewhat mysterious. This theory is dubbed 
M theory \mtheory. It appears that all degrees of freedom, given enough
supersymmetries, are in our possession, and the future effort will be directed
toward finding out a nonperturbative formulation of M theory.
Though abstract and seemingly remote from the real world, M theory
already has found many useful applications, in particular to supersymmetric
gauge theories in various dimensions \gauge, and to quantum properties
of black holes \bhrev.

String theory is the most promising approach to quantum gravity \gsw. The
primary motivation for many string theorists is to understand how the
universally accepted theory in particle physics, called the standard
model, comes about from some deeper principles, and how one eventually
understands some genuine quantum gravity phenomena. On the one
hand, to resolve the
so-called hierarchy problem in scales, supersymmetry is a helpful tool
provided it is broken dynamically. This certainly demands some nonperturbative
treatment of quantum field theory or string theory.
On the other hand, any visible quantum gravity effects must involve
nonperturbative processes, this is because the effective coupling constant
$G_Nm^2$ becomes of order 1 in the quantum gravity regime. String theory
was formulated, prior to the second string revolution, only perturbatively.
Thus, we had little hope to achieve either goal in the past.

Among various dualities in string theory, T-duality was first discovered
\gpr. It can be realized order by order in the perturbation theory.
T-duality has no analogue in field theory, although some novel constructs
such as Nahm transformation does have a link to T-duality \nahm. Strong-Weak
duality, or S-duality, maps a strongly coupled theory to a weakly coupled
one. It is a generalization of Olive-Montonen duality in ${\cal N}=4$ 
super Yang-Mills theory to string theory \om. As such, it requires
certain amount of supersymmetry that is unbroken in the corresponding vacuum.
The checks of S-duality in various situations mostly have been limited
to the stable spectrum (BPS). Of course some nontrivial dynamic
information is already encoded in the BPS spectrum, since many of the
states are bound states of some ``elementary states'', and highly
technical work must be done in order to merely prove the existence of
these bound states. Combination of various T-dualities and S-dualities
generates a discrete nonabelian group called the U-duality group \uduality.
Incidently, these U-duality groups are just discretization of global
symmetry groups discovered long ago in the context of supergravity.
String duality is a highly nontrivial generalization of duality in
field theory. In field theory, the S-duality maps the description
with a weak (strong) coupling constant to a description with a strong
(weak) coupling constant. In string theory, there is no free dimensionless
constant. Rather, the coupling constant is often the vacuum expectation
value of the dilaton field. The collection of the vev's of massless scalar
fields is called the moduli space. Therefore, in many cases, a duality 
transformation maps one point in the moduli space to another in the moduli
space. If these two points can be described in a single theory, then
this duality transformation is a gauge symmetry, unlike that in a field
theory.

The most powerful technique developed for studying string duality is
that of D-branes \dbrane. D-branes are extended objects on which open
strings can end. D stands for Dirichlet, a reference to the boundary
conditions on the string world-sheet. This prescription, with
corrections taking the recoil effects into account, is valid for the whole
range of energies. This property alone singles out D-brane technology
from the others, since most of the other tools are applicable only 
in the  low energy regime. It must be emphasized that D-branes are valid only 
in the weak coupling limit of string theory. However, D-branes
represent states that are invisible in the standard perturbation 
string theory. In fact, most of the heavy solitonic objects in string
theory can be identified with D-branes. Since a D-brane, or a collection
of D-branes, contains an open string sector, there is a field theory
associated to it in the low energy limit. This facilitates the study of bound
states. Bound states can be interpreted as excitations in this low
energy field theory, some at the classical level, and some at
quantum level. Another novel feature of the D-brane physics is that
the low energy D-brane field theory actually describes the short distance
physics of the closed string sector. This is due to the s-t channel
duality of the string interactions \gsw. 

There are many interesting applications of the D-brane technology. We would 
like to single out two of them. One is the application to the
study of quantum field theories. The reason for this possibility is
obvious, that the low energy theory of D-branes is a field theory.
Some ingenious arrangements of intersecting D-branes and M theory fivebranes
make it possible to read off some of the nonperturbative results in a 
field theory directly from D-brane dynamics \gauge. Since this is
a vast and quite independent subject, we will ignore it in these
lectures. Another application is to the quantum physics of black holes.
For the first time, the Bekenstein-Hawking entropy formula is derived,
although for a special class of black holes \bhrev. In string theory 
it is possible
to have extremal black holes with a macroscopic horizon, due to 
many different charges that can be carried by a stable soliton. The 
microscopic degrees of freedom are attributed, in the so-called
D-brane regime, to the appropriate open string sector.
More surprisingly, the Hawking radiation and the grey-body factor can be
reproduced at low energies. This represents tremendous success for
M/string theory.

While much has been learned since 1994, the main goal of developing 
duality for many theorists is still far beyond the horizon, that is to
formulate the M/string theory nonperturbatively and in a background
independent fashion. It is fair to say that nowadays we cannot say
about the nature of spacetime, and the underlying principles
of string theory, much beyond what we could when string theory was
first formulated as a theory of quantum gravity \ssy\ (But see the next
paragraph). It is a
miracle that the fundamental quanta of gravity, the graviton, emerges
naturally in the string spectrum. Moreover, supersymmetry and gauge
principle seems to be codified in string theory too. However, the 
spacetime itself, though secondary as believed by many, has not emerged
naturally thus far. It might be that a certain kind of correspondence
principle is lacking. Here the quanta are gravitons etc., while a
``classical orbit'' is spacetime or other classical backgrounds. By analogy
then, we need a formulation much similar to Dirac's formulation of
quantum mechanics in which the correspondence between quantum mechanical
objects and the classical ones is best spelt out. Thus it appears
that once that goal is achieved, we will have much better understanding
of the relation between quantum mechanics and gravity, and possibly of
quantum mechanics itself. To some people, eventually quantum mechanics
will stand on itself, while a classical object such as spacetime will be 
secondary and
emerge as an approximation. Still, we do not have a framework in which
such an approximation can be readily achieved.

Despite the above disappointment, there is a temporary and quite popular
nonperturbative formulation proposed under the name Matrix theory \matri. 
This proposal makes the best use of various aspects of string duality
we have learned. In particular, the D-brane intuition forms its most
solid foundation. This formulation, though nonperturbative in nature,
works only in the special frame namely the infinite momentum frame. 
As such, it strips away unnecessary baggage such as redundant
gauge symmetries and unphysical states. It also shares many unpleasant 
features of this kind of physical gauges: some fundamental symmetries
including global Lorentz symmetry and local Poincare symmetries are
hard to prove. Since space coordinates are promoted to matrices,
it reveals the long suspected fact that spacetime is indeed noncommutative
at the fundamental level \witten. At present, there are also many technical
difficulties associated to compactifications on curved spaces and
on compact spaces of dimension higher than 5. This might point to
the fundamental inadequacy of this proposal.

Matrix theory has its limited validity. It is therefore quite a surprise
that black holes and especially Schwarzschild black holes in various
dimensions have a simple description in matrix theory. Many of
speculations made on quantum properties of black holes since Bekenstein's
and Hawking's seminal works can now be subject to test. Since the quantum
nature of spacetime becomes very acute in this context, we expect
that further study of black holes in the matrix formulation will teach us
much about the formulation itself.

This article is organized as follows. We will summarize the salient 
features of M theory as the organizing theory underlying various string
theories in the next section. Discussion about U-duality and BPS spectrum 
is presented in sect.3. We then introduce D-branes, first through M-branes
then through the perturbative string theory, in sect.4. Sect.5. is
devoted to a presentation of matrix theory, hopefully in a different
fashion from those of the existing reviews. Sect.6 is devoted 
to a brief description of quantum black holes in M/string theory.
We end this article with the final section discussing the AdS/CFT
correspondence, or known as Maldacena conjecture. This is the subject
being currently under intensive investigation. 

Finally, a word about references of this article. The inclusion
of original research papers only reflects the knowledge or lack of
knowledge, and personal taste of this author. Undoubtedly many important 
contributions are unduly omitted, we apologize for this to many authors.

\newsec{M theory as the theory underlying various string theories}

There is no consensus on the definition of M theory, since nobody 
knows how to define it in the first place. Our current understanding
of it is through rather standard notion of vacua: The (moduli) space
of all possible stable, static solutions in various string theories
is connected in one way or another, therefore there must be a 
unique underlying theory covering the whole range. One of the 
interesting limits is the 11 dimensional Minkowski space with 
${\cal N}=1$ supersymmetry. Its low energy limit is described by
the celebrated 11 dimensional supergravity, discovered before the
first string revolution \maxsugra. Practically, as one confines oneself
in the low energy regime, any point in the moduli space can be regarded
as a special solution to the 11D supergravity. Needless to
say, such a specification of M theory is quite poor. For a quantum
theory of gravity, there is no reason to focus one's attention
on those solutions in which there is a macroscopic Minkowski space.
To this class, one can add solutions containing a macroscopic
anti-de Sitter space, and time-dependent solutions. The latter
is relevant to cosmology. The reason for restricting ourselves
to the usual ``vacua" is that these are the cases we understand
better in ways of a particle physicist: We know how to treat
states of finite energy, and interaction therein. 

During the first string revolution, we learned that in order to make
a string theory consistent, supersymmetry is unavoidable. Further,
these theories automatically contains gravity, and have to live
in 10 dimensional spacetime. There are two closed string theories
possessing ${\cal N}=2$ supersymmetry. These are type II theories.
Type IIA is non-chiral, and hence its super-algebra is non-chiral.
Type IIB is chiral, that is, the two super-charges have the same
chirality. In 10 dimensions, these theories do not contain nonabelian gauge
symmetry. There are three theories with ${\cal N}=1$ supersymmetry,
all contain gauge symmetry of a rank 16 gauge group. The rank
and the dimension of the gauge group are fixed by the anomaly cancellation
conditions. This constitutes the major excitement in the first
revolution, since for the first time the gauge group is fixed by
dynamics. Of the three theories, two are closed string theories
with gauge group $E_8\times E_8$ and $SO(32)$, called heterotic string. 
The third is an open string theory (with closed strings as a 
subsector) of gauge group $SO(32)$.

Numerous ``theories" in lower dimensions can be obtained from the five 
10 dimensional theories, through the compactification procedure.
It is here one discovers that the five theories are not all different
theories. In 9 dimensions, type IIA is related to IIB by T-duality
on a circle \refs{\dbrane,\tdual}. Type IIA on a circle of radius $R$ is 
equivalent to
type IIB on a circle of radius $\alpha'/R$. The moduli space is
the half-line, but one is free to call a point either IIA or IIB.
Similarly, the two heterotic strings are related in 9 dimensions \pgin.
Thus, in the end of the first string revolution, it was known
that there are only three different string theories.
T-duality is an exact symmetry on the world-sheet of strings, namely
the perturbative spectrum and amplitudes are invariant under this
map. It is reasonable to extrapolate to conjecture that this symmetry
is valid nonperturbatively. The most strong argument in support of
this, independent of the the web of various dualities, is that T-duality 
can be regarded as a unbroken gauge symmetry. Since this is a discrete
symmetry, there is no reason for it to be spontaneously or dynamically
broken.

It is the hallmark of the second string revolution that the above string
theories possess strong-weak duality symmetry. First of all, the type IIB
string is self-dual \uduality. This duality is very similar to the self-duality
of ${\cal N}=4$ supersymmetric Yang-Mills theory (SYM) in 4 dimensions.
There is a complex moduli, its imaginary part being $1/g$, $g$ 
the string coupling constant. Without self-duality, the moduli space is
thus the upper-half complex plane. Now the duality group is $SL(2,Z)$
acting on the complex coupling as the rational conformal transformation.
The real moduli space is then the familiar fundamental domain.
This remarkable symmetry was already discovered in the supergravity
era, without being suspected a genuine quantum symmetry at the time.
Another remarkable discovery made three years ago is that IIA string also
has a dual. In the strong coupling limit, it is a 11 dimensional
theory whose low energy dynamics is described by 11 dimensional 
supergravity. Now the new dimension which opens up is due to the appearance
of a Kaluza-Klein worth of light modes, being solitons in the IIA
theory. Relating these states to KK modes implies that the string
coupling is proportional to the radius of the new dimension.

Furthermore, type I string theory contains stringy soliton solutions,
these are naturally related to the heterotic string. Thus type I
string is S-dual to the heterotic string with the gauge group $SO(32)$.
Finally, as Horava and Witten argued, the heterotic string can be
understood as an orbifold theory of the 11 dimensional M theory \hw.
This completes the full web of string theories down to 9 dimensions.

Compactifying to even lower dimensions, more duality symmetries emerge.
For instance, type IIA on a K3 surface is dual to the heterotic string
on $T^4$ \refs{\uduality, \sen}. This is a quite new duality, since 
the heterotic string is
a five-brane wrapped around $K3$ in the IIA theory. The universal feature
is that in lower and lower dimensions, more and more duality symmetries
surface, and this reflects the fact that the spectrum becomes ever 
richer in lower dimensions and various limits can be taken to see
new light degrees of freedom. Again, the U-duality groups, the largest
duality groups, already made their appearance in the supergravity
era as the global symmetry of supergravities. New light degrees of
freedom make it possible to have nonabelian gauge symmetry in the
type II theories. It will be seen how this is closely tied up with
geometric features of compactification and the existence of various
p-brane solitons.

\subsec{A brief review of string theory}

String theory has been defined only perturbatively \gsw.  When a string moves
in spacetime, it sweeps a 2 dimensional world-sheet.
A complete set of
perturbation rules similar to Feynman diagrams is given by specifying
a local form of a two dimensional field theory and summing over all possible
topologies of surfaces. This makes string theory quite different from
a quantum field theory: Surfaces are  smoother objects than Feynman
diagrams. This single fact is the origin of many stringy miracles.
For instance, the high energy behavior of a scattering amplitude is 
much softer \gross. To see this, we only need to know that the string amplitude
is proportional to $\exp (-A)$, where $A$ the area of the world-sheet.
The area of the interaction region is large and smooth in the high
energy limit. Another miracle is the s-t channel duality. This duality
serves as the prime motivation for constructing Veneziano amplitude, whose
discovery predates string theory.

Since string theory is specified only perturbatively, therefore its 
classification is carried out by classifying different types of the
world-sheet theories. The most important symmetry on the world-sheet
is conformal symmetry. Matter fields induce conformal anomaly on the
world-sheet, and this must be cancelled in order to decouple the intrinsic
world-sheet metric. Without additional local symmetry, it is found that
there must be 26 free scalars on the world-sheet, implying that the
bosonic string theory makes sense only when embedded into 26 dimensional
spacetime. However, this theory is ill-defined due to the existence of
a tachyon state.

To improve upon the situation, one has to introduce more gauge symmetry on
the world-sheet. Supersymmetry was discovered in this context. To
implement supersymmetry, for each scalar $X^\mu$, a Majorana spinor
$\psi^\mu$ is introduced. Now, each fermion contributes $1/2$ to the
central charge, and the ghosts of fixing local world-sheet ${\cal N}=1$
supersymmetry contribute $11$ to the central charge, the conformal
anomaly cancellation condition is $3/2 D=26-11$, and the solution is
$D=10$. Thus, for the spinning string the critical dimension is $10$.
The world sheet action, after removing world-sheet metric and gravitino 
field, reads
\eqn\wsa{S={T\over 2}\int d^2\sigma\left(\p_\alpha X^\mu\p_\alpha X^\mu-
i\overline{\psi}^\mu\gamma^\alpha\p_\alpha\psi^\mu\right),}
where $T$ is the string tension, and sometimes is denoted by $1/(2\pi\ap)$,
and $\ap$ is called the Regge trajectory slope.

For the time being we focus on the closed string.
The first quantization is carried out by solving the equations of motion
for $X^\mu$ and $\psi^\mu$. It is easy to see that $X^\mu=X^\mu (t-\sigma)
+\tilde{X}^\mu(t+\sigma)$, the left-moving piece and right-moving 
piece. Similarly, the Majorana spinor $\psi^\mu$ is separated into
a left-moving part and a right-moving part. The component of $\psi^\mu$
with positive chirality is left-moving, and the one with negative
chirality is right-moving. As always with fermions, there are two possible
periodic boundary conditions: $\psi^\mu(\sigma+2\pi)=\pm\psi^\mu(\sigma)$.
The sector in which all $\psi^\mu$ are periodic is called the Ramond sector, 
and the sector in which $\psi^\mu$ are anti-periodic is called the 
Neveu-Schwarz sector. It must be emphasized that the world-sheet 
supersymmetry demands all $\psi^\mu$
to have the same boundary condition. However, since SUSY does not mix the
left-moving and the right-moving parts, therefore there are four
possible pairings, (R,R), (NS, NS), (R,NS), (NS,R). 

In the Ramond sector, there are fermionic zero modes, satisfying the
anti-commutation relations $\{d^\mu, d^\nu\}=\eta^{\mu\nu}$. This
is just the ten-dimensional Clifford algebra. Therefore the ``vacua''
form a spinor representation of dimension $2^5$. There is
a unique vacuum in the NS sector. Consider the left-moving
sector, the world-sheet energy operator is $L_0=\half p^2+N_L/\ap$, where
$N_L$ is the oscillator operator. To demand the Lorentz algebra be
closed, we find that $(L_0-a)|phys\rangle$ for a physical state, 
where $a$ is a constant depending on the boundary conditions of $\psi^\mu$.
$a=0$ for the R sector, and $a=1/2$ for the NS sector. Again there
would be a tachyon mode in the NS sector, if we do not execute 
a certain projection procedure. A consistent projection exists, and is
called GSO projection \gso. To this end, construct an operator $(-1)^{F_L}$
which anti-commutes with $\psi^\mu$ and commutes with $X^\mu$, moreover,
it contains a factor $\gamma_{11}$ when acting on the Ramond sector.
A physical state is defined as a positive eigen-state of $(-1)^{F_{L}}$,
in particular, if one assigns $-1$ to the NS vacuum, this tachyonic
state is discarded. Notice also that only half of the ``vacuum'' states
in the R sector survives, say the half with positive chirality under
$\gamma_{11}$. Similarly, one can define $(-1)^{F_R}$ for the right-moving
sector and exercise the same projection. Now one is free to choose
either $\tilde{\gamma}_{11}$ or $-\tilde{\gamma}_{11}$ that is contained
in this G parity operator. For the first choice, we obtain a chiral
theory, because the surviving spinors in both R sectors have the same
chirality. This is type IIB string theory. For the second choice, the 
theory is nonchiral, since spinors of both chiralities exist. This
is IIA string theory.

Thus, in the NS sector, the first states surviving the GSO projection
have $N_L=1/2$. There are ten such states $\psi^\mu_{-1/2}|0\rangle$.
From the tensor products (R,NS) and (NS,R) we would obtain two sets
of $10\times 2^4$ states. These are two gravitini. On-shell condition
will eliminate more states thus there are only two sets of $8\times 2^3$
physical states. In all, there are $2^7$ massless fermionic states.
States in (R,R) and (NS,NS) sectors are bosonic. At the massless level,
there are total $8\times 8=2^6$ states in (NS,NS) sector. These are
just gravitons $G_{\mu\nu}$, ``axions'' $B_{\mu\nu}$ and dilaton $\phi$.
There are also $2^6$ bosonic states in the (R,R) sector. These states
are bi-spinors. One can use matrices $\gamma_0\gamma_{\mu\nu\dots}$ to 
contract these spinors to obtain anti-symmetric  tensor fields. It is
straightforward to see that in the type IIA case, only tensor fields
of even rank are obtained, and in the IIB case, only tensor fields of
odd rank are obtained. 

It is a curious feature of the Ramond-Neveu-Schwarz formulation that
a tensor field thus constructed corresponds to a field strength,
rather than an elementary field itself \fms. Thus in the IIA theory, there
is vector field $C^{(1)}$, a rank three anti-symmetric tensor field 
$C^{(3)}$, and their duals. In the IIB theory, there is a scalar field
$C^{(0)}$, a rank two anti-symmetric tensor field $C^{(2)}$, a self-dual
rank four anti-symmetric tensor field $C^{(4)}$. All these fields will
play an important role in our discussion on D-branes later.
Spacetime symmetry is hidden in  the RNS formulation. From the existence
of gravitino fields, it is clear that both type II theories possess
${\cal N}=2$ SUSY. One is chiral, another is non-chiral.

The above discussion can be readily generalized to the open string theory.
An open string sweeps a world-sheet with boundary.
In order to obtain the equation of motion from the world-sheet action, it
is necessary to specify appropriate boundary conditions. Here a Lorentz
invariant boundary condition is the Neumann boundary condition. This
implies that the two ends of an open string move with the speed of light.
For both the bosonic fields $X^\mu$ and the fermionic field $\psi^\mu$,
the left-moving modes are related to the right-moving modes through
the boundary conditions. Thus, there are only two sectors, the R sector
and the NS sector. Again one has to apply the GSO projection in order
to get rid of tachyon. In the NS sector, there are 8 on-shell massless
states and they correspond to a vector field. There are 8 massless
fermionic states in the R sector, corresponding to a Majorana-Weyl
fermion. These two fields form a ${\cal N}=1$ vector super-multiplet,
and the action is that of the ${\cal N}=1$ U(1) SYM. This construction
is generalized to the nonabelian case by assigning the so-called 
Chan-Paton factor to the ends of a string. It turns out that the open
string is nonorientable and the only consistent gauges groups are
$SO(N)$ and $Sp(N)$.

An open string loop amplitude contains some poles which can be interpreted
as closed string states. This is due to the fact that a string loop
diagram can be deformed in such a way that it contains an intermediate closed 
string state explicitly. To ensure unitarity of the S-matrix, closed
string states must be included in the spectrum. In particular, an open
string theory necessarily contains graviton and 
dilaton. For a generic
gauge group, there is a tadpole for the R-R ten form field. To have this
tadpole canceled, the gauge group must be of rank 16 and 496 dimensional.
There are two possible such groups, $SO(32)$ and $E_8\times E_8$.
The latter cannot be generated by the Chan-Paton factor. Note that
when Green and Schwarz first discovered this, they demanded the gauge
anomaly to be canceled. This cancellation is equivalent to the vanishing
of the R-R ten form tadpole.

The construction of heterotic string was based on the basic observation that
in a consistent string background, the left-moving modes on the world-sheet
are decoupled from the right-moving modes \ghmr. To have a consistent 
theory, either
sector must be embedded into a consistent, anomaly free theory. For instance,
when the left-moving sector is embedded to that of type II theory, and
the right-moving mode embedded into the bosonic string theory, the standard
heterotic string is obtained. There are 26 scalars in the right-moving
sector, 10 of them are paired with those in the left-moving sector in order
to have 10 noncompact scalars. These give rise to 10 macroscopic spacetime
dimensions. The remaining 16 scalars can not be arbitrarily chosen. The
one-loop modular invariance forces them to live on a torus constructed
by $R^{16}/\Gamma_{16}$, where $\Gamma_{16}$ is a 16 dimensional even self-dual
lattice. There are only two such lattices, one is given by the root
lattice of $SO(32)$, the other is the root lattice of $E_8\times E_8$.
States constructed in the NS sector include gauge bosons of the
corresponding group. It is not surprising that the one-loop modular 
invariance is closely related to the anomaly cancellation condition,
thus these groups were anticipated by Green and Schwarz \gsm.

\subsec{Low energy effective actions}

We start with the type IIA action, since this theory is closely related
to the 11 dimensional supergravity. In a string theory, there are two
basic scales. The fundamental one is the string scale, defined by
$M_s^2=T$, or $l_s^2=\ap$. All the massive string states are graded
by this scale. The second scale is the Planck scale, it is determined 
by the Newton constant $G_{10}$. The gravity strength
is proportional to $g^2$, where $g$ is the string coupling constant.
The Newton constant has a dimension $L^8$, and indeed $G_{10}=g^2l_s^8$.
The Planck length is then $l_p=g^{1/4}l_s$. Now in a string theory,
$g$ is not a free parameter, it is determined by the vacuum expectation
value of dilaton, $g=\exp (\phi)$. Therefore the Planck length is
not as fundamental as the string scale, as viewed in string theory.
We shall soon see that there is a third scale in IIA theory, it is
the 11 dimensional Planck length. Since all massive states are graded
by $M_s$, one can integrate them out to obtain a low energy effective
action for massless fields. In the bosonic sector, there is a metric
$G_{\mu\nu}$, an antisymmetric field $B_{\mu\nu}$, a dilaton, a vector
field $C^{(1)}_\mu$, a rank three tensor field $C^{(3)}_{\mu\nu\rho}$.
In the fermionic sector, there are two gravitino fields $\psi_\mu$
with opposite chiralities. The field content forms the ten dimensional
type IIA supergravity multiplet. In the low energy limit ($E\ll M_s$),
it was shown by taking a direct zero slope limit of string scattering
amplitudes that the effective action coincides with that of the IIA 
supergravity. Since we are not concerned with local supersymmetry yet, it
is enough to write down the bosonic part of the action
\eqn\iia{\eqalign{S&={1\over 8\pi^2l_s^8}\int d^{10}x\sqrt{g}[
e^{-2\phi}\left(R+4(\p_\mu\phi)^2
-{1\over 2\times 3!}H_{\mu\nu\rho}H^{\mu\nu\rho}\right)-{1\over 4}
F_{\mu\nu}F^{\mu\nu}\cr
&-{1\over 2\times 4!}F_{\mu_1\dots
\mu_4}F^{\mu_1\dots \mu_4}],}}
where we denote $2\pi\ap$ by $l_s^2$. $H$ is the field strength of
$B_{\mu\nu}$, $F_{\mu\nu}$ is the field strength of the vector field
$C^{(1)}$ and $F_{\mu_1\dots \mu_4}$ is th field strength of $C^{(3)}$.
We adopt the definition $F_{\mu_1\dots\mu_{p+1}}=\p_{\mu_1}C^{(p)}_{\mu_2
\dots\mu_{p+1}}\pm \hbox{cyclic permutations}$.

One crucial feature of the low energy effective action is that the action
of the R-R fields are not weighted by the factor $e^{-2\phi}$. One certainly
can redefine these fields to have this weighting factor, then the simple gauge
symmetry $C^{(p)}\rightarrow C^{(p)}+d\epsilon^{(p-1)}$ is lost. This feature
is reflected in the world-sheet technique for calculating scattering
amplitudes. The vertex operator for an on-shell R-R state corresponds
directly to the field strength, therefore perturbative string states
are not charged with respect to these long range fields: There is no the
analogue of the Aharonov-Bohm effect, therefore there is no R-R charge 
perturbatively. This is to be contrasted with the $B_{\mu\nu}$ field.
The fundamental string is charged with respect to it. As we shall see,
the form of the action for an R-R field is responsible for the fact that
a solitonic state charged under this field has a mass (tension) scaling
as $1/g$.

To see that the IIA effective action is a dimensional reduction of
the 11 dimensional supergravity, we need to identify the field content.
Compactifying the 11 dimensional theory on a circle of radius $R$,
we obtain a metric, and scalar field $\phi$ from $g_{11,11}$, and 
a vector field through the standard Kaluza-Klein mechanism. This vector
field is identified with $C^{(1)}$. There is a rank three antisymmetric
tensor field $A$ in the bosonic sector of the 11D supergravity. It gives
rise to $C^{(3)}$ when all three indices are restricted to 10 dimensions.
the components $A_{11,\mu\nu}$ is identified with $B_{\mu\nu}$.
This completes the identification of the bosonic sector. The 11D gravitino
is a 11D Majorana fermion, and decomposes into a 10D fermion of positive
chirality and a 10D fermion of negative chirality. This is exactly
the fermionic sector of the type IIA supergravity. Concretely, we have
\eqn\elevend{ds_{11}^2=e^{4\phi/3}(dx_{11}-C^{(1)}_\mu dx^\mu)^2 +e^{-2\phi/3}
G_{\mu\nu}dx^\mu dx^\nu,}
then the 11D supergravity action reduces to the IIA effective action
when all massive KK states are discarded. The above decomposition
implies that in the 11D Planck unit, there is the relation $R^2=
g^{4/3}$ or $g=(R/l^{11}_p)^{3/2}$. Further, the two Newton constants
are related by $G_{10}=G_{11}/R=(l_p^{11})^9/R$. The above two relations
combined yield $l_s^2=(l_p^{11})^3/R$. We will soon see the physical
meaning of this relation.

A KK mode has a energy $E=n/R$, with an integer $n$. From the above
relations between the compactification scale and the string coupling
constant, we deduce $R=gl_s$, so $E=n/(gl_s)$. This state carries
$n$ units of charge of $C^{(1)}$, and it must be a nonperturbative
state in string theory, since its mass is proportional to $1/g$. This, 
as will be seen,
is a generic feature of a R-R charged state. The KK mode with $n=1$
is called a D0-brane, as will be explained later. In the string theory
framework, other higher KK modes can be regarded as bound states of
the fundamental D0-branes, or bound states of anti-D0-branes (for
a negative $n$). In the strong coupling limit, $R$ becomes much
larger than the 11D Planck scale, and a new dimension opens up. It is
no longer possible to ignore KK states since they become light in 10
dimensions and start to propagate in the full 11 dimensions. This
is one of the most striking results in the past four years \refs{\towns,
\uduality}.

The low energy NS-NS sector of type IIB string theory is identical
to that of IIA theory, and the low energy effective action of this
part is the same as that in \iia. In addition to $\phi$, there is a
second scalar field $C^{(0)}$ in the R-R sector. This can be combined with 
$\phi$ to form a complex field $\tau =C^{(0)}+ie^{-\phi}$. There are two 
more fields in the R-R sector, $C^{(2)}$ and $C^{(4)}$. The condition
on $C^{(4)}$ is that its field strength $dC^{(4)}$ is a self-dual
5 form. There is no simple action for this field, so we will not attempt
to write down an action for it. The field strength of $C^{(2)}$, call
it $H'$, together with the field strength $H$ form a doublet of 
$SL(2,R)$. In fact, one can write down a $SL(2,R)$ invariant effective
action as follows.
\eqn\iib{S={1\over 16\pi G_{10}}\int d^{10}x\sqrt{g}\left(R-{1\over 12}
H^T_{\mu\nu\rho}MH^{\mu\nu\rho}+{1\over 4}\tr (\p_\mu M\p^\mu M^{-1})
\right),}
where the new metric is $G^E_{\mu\nu}=e^{-\phi/2}G_{\mu\nu}$, the Einstein
metric. This metric is invariant under $SL(2,R)$. $H^T$ is the doublet
$(H,H')$ and the two by two matrix 
\eqn\twoby{M={1\over\Im \tau}\left(\matrix{|\tau |^2\ &\Re\tau\cr
\Re\tau &1}\right).}
The action of a $SL(2,R)$ element
$$\Lambda= \left(\matrix{a&b\cr c&d}\right)$$
on $\tau$ is
\eqn\taut{\tau\rightarrow {a\tau +b\over c\tau +d},}
and the action on $H$ is $H\rightarrow (\Lambda^T)^{-1}H$. 

Although we did not write down an action for $C^{(4)}$, it must be
noted that $C^{(4)}$ is invariant under $SL(2,R)$. The equation of
motion is a first order differential equation, the self-dual condition
\schwarz.
Since only the Einstein metric is invariant under $SL(2,R)$, the
self-duality is imposed with the use of the Einstein metric. The whole
set of equations of motion including fermions is $SL(2,R)$ invariant.
This group is broken at the quantum level to $SL(2,Z)$, due to the
existence of solitonic objects. The weak-strong coupling duality
is a special element of $SL(2,Z)$: $\tau\rightarrow -1/\tau$.
The fundamental string is charged under
$H$, then simply due to symmetry, there must be a string-like solution
charged under $H'$. This string is a D-string, and its tension is given
by $T/g$, where $T$ is the fundamental string tension. Moreover, there
are infinitely many bound states of these strings, called $(p,q)$ strings
\jhs.
It carries $p$ units of $H$ charge, $q$ units of $H'$ charge. For
such a string to be stable, $p$ and $q$ must be coprime in order to
prevent the bound state to disintegrate into pieces. The tension formula
for the $(p,q)$ string will be given later.

When IIA theory is compactified to 9 dimensions, more massless fields
appear. In addition to $\phi$, there is one more scalar from $G_{99}$.
These two scalars form a complex scalar, just as $\tau$ in IIB theory.
There are three vector fields, one inherited from $C^{(1)}$, another from
$G_{9\mu}$. They form a doublet of $SL(2,R)$. The third comes from
$B_{9\mu}$ and is a singlet under $SL(2,R)$. There are two rank two 
anti-symmetric fields, one from $B$, another from $C^{(3)}_{9\mu\nu}$,
these form a doublet of $SL(2,R)$. Finally, there is a rank three tensor
field. The $SL(2,R)$ symmetry becomes explicit when the IIA in 9 dimensions
is regarded as the compactification of the 11D theory on a torus $T^2$.
For instance, the doublet vectors fields are just $G_{11\mu}$
and $G_{9\mu}$. So $SL(2,R)$ is the symmetry group acting on $T^2$.
$SL(2,R)$ breaks to $SL(2,Z)$ simply for geometric reason. Thus we
have seen that just like IIB theory, there is an $SL(2,Z)$ duality symmetry
in the 9 dimensional IIA theory. This is not surprising, since we already
mentioned that IIA is T-dual to IIB in 9 dimensions.

The bosonic content of the massless spectrum of IIB in 9 dimensions is
identical to the above. Still there is a complex scalar $\tau$. There are
three vector fields, one from $G_{9\mu}$, a singlet of $SL(2,R)$;
the other two from $B_{9\mu}$
and $C^{(2)}_{9\mu}$ forming a doublet. Notice that $G_{9\mu}$ in IIB
is not to be identified with $G_{9\mu}$ in IIA, since the latter is
in the doublet. Thus under T-duality, $G_{9\mu}$ is exchanged with
$B_{9\mu}$, a well-known fact. The two rank two fields still form a 
doublet. Finally, one gets a rank three field from $C^{(4)}_{9\dots}$.
The rank four field is dual to the rank three field due to the self-dual
constraint on the original field $C^{(4)}$. For more detailed
discussion on the effective action with $SL(2,Z)$ symmetry, see \bhr.

We conclude that in 9 dimensions, the type II theories are unified, and
possess $SL(2,Z)$ duality symmetry. This duality group is the geometric
symmetry group of the two torus on which the 11D theory is compactified
\pajs.

\subsec{Horava-Witten construction and type I/heterotic string theory}

The type I string theory is dual to the heterotic string theory, with 
gauge group $SO(32)$. The low energy effective actions are identical, 
provided we switch the sign of the dilaton field when switch from one
string theory to another. Since $g=\exp (\phi)$, the duality map is
a weak-strong duality. If one theory is weakly coupled, then the other
is strongly coupled. This helps to avoid an immediate contradiction:
In the heterotic string perturbative spectrum there is no sign of open
strings; and in the open string theory although there is a closed
string sector, there is no sign in the perturbative spectrum
of heterotic strings carrying $U(1)$ currents. Heterotic string will
appear as solitonic solution in the open string theory. On the other
hand, since there is no stable macroscopic open string, thus open 
string does not emerge as a solitonic solution in heterotic string
theory. We will see the origin of open strings when an M theoretical
interpretation of type I/heterotic string becomes available.

In the closed string sector of type I theory, there is a metric, an
antisymmetric field $C^{(2)}$ and a dilaton. Note that an open
string is not charged under $C^{(2)}_{\mu\nu}$, since an open string
is non-orientable and thus there is no coupling $\int C^{(2)}$ in the
world-sheet action. The corresponding closed string is also non-orientable,
thus uncharged against the $C$ field. The super-partner of these
massless closed string states is a gravitino with $2^6$ degrees of
freedom, and they together form the ${\cal N}=1$ supergravity multiplet.
In the open string sector of type I theory, there are nonabelian gauge
fields and their super-partners, gauginos. Now in the heterotic string
theory, there is no open string sector. The massless states are gravitons,
$B_{\mu\nu}$ quanta, dilaton, a gravitino, gauge bosons and gauginos.
The content is exactly the same as that of type I theory. Since $B$
is identified with $C^{(2)}$, and the heterotic string is charged
under $B$, thus it must appear as a solitonic state in type I theory,
and the string tension is proportional to $1/g_I$ as $C^{(2)}$ is a
R-R field. The low energy effective action of the closed string
states has the same form of type II theories. The effective action
of the super Yang-Mills part is of interest, and is just
\eqn\sym{S={1\over l_s^6}\int d^{10}\sqrt{g}e^{-\phi}\tr\left(-{1\over 4}
F_{\mu\nu}F^{\mu\nu}
+{i\over 2}\overline{\psi}\gamma^\mu\p_\mu \psi\right),}
where $\psi$ is in the adjoint representation of the gauge group, and 
is a Weyl-Majorana spinor. The above action is written in the type I 
language. The gauge coupling constant is 
$g_{YM}^2=gl_s^6$, $g$ the string coupling constant. In the heterotic side,
the gauge coupling is given by $g^2_{YM}=g_h^2l_h^6$, where $g_h$ is the
heterotic string coupling constant, and $l_h^{-2}$ the tension of the 
heterotic string. This is consistent with the fact that even the vector 
supermultiplet is interpreted as a closed string excitation in heterotic
string theory. Using $g_h=1/g$ and $l_h^2=gl_s^2$,
it can be checked the two definitions of the gauge coupling constant
agree.

Now, both type I and heterotic theories are chiral, there is a potential
gauge anomaly as well as gravitational anomaly. The anomaly gets 
canceled only when the gauge group is $SO(32)$ or $E_8\times E_8$.
For type I string, only $SO(32)$ is possible, and it is dual to
heterotic string with the same gauge group.

We are left with the heterotic string with gauge group $E_8\times E_8$.
It is related to the other heterotic string theory by T-duality only
when it is compactified to 9 dimensions. Does it have a dual theory
already in 10 dimensions? The Horava-Witten construction answers this
question positively \hw.

M theory on $R^{10}\times S^1$ is just IIA string theory. Since M theory
is invariant under parity reflection, it is natural to ask whether it makes
sense to construct orbifolds of this theory. The simplest possibility is
$R^{10}\times S^1/Z_2$. Here $Z_2$ acts on $S^1$ by the reflection:
$X^{11}\rightarrow -X^{11}$. Now, the three form field $A$ is odd under
the parity reflection, so only the components $A_{11,\mu\nu}$ are even
and survive the projection. There will be no three form in 10 dimensions
after the $Z_2$ projection. The $Z_2$ projection acts on fermions as
$\psi\rightarrow \gamma_{11}\psi$. Thus, only half of gravitino which
satisfies $\psi=\gamma_{11}\psi$ is left. Further, $G_{11\mu}$ is odd, and
thus projected out. It is not hard to see that the massless spectrum 
in the 11D supergravity multiplet left after the projection coincides 
with that of the heterotic supergravity multiplet, the 10 dimensional
${\cal N}=1$ supergravity multiplet. 

The novelty of Horava-Witten construction is the way to produce the vector
supermultiplet of the gauge group $E_8\times E_8$. This gauge sector comes
into play by the requirement that the gravitational anomaly must be
canceled. Now the supergravity multiplet on $R^{10}\times S^1/Z_2$ is
chiral viewed in 10 dimensions, therefore the gravitational anomaly will
arise. The diffeomorphisms to be considered are those of $R^{10}\times
S^1$ commuting with $Z_2$. Under a diffeomorphism generated by
$\delta X^I=\epsilon v^I$ ($I=\mu, 11$), we postulate that the anomaly have
a local form
\eqn\anom{\delta\Gamma =\epsilon\int d^{11}x\sqrt{g}v^IW_I,}
where the integral is taken over the manifold $R^{10}\times S^1/Z_2$.
Apparently, if $x$ is a smooth point within the bulk of the manifold,
there should be no local contribution to the anomaly, since there is
no anomaly in 11 dimensions. So $W_I$ must be supported at the orbifold
points of $S^1/Z_2$, $X^{11}=0, \pi$. The above integral reduces to 
integrals over the two 10 dimensional boundaries
\eqn\banom{\delta\Gamma =\epsilon\int d^{10}x\sqrt{g}v^IW_I(x^{11}=0)
+\epsilon\int d^{10}x\sqrt{g}v^IW_I(x^{11}=\pi ).}
This form implies that there is an anomaly inflow toward the two walls.
And the walls are thus some kind of defect. $W_I(x^{11}=0,\pi )$ must
be given by the standard gravitational anomaly in 10 dimensions.

The anomaly must be canceled by introducing massless fields living only
on boundaries. Without much ado, we know that the only consistent way
is to introduce a gauge supermultiplet on each boundary. The usual
Green-Schwarz mechanism is applicable here, thus the gauge group must have
rank 16, and be 496 dimensional. $SO(32)$ is not a good candidate, since
it can not be equi-partitioned to the two walls. The reasonable choice
is $E_8\times E_8$. We shall not run into details of anomaly cancellation,
but only mention that the way to cancel the anomaly, although similar
to that in the heterotic string theory, has an interesting twist, because
here all gravitational fields actually live in 11 dimensions,
and everything must be written in an 11D integral form.

The relation of the string coupling to the size $R$ of $S^1/Z_2$ is the
same as in the IIA case, $g=(R/l_p)^{3/2}$, where $l_p$ is the 11 dimensional
Planck length. And the relation of the string tension to $R$ is
$l_s^2=l_p^3/R$. 
This can be seen by a similar analysis of the low 
energy effective action. Another way to see this is through the the
mechanism of generating strings from membranes. In the IIA case, a closed
string is just a membrane wrapped around $X^{11}$. In the present context,
an open membrane with ends attached to the walls appears as a closed 
string. A stable string comes from a stretched membrane between the
two walls. It is interesting to note that this mechanism is quite
similar to the Chan-Paton mechanism to generate gauge symmetry by assigning
colors to the ends of a string.

The strong coupling regime of the heterotic string is better described by
a yet unknown 11D supersymmetric theory, whose low energy limit 
is supergravity.
This connection between the M theory on $R^{10}\times S^1/Z_2$ and
the $E_8\times E_8$ string theory sheds light on the duality between 
the $SO(32)$ heterotic
string and type I string. Compactifying further the M theory on
$R^9\times S^1\times S^1/Z_2$, we obtain a 9 dimensional $E_8\times E_8$
heterotic string. It is possible to switch on Wilson line along $S^1$,
thus change the unbroken gauge group.
By T-duality, this theory is related to the $SO(32)$ heterotic string
in 9 dimensions. Now an open membrane wrapped on the cylinder $S^1\times
S^1/Z_2$ can be either interpreted as a closed string, as on the heterotic 
side when $S^1/Z_2$ has a small size, or an open string on the type I side,
when $S^1$ has a small size. We thus see that both open string and
heterotic string have a common origin in 11 dimensions. Because the
geometric truncations are different, just as in the case when M theory
is compactified on $T^2$, the string couplings are related by the
reciprocal relation.

The relation of 5 string theories to M theory is summarized in the following 
diagram.
\bigskip
{\vbox{{\epsfxsize=3.0in
        \nobreak
    \centerline{\epsfbox{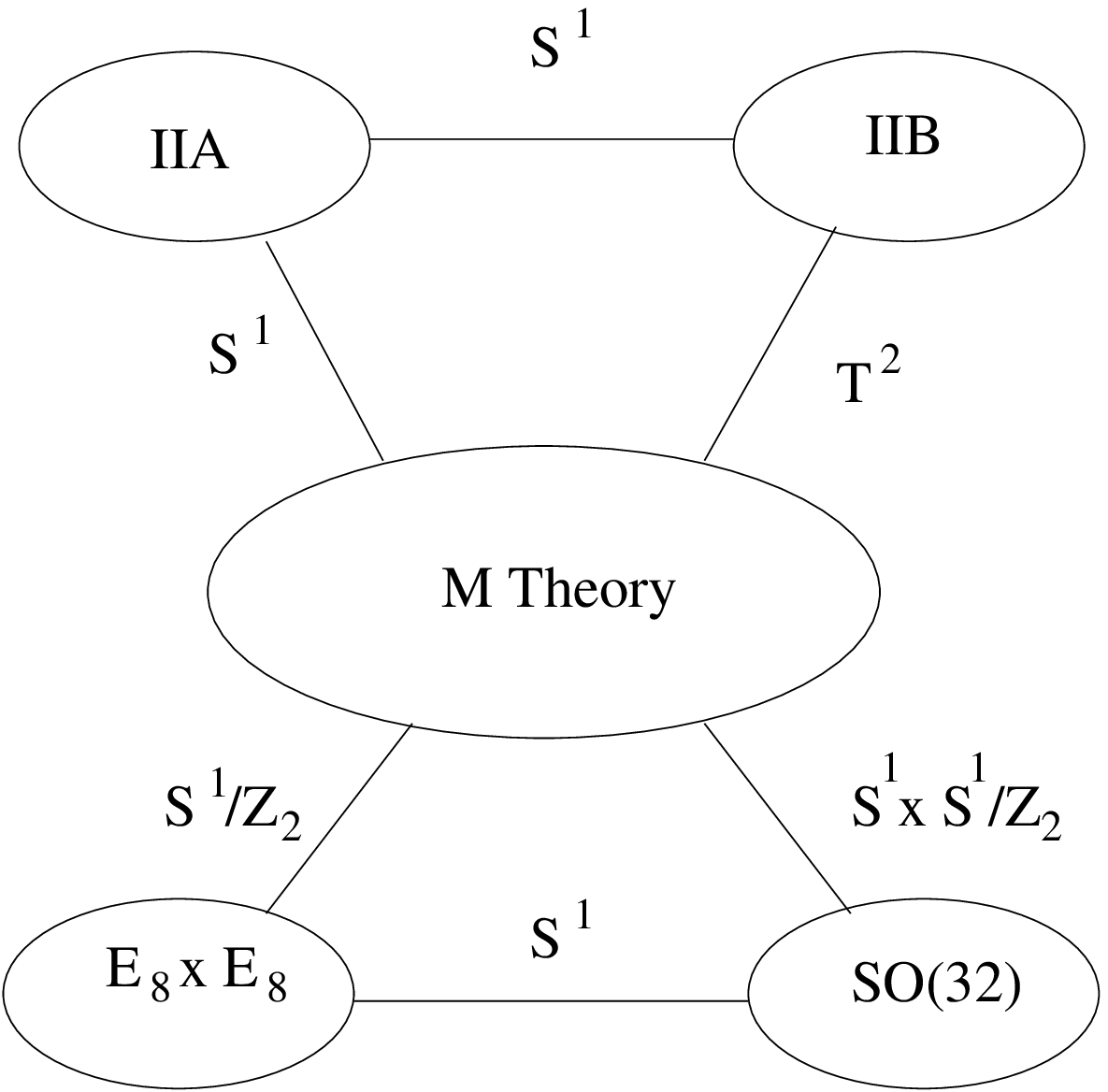}}
\nobreak\bigskip
    {\raggedright\it \vbox{
{\bf Figure 1.}
{\it A schematically representation of M theory and its descendants.}
 }}}}}

\newsec{BPS spectrum and U-duality}

The most evidence in support of various duality relations by far 
comes from the so-called BPS spectrum and the low energy effective 
actions. A BPS state, by definition, is a stable state often carrying
different charges. Being stable, it cannot decay into other states,
thus its stability is independent of the coupling constant and other moduli 
parameters in the theory. Admittedly, the identification of two low energy
effective actions after certain field redefinitions is a rather weak
condition for the two theories in question to be dual. The BPS spectrum
provides rather strong evidence, since some states in one theory are
nonperturbative bound states, their existence puts strong constraints
on the dynamics.

A BPS state often preserves a certain mount of supersymmetry. Thus a powerful
tool to analyze these states is the super-algebra. We have seen that
all theories down to 9 dimensions have the same origin in M theory,
it is then economic to directly work with the super-algebra of M
theory.

\subsec{BPS states in 11 dimensions}

The superalgebra of M theory in 11 dimensions is the super Poincare
algebra. As such, there are total $32$ supercharges $Q_\alpha$. It is
possible to choose a Majorana representation of gamma matrices such
that all $Q_\alpha$ are Hermitian. The anticommutators are
given by
\eqn\antic{\{Q_\alpha, Q_\beta\}=(C\gamma_\mu)_{\alpha\beta}P^\mu,}
where the index $\mu$ runs over $0,\dots, 9, 11$. In the Majorana
representation, all $\gamma$ matrices are real, the $\gamma_i$ are 
symmetric, while $C=\gamma^0$ is anti-symmetric.

It is possible to generalize the anti-commutation relations to include
more central charges. The anticommutators are symmetric in indices
$\alpha$ and $\beta$, one must add symmetric matrices to the
R.H.S. of \antic. 
In addition to $C\gamma_\mu$, only $C\gamma_{\mu\nu}$
and $C\gamma_{\mu_1\dots\mu_5}$ are symmetric, the maximally generalized
algebra is then
\eqn\ganti{\{Q_\alpha, Q_\beta\}=(C\gamma_\mu)_{\alpha\beta}P^\mu
+\half (C\gamma_{\mu\nu})_{\alpha\beta}Z^{\mu\nu}+{1\over 5!}
(C\gamma_{\mu_1\dots\mu_5})_{\alpha\beta}Z^{\mu_1\dots\mu_5},}
As we shall see shortly, the objects carrying charge $Z^{\mu\nu}$
are membranes, and the objects carrying charge $Z^{\mu_1\dots \mu_5}$
are fivebranes.

Consider a state with nonvanishing $P$ only. The L.H.S. of \antic\
is a Hermitian matrix. When sandwiched by a physical state, say
$\langle P|\{Q_\alpha, Q_\beta\}|P\rangle$, we obtain a matrix whose
eigenvalues are either positive or zero. A zero eigenvalue is possible
only when $|P\rangle$ is annihilated by a linear combination of 
32 charges $Q_\alpha$. This particular supersymmetry is unbroken in
the presence of this state. Whenever a zero eigenvalue is present,
the determinant of this matrix vanishes. On the other hand, the determinant
is easily computed using the R.H.S. of \antic, and is 
$\det (C\gamma_\mu P^\mu)=\det (\gamma_\mu P^\mu)=(P^2)^{16}$. 
Only when the on-shell condition $P^2=0$ is satisfied, there is a 
zero eigenvalue.  For a single particle state, this is a supergraviton
in 11 dimensions. In case there is no zero eigenvalue, $E^2>P_iP_i$,
this is the familiar BPS bound. When this bound is not saturated, the
state can decay into, for example, a bunch of supergravitons.

To see how many supersymmetries are unbroken with a supergraviton state,
we need to examine without loss of generality, the case $P_{11}\ne 0$.
Now $C\gamma_\mu P^\mu=P_{11}(1-\gamma^0\gamma_{11})$. The matrix
$\gamma^0\gamma_{11}$ has 16 eigenvalues $1$ and 16 eigenvalues $-1$,
so the matrix $C\gamma_\mu P^\mu$ has half of eigenvalues equal to zero,
corresponding to the condition $\gamma^0\gamma_{11}=1$. Therefore, there
are 16 supersymmetries unbroken by this supergraviton, and they satisfy
the condition $\gamma^0\gamma_{11}\epsilon =\epsilon$.

A membrane carries charge $Z^{\mu\nu}$. Due to Lorentz invariance, 
$Z^{\mu\nu}$ can be rotated into a nonvanishing component with
two spatial indices, if it is space-like, $Z_{\mu\nu}Z^{\mu\nu}>0$,
or into a nonvanishing component with a time index and a space index,
if it is time-like. Consider the first case, when $Z^{ij}\ne 0$.
If all $P$ except $E$ are vanishing, the R.H.S. of \ganti\ reduces to
\eqn\mem{E-\gamma^0\gamma_{ij}Z^{ij}.}
Again, half of eigenvalues of $\gamma^0\gamma_{ij}$ are $1$, and half 
are $-1$. The above matrix has 16 zero eigenvalues if $E=|Z^{ij}|$,
and $sgn(Z^{ij})\gamma^0\gamma_{ij}=1$. In other words, the
unbroken supersymmetry satisfies
\eqn\unb{sgn(Z^{ij})\gamma^0\gamma_{ij}\epsilon =\epsilon.}

However, in a noncompactified 11 dimensional spacetime there is
no stable, finite energy membrane configuration. The superalgebra
contains much information about the BPS spectrum, but some dynamics
is to be imported. Here explicitly, the membrane charge $Z^{ij}$
is generated by a membrane stretched over the $(ij)$ plane
\eqn\memc{Z^{ij}=Q\int dX^i\wedge dX^j,}
and there can be no boundary on this membrane, thus $Z^{ij}$
is infinite, and $E$ is infinite too. A finite, stable membrane
configuration can be obtained by compactifying at least two
spatial dimensions, say $X^i, X^j$ on a flat torus.

A membrane appears as a solitonic solution in the low energy
effective action, with the long range three form $A_{\mu\nu\rho}$
nonvanishing. Indeed, a membrane is directly coupled to this
field through the following coupling
\eqn\memc{Q\int A_{\mu\nu\rho}dX^\mu\wedge dX^\nu\wedge dX^\rho.}
A membrane is dynamical, that is, it propagates in spacetime.
The world-volume theory of a single membrane, the ``fundamental"
membrane, is described by a free $2+1$ supersymmetric theory with
16 supercharges. The world-volume theory of multiple membranes
is unknown.

Next, what object can carry a time-like membrane charge, say
$Z^{0i}$? Generalizing the above analysis, it is easy to find
that the unbroken supersymmetry is
\eqn\nine{sgn(Z^{0i})\gamma_i\epsilon=\epsilon.}
Since $\gamma^0\gamma^1\dots\gamma^{11}=1$, the above condition
is equivalent to 
\eqn\nint{\gamma^0\gamma^1\dots\hat{\gamma}^i\dots\gamma^{11}
\epsilon=\pm \epsilon.}
Townsend then conjectured that this is given by a 9-brane,
whose world-volume is orthogonal to $X^i$. Indeed, the 9-brane
walls of Horava-Witten preserve supersymmetry $\gamma_{11}
\epsilon=\epsilon$, and the walls are orthogonal to $X^{11}$.
However, these walls are not dynamic, in the sense that their
position in $X^{11}$ does not fluctuate, unlike most of branes
we are discussing. We leave this speculation as a curiosity.

The object carrying charge $Z^{i_1\dots i_5}$ is a fivebrane.
The unbroken supersymmetry is
\eqn\fsusy{sign(Z^{i_1\dots i_5})\gamma^0\gamma_{i_1\dots
i_5}\epsilon =\epsilon.}
Again it contains 16 components. A fivebrane is magnetically
charged with respect to $A_{\mu\nu\rho}$. One can define
a 6-form field $A^{(6)}$ dual to $A$ through $dA^{(6)}= ^*dA$.
A fivebrane is coupled to $A^{(6)}$ 
\eqn\fcoup{Q_5\int A^{(6)},}
where the integral is taken over the fivebrane world volume.

Since a fivebrane carries a magnetic charge of $A$ while a membrane
carries an electric charge of $A$, it is natural to ask whether
there is a Dirac quantization condition for the these charges. 
There is one, and the way to obtain it is similar to the way to 
obtain the original
Dirac quantization on the electric charge and the magnetic charge
in 4 dimensions, although both objects in question are extended
objects. We shall not try to derive this condition here, but will
give it in the following section. Note also that the membrane charge
and the fivebrane charge satisfy the minimal Dirac quantization
condition \dkl.

The world-volume of a fivebrane is a free theory with 16 supersymmetries.
As always with a state which breaks 16 bulk supersymmetries, there are
16 Goldstinos induced on the world-volume. It is easy to identify part
of the bosonic sector related to the breaking of translational invariance.
There are 5 scalars on the world-volume corresponding to Goldstones 
of the breaking translational invariance, namely there are 5 transverse
directions to the fivebrane. There is a shortage of three bosonic
degrees of freedom, compared to 8 on-shell fermionic degrees of freedom.
This is supplemented by a self-dual two-form field. The whole 
supermultiplet is called the tensor multiplet in $5+1$ dimensions.
Due to the self-duality of the tensor field, the theory is chiral.
This can also be seen by examining the unbroken SUSY in \fsusy. The
world-volume theory of multiple fivebranes is currently unknown, although
the limit in which all fivebrane coincide is understood to be described
by a super conformal field theory.

There are BPS states carrying two kinds of charges yet preserving half of
supersymmetry. For example, one can boost a membrane either in a direction
orthogonal to it or in one of its longitudinal directions. In the first
case, the total energy is given by the standard relativistic formula
for boosting a massive object. The energy is the sum of the momentum
and its rest mass in the second case. For this case
the boosted membrane is sometimes called a threshold bound state of a membrane
and supergravitons, a fancy name.

So far all the BPS states we have discussed preserve half of 
supersymmetry, namely there are 16 unbroken supercharges. The simplest
example of states breaking more supersymmetries is provided by a
``bound state'' of a membrane and another membrane. Consider the situation
when, say, $Z^{12}\ne 0$ and $Z^{34}\ne 0$. $\gamma^0\gamma_{12}Z^{12}$
commutes with $\gamma^0\gamma_{34}Z^{34}$,  and they can be diagonalized
simultaneously. In this case if both $sign(Z^{12})\gamma^0\gamma_{12}
\epsilon =\epsilon$ and $sign(Z^{34})\gamma^0\gamma_{34}
\epsilon =\epsilon$ are satisfied, the R.H.S. of \ganti\ has zero eigenvalues.
This bound state breaks $3/4$ of whole supersymmetry.

There are less trivial examples of BPS states preserving $1/4$ of 
supersymmetry. For example, an open membrane stretched between two 
parallel fivebranes \andy, and 
states with two central charges whose corresponding matrices in \ganti\
anti-commute, say when a membrane is trapped in a fivebrane.

\subsec{Compactification and U-duality}

It is in this subsection we run into the most technical of all subjects
reviewed in this article.

A low dimensional string theory is often obtained by compactifying a 10
dimensional string theory on a compact space. The simplest compact spaces
are tori. If the metric on the torus is flat, no supersymmetry is broken,
and the low dimensional theory has as many unbroken supercharges as in the
original 10 dimensional theory. For instance, if we consider type II
string theory on $T^6$, the low energy theory in four dimensional spacetime
is the ${\cal N}=8$ supergravity theory which automatically contains
$28$ abelian vector fields. The gauge theory is always abelian 
on the moduli space. This theory can be viewed as 
compactification of M theory on $T^7$. Now it is straightforward to count
the dimension of the moduli space. There are $28$ scalars of the form
$G_{mn}$, where $m,n$ are tangential indices on $T^7$. There are 
$35$ scalars of the form $A_{mnp}$. So the dimension of the moduli space
is $63$. Globally, the moduli space is the coset space
\eqn\cose{E_{7(7)}(Z)\backslash E_{7(7)}/SU(8),}
where the group $E_{7,(7)}$ is a noncompact version of the exceptional
group $E_7$, and its dimension is $126$. The discrete group $E_{7(7)}(Z)$
is a integral version of  $E_{7,(7)}$. This is just the U-duality group
\uduality.

The appearance of the U-duality group can be understood as follows. There
are $28$ abelian gauge fields, and solutions which are either electrically
charged or magnetically charged exist. The Dirac-Schwinger-Zwanziger
quantization condition for $28+28$ charges is invariant under
a general $Sp(28,Z)$
transformation. The theory is not symmetric under the full group 
$Sp(28)$, but only under $E_{7(7)}$, therefore the discrete symmetry must
be $E_{7(7)}(Z)=E_{7(7)}\cap Sp(28,Z)$. From the string theory perspective,
$E_{7(7)}(Z)\supset SO(6,6,Z)\times SL(2,Z)$. $SO(6,6,Z)$ is the
T-duality group of $T^6$, while $SL(2,Z)$ is the S-duality group. The full
U-duality group is much larger than the simple product of the other
two smaller groups.

We are somehow cavalier when we write down the product $SO(6,6,Z)
\times SL(2,Z)$, since the T-duality group does not commute with
the S-duality group of type IIB. This is quite obvious in the
geometric context of the M theory compactification on $T^7$. Take
a $T^2$ out of $T^7$, one may identify group $SL(2,Z)$ with the
geometric symmetry group of this torus. Now the T-duality group
mixes one of the circle of $T^2$ with the remaining $T^5$, that is,
the representation space of $SL(2,Z)$ is not invariant under 
$SO(6,6,Z)$. It is not hard to see, by a simple group decomposition
of $E_{7(7)}(Z)$, that the T-duality group and the S-duality group
together generate the whole U-duality group.

We have avoided talking about how the electric charges and magnetic
charges are generated. In general, there are various dyons, and
the complete statement is that an integral lattice of 56 dimensions
is generated by all possible PBS states of various charges. To see
how the elementary charges come about, we take a look at how the 28 
abelian gauge fields are inherited upon compactification. $7$ vector
fields come from the standard KK scheme on $T^7$, $g_{m\mu}$.
Thus the electric charges are just those KK modes. Magnetic charges
are carried by the so-called  KK monopoles. The remaining $21$
vectors fields are $A_{mn\mu}$. We already learned that in 11 dimensions
membranes are electrically charged under $A$. Now an electric, 
``point-like" charge in 4 dimensions with respect to $A_{mn\mu}$ is just 
a membrane
wrapped on the corresponding two circles. Further, we also learned
that a fivebrane is magnetically charged with respect to $A$. Now,
it is straightforward to wrap a fivebrane on the $T^5$ orthogonal
to the two circles to generate a corresponding monopole. 
(We simplified the context to consider a rectangular $T^7$.) Dyonic
states are various bound states of KK modes, KK monopoles, membranes
and fivebranes.

There can be no nonperturbative gauge symmetry at any point on the moduli
space. This is simply prevented by ${\cal N}=8$ supersymmetry. This
large amount of supersymmetry necessarily mixes vector bosons with
a spin-two state. The only spin-two state is graviton, and its vector
partners are just those $28$ abelian fields.

Higher dimensional situation can be derived by decompactifying some
circles, and we shall not endeavor to be complete here.

A 4 dimensional string theory with less supersymmetry, say ${\cal N}
=4$ SUSY can be obtained using either $T^6$ compactification of 
heterotic/type I string, or compactification of type II string on
$K3\times T^2$, where $K3$ is a two dimensional complex manifold of 
holonomy $SU(2)$ \aspinwall. It is known that for a Majorana spinor there are
two covariant constant modes on $K3$, thus from each 10D Weyl-Majorana
spinor survive two 6D Weyl-Majorana spinors, which in turn can be
regarded as two 4D Majorana spinors. Since there is a factor $T^2$
in the compact manifold, one does not have to distinguish between
type IIA and type IIB. There are therefore two ${\cal N}=4$ string
theories in 4 dimensions. We shall argue a little later that the
two theories are actually one theory, they are dual to each other.

Consider heterotic/type I string on $T^6$ first. In the heterotic 
language, it is readily seen that the moduli space is 
\eqn\hmodu{[SO(6,22,Z)\backslash SO(6,22)/(SO(6)\times SO(22))]
\times [SL(2,Z)\backslash SL(2)/U(1)].}
The first factor is the Narain space in 4 dimensions. $6$ in $SO(6,22)$ 
is associated to the number of left-moving scalars on the heterotic 
world-sheet, and $22$ is associated to the number of right-moving scalars.
$SO(6,22)$ is the T-duality group. The second factor is associated
to the complex scalar containing the axion (dual to $B_{\mu\nu}$)
as the real part, and $e^{-2\phi}$ as the imaginary part. Naively,
one would expect this moduli space be the upper-half plane, namely
$SL(2)/U(1)$. However, this string theory has a self-duality group
$SL(2,Z)$, generalizing the Olive-Montonen duality of ${\cal N}=4$ 
super Yang-Mills theory. This duality was first seriously investigated
by Schwarz and Sen \ss, and much solid evidence was collected adjoining
the Olive-Montonen conjecture. The gauge
group is $U(1)^{28}$ at a generic point on the moduli space. 
As is well-known, at many special points enhanced nonabelian gauge
symmetry appear, and the corresponding gauge bosons are already
present in the perturbative spectrum.

Heterotic/type I theory on $T^6$ can be regarded as the M theory 
compactification on $(S^1/Z_2)\times T^6$, according to Horava-Witten
construction. As the type II string theory in 4 dimensions, many
electric charges and magnetic charges originate from KK modes, wrapped
membranes and five-branes. Some other charges, however, must be derived
from the generalized Chan-Paton factors associated to open membranes
attached to Horava-Witten walls. 

Type II theory on $K3\times T^2$ has $16$ unbroken supercharges. Naturally
one wonders whether this theory is a different manifestation of the
heterotic theory on $T^6$. The answer to this question is yes. Indeed,
type IIA compactified on $K3$ is dual to heterotic string on $T^4$. It is
better to start with 6 dimensions in order to understand this duality
better, and the self-duality of the ${\cal N}=4$ theory in 4 dimensions
better. The moduli space of the heterotic string on $T^4$ is given by the
Narain space
\eqn\narain{{\cal M}_{4,20}=SO(4,20,Z)\backslash SO(4,20)/(SO(4)\times
SO(20)),}
this agrees with the moduli space of the type IIA on $K3$. This is quite
nontrivial, since the moduli space of the latter theory has a complicated
geometric origin, and its global structure is subtly related to algebraic
geometric features of the $K3$ surface \sam. Here we will be content with
counting the dimension of the moduli space of IIA on $K3$. First of all
these moduli all come from the NS-NS sector, since the odd cohomology
of $K3$ is empty. There are 22 moduli from the $B$ field, since the
second cohomology group is $22$ dimensional. The moduli space of the 
deformation of Ricci flat metric is
\eqn\ricci{SO(3,19,Z)\backslash SO(3,19)/(SO(3)\times SO(19)),}
(its geometric origin is quite complicated, we skip it here.) it has
dimension $19\times 3=57$. Finally a real moduli comes from the dilaton.
The total dimension of the IIA moduli space is then $80$, exactly the same
as that of \narain.

The rank of gauge group of the heterotic string in 6 dimensions is 
$20+4=24$. The group is abelian at a generic point on the moduli space
\narain. Again there are enhanced gauge symmetry groups at some 
special points. On the IIA side, there is a gauge field from $C^{(1)}$,
$22$ gauge fields from $C^{(3)}_{mn\mu}$, since this number is equal
to the dimension of the second cohomology group of $K3$. Finally
$C^{(3)}_{\mu\nu\rho}$ in 6 dimensions is dual to a vector field, thus
there are total $24$ abelian gauge fields. If this theory is really dual
to the heterotic string, there must be a mechanism to generate enhanced
gauge symmetry. The natural place to look for new vector multiplets
is by examining which solitonic states become light in special situations.
Indeed, membranes can be wrapped on various homologically nontrivial
surfaces in $K3$, and some of these surfaces degenerate to a point
when $K3$ is deformed to the corresponding special point on the
moduli space. It is quite requisite that the membranes indeed
form vector supermultiplets.

The duality between the two theories in 6 dimensions is strong-weak
duality. By examining the low energy effective actions, one finds that
the map between two dilaton fields is $\phi\rightarrow -\phi$, thus
inverting the string coupling. The heterotic string appear in the IIA theory
as fivebranes wrapped on $K3$. It was checked that indeed when the
volume of $K3$ is small, the effective world-sheet symmetry of a wrapped
fivebrane agrees with that of the heterotic string \sen.

Compactifying both theories on a further $T^2$, we obtain the duality
between the two ${\cal N}=4$ string theories in 4 dimensions. For each 
theory
the T-duality group on $T^2$ is $SO(2,2,Z)=SL(2,Z)_U\times SL(2,Z)_T$.
The first factor acts on the complex structure and the second on the
Kahler structure. It is quite interesting that the second factor
in the IIA theory is mapped to the S-duality group in the heterotic 
string theory. This is the origin of the S-duality group \uduality.

We have described the two most fundamental U-duality groups in 4 dimensions,
and these are related to various higher dimensional U-duality groups.
Upon compactification on more complicated Calabi-Yau spaces, theories
with less supersymmetry can be obtained. These theories are still under
control in 4 dimensions, if the SUSY is ${\cal N}=2$ \sgms.
The phenomenologically
interesting situation is ${\cal N}=1$, and unfortunately much less
is known for these theories. Another interesting direction is spacetime
of fewer dimensions. One expects the U-duality get ever richer in
lower dimensions \lowd. Again this is a regime we are currently lacking
useful tools to explore.

\newsec{ D-branes as a powerful nonperturbative tool}

Membranes and fivebranes in M theory are important objects for realizing
various dualities. The world-volume theory of multiple M-branes are not
well understood. String theory is obtained from M theory compactification.
The perturbation of a string theory is well-defined. It is then a good
question to ask whether one can describe brane dynamics better in the
string context. For a wide class of branes, the answer is surprisingly
simple, that indeed these branes can be described in a perturbative
string theory, their existence induces a new sector, an open string sector.
The ends of these open strings are attached to D-branes, here $D$
stands for Dirichlet since the world-sheet boundary conditions for
open strings are Dirichlet \dbrane.

\subsec{D-branes from M-branes}

M theory on manifold $R^{10}\times S^1$ is the IIA string theory on 
$R^{10}$ with coupling constant $g=(R/l_p)^{3/2}$. IIA strings are
just membranes wrapped on $S^1$, thus the tension of the string is
related to tension of membrane $T_2$ through $T=T_2R$. The membrane
tension can be determined by the Dirac quantization condition
on membrane and fivebrane, and is just $l_p^{-3}$, so we have the
relation $l_s^2=l_p^3/R$.

A membrane stretched along a two plane in $R^{10}$ is a solitonic solution
in the string theory carrying R-R charge of $C^{(3)}$. Its tension
$l_p^{-3}=l_s^{-3}/g$. That is, if one holds the string scale fixed,
the brane tension goes to infinity in the weak coupling limit. This 
is typical of a soliton solution, but the power $g^{-1}$ is atypical.
As we shall soon see, this behavior is what exactly one expects of a
D-brane. That is, a membrane in $R^{10}$ is a D2-brane. As we mentioned
earlier, a KK mode has a mass $l_s^{-1}/g$, and this is a D0-brane.

A fivebrane wrapped around $R$ is regarded as a fourbrane in 10 dimensions.
Up to a numeric factor, the fivebrane tension is $l_p^{-6}$, therefore
the fourbrane tension is $l_p^{-6}R=l_s^{-5}/g$. Again it scales as 
$1/g$, the generic feature of a D-brane. Thus, a wrapped fivebrane on the
M circle is a D4-brane. It is dual to D2-brane, as a consequence of
the membrane-fivebrane duality.
A KK monopole gets interpreted as a sixbrane.
Take $S^1\times R^3$ out of $S^1\times R^{10}$, the Taub-NUT solution
on  $S^1\times R^3$ carries the magnetic charge with respect to $C^{(1)}_\mu
=g_{11\mu}$. The solution is Lorentz invariant on the remaining spacetime
$R^7$, thus it is a sixbrane. It is dual to D0-brane, so it is a
candidate for D6-brane. Indeed, the tension of this sixbrane is just
the monopole mass on $S^1\times R^3$, and is given by $R^2/G_{11}
=l_s^{-7}/g$, again the right behavior for a D-brane.

To conclude, a D-brane in IIA string theory always has an even spatial
dimension.

To get to IIB string, M theory must be compactified on $T^2$.
If one of the circle is taken as the M circle, a membrane wrapped
around this circle become the fundamental string. If this fundamental
string is further wrapped around the second circle $w$ times, we 
obtain a winding string in the IIA theory, which becomes, according 
to T-duality, a string carrying $w$ unit momentum along the T-dual
circle. Similar, a string carrying momentum in the IIA picture
is interpreted as a string wrapped on the T-dual circle in the IIB
picture.

A membrane transverse to the M circle, as we explained before, is
a D2-brane in the IIA theory. This D2-brane can be wrapped around
the second circle, or transverse to the second circle. In the
first case, it appears as a string in 9 dimensional open spacetime.
Let the radius of the M circle be $R$, and the radius of the
second circle be $R_1$. The tension of this string is $l_s^{-3}
g_A^{-1}R_1$, where $g_A$ is the IIA string coupling constant.
According to the T-dual formula, $g_A^{-1}=g_B^{-1}l_s/R_1$,
where $g_B$ is the IIB string coupling constant, we find the
string tension be $l_s^{-2}g_B^{-1}$.  According to the $SL(2,Z)$
duality invariance, this tension is just the tension of the
string which carries the R-R charge of field $C^{(2)}$. 
Again the behavior $1/g_B$ is that of a D-brane. This object
is a D1-brane, or a D-string. In order to go to the 10D IIB string
limit, $R_1\rightarrow 0$, since in this case the radius of
the T-dual circle $l_s^2/R_1\rightarrow \infty$. Now  $g_B
=(g_Al_s)/R_1=R/R_1$, for a fixed $g_B$,  $R\sim R_1\rightarrow 0$.
The 10D IIB limit is obtained by shrinking both radii.

The second case mentioned in the previous paragraph gives rise
to a 2-brane in 9 noncompact spacetime. The tension of this 2-brane is
$l_s^{-3}g_A^{-1}=l_s^{-4}g_B^{-1}\tilde{R_1}$, where $\tilde{R}_1
=l_s^2/R_1$ is the radius of the T-dual circle in the IIB theory.
This can be interpreted as a D3-brane wrapped around this circle
in the IIB picture. How can one obtain a unwrapped D3-brane?
This must be a D4-brane in the IIA theory wrapped around $R_1$.
Indeed the 3-brane tension is $l_s^{-5}g_A^{-1}R_1=l_s^{-4}g_B^{-1}$,
the same formula we obtained before. Eventually, this is a fivebrane
wrapped on the torus on which M theory is compactified.

Furthermore, a D4-brane transverse to the $R_1$ circle has a tension
formula whose correct interpretation is a D5-brane wrapped around
the T-dual circle. A unwrapped D5-brane is not a M-fivebrane
transverse to the torus, as we already learned that the latter is a 
NS-fivebrane in the IIA theory, and its tension does not have the
right scaling in the string coupling. However, one can take a 
D6-brane wrapped around $R_1$, which is just a KK monopole on the
M-circle. A simple calculation shows its tension has the correct
scaling behavior in $g_B$, thus it is a D5-brane transverse to
the $\tilde{R}_1$ circle.

The above discussion clearly shows that there is a simple relation
between D-branes in the IIA theory and D-branes in the IIB theory.
The D-brane grows one more dimension if it is transverse to the
circle on which T-duality is performed, and loses one dimension
if it is wrapped around this circle. As a consequence, the a D-brane
in IIB string theory always has odd spatial dimensions.

There are higher dimensional D-branes in both type II theories.
A D-brane with dimensions higher than 6 necessarily induces some 
unusual geometry in the transverse space, and thus its nature is
more complicated.

\subsec{D-branes as a consequence of T-duality}

Take the bosonic string as a simple example. Let $X$ be the scalar
compactified on a circle of radius $R$. For a closed string, the
solution to the world-sheet action is $X(z,\bar{z})=X(z)+\tilde{X}(z)$,
where
\eqn\expan{\eqalign{X(z)&=x+i\sqrt{{\ap\over 2}}(-\alpha_0\ln z
+\sum_{n\ne 0}{\alpha_nz^n\over n}), \cr
\tilde{X}(\bar{z})&=\tilde{x}+i\sqrt{{\ap\over 2}}(-\tilde{\alpha}_0\ln 
\bar{z} +\sum_{n\ne 0}{\tilde{\alpha}_n\bar{z}^n\over n}),}}
where we used the complex coordinate on the world-sheet which is a cylinder.
Explicitly, we have $z=\exp(t+i\sigma)$, $t$ is the Euclidean world-sheet
time. On a circle, this solution is also specified by the total momentum
and winding number. The momentum along the circle is proportional to
$\int d\sigma\p_tX\sim \alpha_0+\tilde{\alpha}_0$, this determines
$$\alpha_0+\tilde{\alpha}_0={2m\over R}\sqrt{{\ap\over 2}}.$$
The winding number is encoded in $\Delta X=\sqrt{2\ap}(\alpha_0-
\tilde{\alpha}_0)\pi =2\pi wR$. Thus
\eqn\zerm{\eqalign{\alpha_0 &=({m\over R}+{wR\over\ap})
\sqrt{{\ap\over 2}},\cr
\tilde{\alpha}_0 &=({m\over R}-{wR\over\ap})\sqrt{{\ap\over 2}}.}}

The T-duality symmetry is readily seen in the above mode expansion.
Exchanging $R$ and $\ap/R$, $m$ with $w$, this maps $\alpha_0$ to
$\alpha_0$ and $\tilde{\alpha}_0$ to $-\tilde{\alpha}_0$. Furthermore,
if the oscillator part $X(z)$ is kept the same, and the sign of the 
oscillator part of $\tilde{X}(\bar{z})$ is reversed, not only the spectrum
is invariant, the correlation functions of vertex operators are
also invariant. This is just the T-duality map. The new coordinate
compactifies on a circle of radius $\ap/R$.

A usual open string does not have a winding number, since it is not 
well-defined. To see this directly, one solves the equation of motion
with Neumann boundary conditions imposed on the ends. $\tilde{X}$ is
no longer independent of $X$, actually one must identify $\tilde{\alpha}_n$
with $\alpha_n$ in order to satisfy Neumann boundary conditions.
Upon T-duality, since $\tilde{X}\rightarrow -\tilde{X}$, the Neumann
boundary condition is longer satisfied. Instead, the Dirichlet boundary
conditions $\p_tX(z,\bar{z})=0$ are satisfied on the both boundaries.
This means that the ends of the new open string are fixed at a certain
value of $X$.
Thus, in a theory containing open strings, the Dirichlet boundary conditions
can not be avoided, since it is a consequence of T-duality map.

The zero mode $x$ in the open string mode expansion has no canonical value 
under T-duality map, it can be anywhere. If we start with type I string
theory, then formally one can associate a 9-brane to a Chan-Paton factor.
The fact the the world-volume of a 9-brane fills the whole 10D spacetime
means that the ends of an open string can move freely in 10D spacetime.
After T-duality, the ends of new open strings must be fixed in the $X$
direction, this means that open strings are attached to a 8-brane whose
world-volume is transverse to $X$. This is a D8-brane. To get a D7-brane,
compactify a spatial dimension along the world-volume of the D8-brane,
and perform T-duality. D-branes of various dimensions can be obtained this
way. This T-duality map between a D(p+1)-brane and a Dp-brane is just what
we predicted using M-branes.

The world-sheet supersymmetry implies that the boundary condition for
the fermions must be same. The situation is similar to the case of
Neumann boundary conditions. In the RNS formalism, one is free to set
$\psi^\mu(\sigma=0)=\tilde{\psi}^\mu(\sigma=0)$, then there are two
choices at the other end: $\psi^\mu(\sigma=\pi)=\pm \tilde{\psi}^\mu
(\sigma=\pi)$. We obtain in the open string sector either Ramond sector
or NS sector. In the Green-Schwarz formalism, the fermionic fields
are spinors. Specify to the IIB theory, the spinors are
$S^a (z)$ and $\tilde{S}^a (\bar{z})$ of the same chirality of the
Clifford algebra of $SO(8)$. Take the light-cone coordinates as 
two coordinates tangent to the D-brane. There $9-p$ transverse directions
to the brane, denoted by $X^{p+1},\dots, X^9$. These indices are part of
$SO(8)$. The boundary conditions
for these spinors are $S(\sigma=0)=\tilde{S}(\sigma=0)$, 
$S(\sigma=\pi)=\gamma^{p+1,\dots,9}\tilde{S}(\sigma=\pi)$. Of course
for this condition to be consistent, $p$ must be odd. A similar statement
for the boundary conditions for spinors in the IIA theory holds, and
in this case $p$ is always even and the two spinors have the opposite
chirality of $SO(8)$.

It is now easy to see which part of SUSY is preserved by the presence of
the D-brane, and which part is broken. The SUSY generators are constructed
in string theory by using contour integrals involving world-sheet spinors.
From each spinor $16$ SUSY generators can be constructed. Each set forms
a Majorana-Weyl spinor of $SO(9,1)$. It follows from the spinor boundary
conditions that only one set of SUSY survives, in other word, there is
a constraint
\eqn\bsusy{Q=\gamma^{0,\dots,p}\tilde{Q},}
on the two sets of SUSY generators. the above agrees with the results obtained
from M-brane considerations. This result can also be derived using
T-duality. We saw that under T-duality transformation, $\tilde{X}^i(
\bar{z})\rightarrow -\tilde{X}^i(\bar{z})$. In order to preserve the 
world-sheet supersymmetry (which is gauged and should not be broken),
$\tilde{\psi}^i(\bar{z})\rightarrow -\tilde{\psi}^i(\bar{z})$. Upon 
quantization, the zero modes of $\psi$ and $\tilde{\psi}$ become two 
sets of gamma matrices. The effect of changing the sign of the gamma matrix 
$\tilde{\gamma}^i$ on the conserved supercharge is to add or remove
the corresponding factor in \bsusy, depending on whether the new D-brane
is wrapped around the T-dual circle or not.

\subsec{Some exact formulas}

To check whether the definition of D-branes will result in the formulas
for the brane tension
we deduced using M-branes, one needs to calculate interactions between
two parallel D-branes. The force is mediated by open strings stretched
between two D-branes. More precisely, one needs to compute the one-loop
vacuum amplitude, see fig.2. By the s-t channel duality, this cylinder
diagram can be viewed as the tree diagram for closed strings. Physically,
this diagram represents the process of emission of a closed string by
one D-brane and the subsequent absorption of this closed string by the
other D-brane.

{\vbox{{\epsfxsize=2in
        \nobreak
    \centerline{\epsfbox{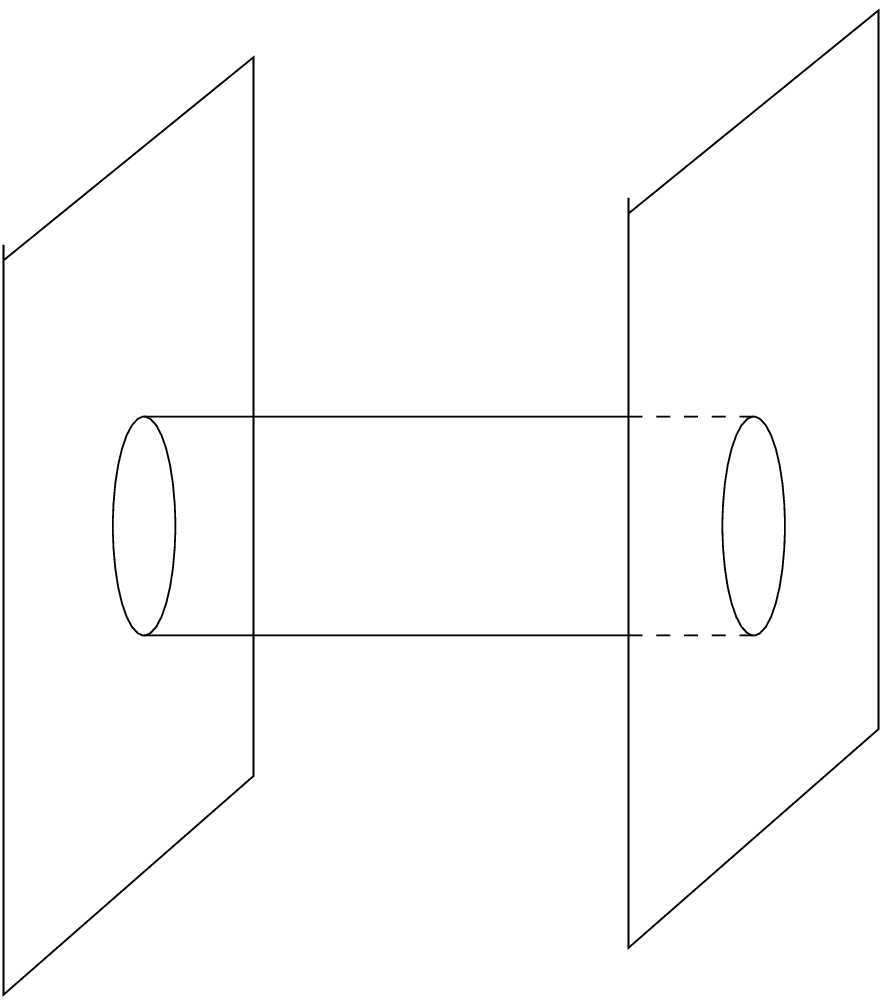}}
        \nobreak\bigskip
    {\raggedright\it \vbox{
{\bf Figure 2.}
{\it The one-loop vacuum diagram of an open string stretched between two
parallel two D-branes. This is the cause of the interaction between these
branes .}
 }}}}}
\bigskip

The one-loop amplitude for open strings stretched between two Dp-branes
is 
\eqn\ampl{A=V_{p+1}\int{d^{p+1}k\over (2\pi)^{p+1}}\int {dt\over 2t}\sum
\pm e^{-2\pi\ap t(k^2+M^2)},}
where $V_{p+1}$ is the volume of the world volume, and can be set to be
finite by an infrared cut-off. The sum is taken over all possible states
of open strings, for a boson, the plus sign is taken, and for a fermion
the minus sign is taken. All the bosons live in the NS sector, and all the
fermions live in the R sector. Note the factor $2\pi\ap$ in the
exponential is chosen for convenience.

The open string spectrum can be determined by imposing the standard
one-shell conditions $L_0-a=0$, where the constant $a$ depends on the
sector. The on-shell condition then implies $M^2=X^2/(2\pi\ap)^2+
\hbox{oscillators}$, where $X$ is the separation between the two branes.
The first term reflects the fact that for
a stretched string without oscillator modes, the mass is given by
$TX$. One also need to execute the GSO projection. 
After a little calculation one finds
\eqn\amp{A=V_{p+1}\int {dt\over 2t}(8\pi^2\ap t)^{-(p+1)/2}
e^{-X^2/(2\pi\ap )}f_1^{-8}(q)\left(-f_2^8(q)+f_3^8(q)-f_4^8(q)\right),}
where $q=e^{-\pi t}$ and
\eqn\ellip{\eqalign{f_1(q)&=q^{1/12}\prod_{n=1}(1-q^{2n}),\cr
f_2(q)&=\sqrt{2}q^{1/12}\prod_{n=1}(1+q^{2n}), \cr
f_3(q)&=q^{-1/24}\prod_{n=1}(1-q^{2n-1}), \cr
f_4(q) &=q^{-1/24}\prod_{n=1}(1+q^{2n-1}).}}
Due to the Jacobi identity for theta functions, the sum in the parenthesis
of \amp\ vanishes identically. This is not surprising, since we know
that the parallel D-branes break only half of supersymmetry, therefore
the standard no force condition between two BPS state is satisfied.

Although the total one-loop amplitude vanishes, one still can glean some
nontrivial information from \amp. This is because, as we pointed out
earlier, that the dual closed string channel contains exchange of many
closed string states. When the separation $X$ is large, the exchange of
massless closed string dominates. As in a gravitation theory, there is
exchange of graviton, plus dilaton in string theory, these are states
in the closed string NS-NS sector. If the D-branes are charged with
respect of R-R tensor field, there is exchange of this massless state.
In fact, after switching to the closed string channel, the term $f_4$
in \amp\ corresponds to contribution of the R-R sector. The large
separation behavior of \amp\ is governed by the small $t$ region. Using
the asymptotics of theta functions, one finally finds
\eqn\cch{A_{NS-NS}=-A_{R-R}=V_{p+1}2\pi(4\pi^2\ap)^{3-p}G_{9-p}(X^2),}
where $G_{9-p}$ is the Green's function in the $9-p$ dimensional 
transverse space.

The Dp-brane is coupled to the R-R field $C^{(p+1)}$ in the form
\eqn\rcoup{\int d^{10}x{1\over 2\times (p+2)!}F_{\mu_1\dots
\mu_{p+2}}F^{\mu_1\dots\mu_{p+2}}+\mu_p\int C^{(p+1)},}
that is, the $p+1$ form can be integrated over the $p+1$ dimensional
world-volume. Because of this coupling, there is a Coulomb like
potential between two parallel Dp-branes induced by the R-R field.
Comparing this effect with the direct calculation of one-loop
amplitude, we deduce
\eqn\char{\mu_p^2=2\pi (4\pi^2\ap)^{3-p}.}
As an immediate consequence, $\mu_p\mu_{6-p}=2\pi$, the minimal
Dirac quantization condition.

The brane tension determines the interaction strength caused by
the exchange of graviton and dilaton. The action of the graviton and
dilaton was given in sect.2. The exchange of a graviton takes
the form of Newton potential in the limit of large separation,
thus its strength is proportional to $\kappa^2T_p^2$, where
$\kappa^2$ is proportional to the Newton constant. The exchange
of a dilaton is also an attractive force, and practically doubles
the above effect, so we have $2\kappa^2T_p^2=\mu_p^2$. Finally,
the relation between $\kappa^2$ and the string coupling constant
in 10 dimensions is $\kappa^2=2^6\pi^7g^2\ap^4$. We obtain
\eqn\tension{T_p={2\pi\over g}(4\pi^2\ap)^{-(p+1)/2}.}
The scaling in the string tension agrees with what we expected.
We have set the convention for tension such that for a D-string
$T_1=T/g$, $T$ is the string tension. For a D0-brane, the tension
is the mass $M=T_0=1/(\sqrt{\ap}g)$.

\subsec{D-brane world-volume theory}

In type I string theory, the vertex operator for an abelian gauge
field is defined by
\eqn\vop{V_\xi =\xi_\mu\p_t X^\mu e^{ik\cdot X},}
where the operator is defined on the boundary of the world-sheet.
$\p_t$ is the tangent derivative along the boundary, given the
Neumann boundary condition $\p_nX^\mu =0$. $\xi_\mu$ is the 
polarization vector.

For open strings attached to a Dp-brane, it is still possible to define
a vertex operator as in \vop, provided $X^\mu$ coincide with the
world-volume coordinates, the longitudinal directions. The only difference
is that $k^\mu$ must lie along the longitudinal directions too, since
only these world-sheet scalars have zero modes. Thus for
a Dp-brane, we know there is an abelian gauge field living in the
world-volume. This is the case in type II theories, since the only
consistent Chan-Paton factor associated to a single D-brane is the
$U(1)$ factor. In type I theory, the story is a little more complicated,
and we shall not attempt to explain it here.
If $X^i$ is one of the transverse coordinates, it is no longer possible 
to define a vertex operator as in \vop, since $\p_tX^i=0$ according
to the Dirichlet boundary condition. However, the following vertex
operator still has the correct conformal dimension
\eqn\hop{V=\xi_i\p_n X^ie^{ik\cdot X}.}
It represents a quanta of a scalar field $\phi_i$ on the world-volume.

In fact the above exhausts all the bosonic massless states on a single
D-brane. Since the D-brane preserves half of supersymmetry, therefore
there must be fermionic parters of these bosonic fields. In type
I theory, these are gauginos, quanta of a Majorana-Weyl fermion.
For open strings attached to a D-brane, there is no essential 
modification for the boundary conditions of world-sheet fermions,
and we expect to have the same content of fermions. These fermions
are just the dimensional reduction of the 16 component gaugino
field in 10D. Thus, for a D-brane in type II theory, the massless
supermultiplet living on the brane is the dimensional reduction
of the 10 dimensional $U(1)$ super Yang-Mills multiplet.

This vector multiplet can be deduced based on the general Goldstone
theorem. The presence of a Dp-brane breaks the translational invariance
in the $9-p$ transverse directions, there must be corresponding Goldstone
modes. These modes must be localized on the brane, since a local
fluctuation of these modes represents the local transverse position
of the brane. If there were such modes propagating in the bulk, these
modes would have to be included in the theory without the presence of
the D-brane. The presence of the brane also breaks $16$ SUSY's, therefore
there are $16$ fermionic Goldstone modes. When on-shell, these modes
obey the Dirac equation, so there are only $8$ on-shell such modes.
However, $8$ fermionic
modes and $9-p$ bosonic modes can not furnish a representation of the
unbroken SUSY's, there must be $p-1$ additional bosonic modes, and this
number is just the number of degrees of freedom  encoded in a massless 
vector field in $p+1$ dimensions.
\bigskip
{\vbox{{\epsfxsize=2.5in
        \nobreak
    \centerline{\epsfbox{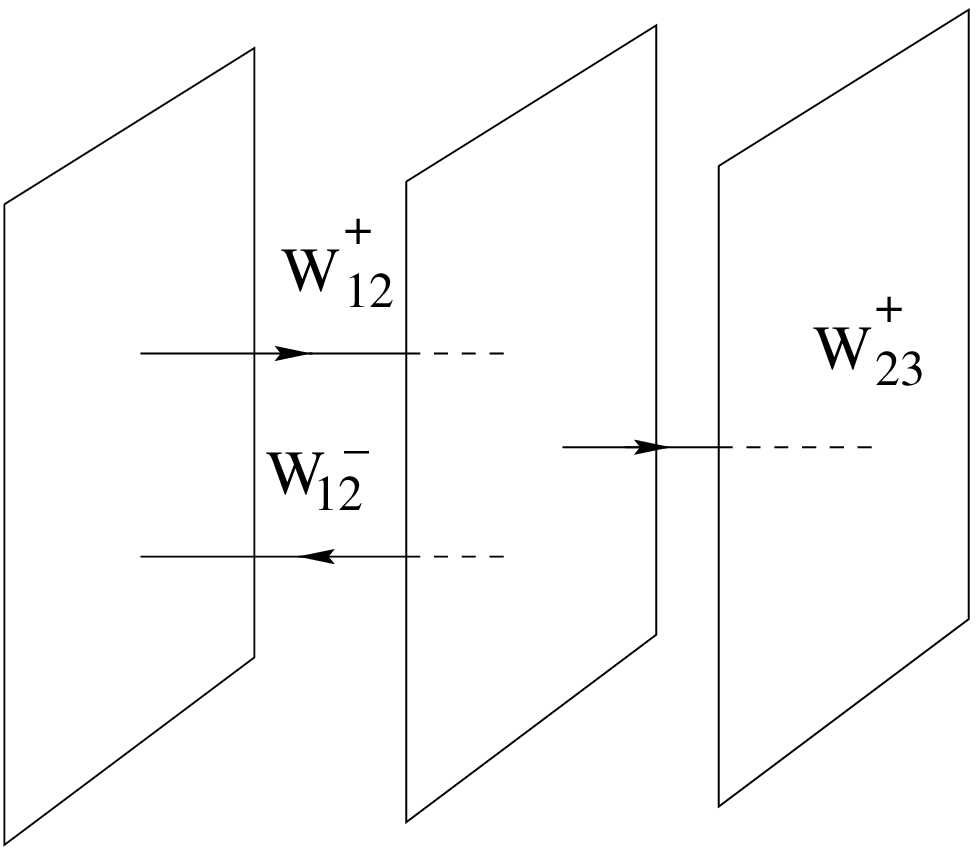}}
        \nobreak\bigskip
    {\raggedright\it \vbox{
{\bf Figure 3.}
{\it The geometric realization of Higgs mechanism by parallel D-branes. 
Massive W bosons are just stretched strings between D-branes, such a string
is charged under the difference of two U(1) groups corresponding to the two
D-branes.}}}}}}
\bigskip

The low energy effective action is similar to \sym, except that we should
replace $2\pi\ap A_i$ by $\phi_i$. Before we write down such an action,
we turn to the case of multiple parallel D-branes. Each D-brane contributes
a $U(1)$ Chan-Paton factor, therefore there are at least a gauge group
$U(1)^N$, $N$ is number of branes. This can not be the whole story,
as we have seen in between each pair of branes, there is a new open 
string sector which is responsible for interaction between these two branes.
The open string in a type II theory is oriented. For one orientation,
the end of an open string is positively charged under the $U(1)$ factor of the
corresponding D-brane, the other end is negatively charged under the
other $U(1)$ factor, and one can regard this as a $W^+$ mode. An open
string with the opposite orientation can be regarded as a $W^-$ mode.
All these modes can be checked to form a vector supermultiplet of the
unbroken supersymmetry. since the mass of these modes is proportional to
the separation of the two D-branes, they become massless modes when two
D-branes coincide, and we conclude that the gauge symmetry is enhanced.
In fact, there are total $N(N-1)$ such vector supermultiplet, together
with the $N$ $U(1)$ vector multiplets, they form the gauge field of
$U(N)$ group. What we have described above is a geometric realization
of Higgs mechanism: When the a pair of branes are separated, a Higgs
vev is given to the corresponding sector, and the group $U(2)$ is
broken to $U(1)\times U(1)$. The W boson has a mass proportional to
the Higgs vev which in turn is just the separation of the two D-branes.

The effective action of the abelian part, for small field strength,
is given by the same action of \sym. For a large field strength, one
has to use the so-called Born-Infeld action \rl
\eqn\bia{S_{BI}=-T_p\int d^{p+1}xe^{-\phi}\sqrt{\det (G_{\mu\nu}+\p_\mu\phi_i
\p_\nu\phi_i+2\pi\ap F_{\mu\nu})}.}
It is easy to verify that for small fluctuations the expansion of the
above action to the quadratic order is the Maxwell action plus free scalars.
It is possible to supersymmetrize the above action to include the
gaugino field. Since we are writing the action in the so-called static
gauge (the world-volume coordinates $\sigma^\mu$ are identified with 
spacetime coordinates $X^\mu$), supersymmetrization is simple. It is
also possible to work with a covariant form.
The procedure involves $\kappa$ symmetry and is quite complicated.

The Born-Infeld action gets modified when a background field $B_{\mu
\nu}$ is turned on. It is straightforward to see that, due to the coupling
$\int B$ on the world-sheet, the gauge invariance $B\rightarrow B+d\alpha$
is broken. This symmetry is restored if a gauge field $\alpha$ is 
switched on, since this field couples to the world-sheet boundary
as $\oint \alpha$. Thus, the combination $F_{\mu\nu}-B_{\mu\nu}$
is invariant under the combined gauge transformation: $B\rightarrow
B+d\alpha$, $A\rightarrow A+\alpha$. Components $B_{ij}$ decouple on
the world-sheet due to the Dirichlet boundary conditions. $F$ in \bia\
must be replaced by $F-B$.

A Dp-brane is coupled to the R-R field $C^{p+1}$. When $F$ is switched
one. some lower rank R-R fields are induced, this is verified by a 
direct calculation using the boundary state technique. We consider a
nonabelian situation. When all R-R
fields are switched on, the full bosonic action is \mld,
\eqn\nona{\eqalign{S&=-T_p\int d^{p+1}x\hbox{STr}\sqrt{\det (\eta_{\mu\nu}
+\p_\mu\phi_i\p_\nu\phi_i+2\pi\ap F_{\mu\nu}-B_{\mu\nu})}\cr
&+\mu_p\int C\wedge \tr \exp(2\pi\ap F-B),}}
where in the second term, the Chern-Simons like coupling, $C$ is the sum of
all possible R-R forms in the theory, and $F-B$ is a two form. We omitted
the nontrivial dependence on $[\phi_i,\phi_j]$. The symbol STr implies that
before the trace is taken, any term involving a product of matrices must be
symmetrized over all matrices.

\subsec{Some applications}

There are many applications of D-brane technology. The most important
is to use them to realize states predicted by various string dualities.
In many cases the predicted BPS states are bound states of D-branes
of various types.

\noindent {\it 1. D0-branes and IIA/11D supergravity duality}

Historically, the first piece of evidence for 11th dimension in the 
strongly coupled IIA string is the analysis of solitonic states
charged under $C^{(1)}$. These are super particles in 10 dimensions,
furnishing short supermultiplets of the type IIA supersymmetry.
The mass is given by $n/(gl_s)$ and is protected by SUSY. In the large
$g$ limit, these masses become light, and there is no reasonable
10 dimensional theory accommodating infinitely many massless spin
2 particles. The only natural scenario, as we have seen, is to
interpret them as KK modes of a 11 dimensional theory compactified
on a circle of radius $R=gl_s$. 

The state with $n=\pm 1$ is a D0-brane or an anti-D0-brane. The system
of N D0-branes is described, in the low energy and small $R$ limit,
by the nonabelian quantum mechanics, which is the dimensional
reduction of 10D SYM to $0+1$ dimensions. There are $9$ Hermitian
bosonic matrices. When all of them are nearly commuting, their
diagonal part can be interpreted as the positions of N D0-branes.
The new ingredient is the off-diagonal elements which have no
clear geometric interpretation. In a loose sense the system can be
said as a realization of noncommutative geometry.

The existence of the single particle state of charge $N/R$ predicts 
that there is a normalizable bound state (and super-partners)
in the N D0-brane quantum mechanics. Since the total energy of
the bound state is the sum of individual masses of D0-branes,
the binding energy vanishes. A state with vanishing binding energy
is called a threshold bound state. Logically one can not exclude
a state with the same R-R charge meanwhile is totally independent of
D0-branes, although nothing like this has been found.
From our experience with
quantum mechanics we know that the spectrum of the N D0-brane
system must be continuous, in order to have a threshold bound state. 
This fact was proven long time ago
in the context of a discretized membrane, whose dynamics happens to
coincide with the nonabelian quantum mechanics in question.
A proof of the existence of threshold bound states was found
only recently, first for $N=2$, later for a prime N. Thus there
can be no other single particle state of the same R-R charge.
However, the wave function of a threshold bound state 
constitutes a very interesting open problem. 

\noindent {\it 2. $(p,q)$-string and IIB S-duality}

The $SL(2,Z)$ duality of IIB string predicts the existence of
$(p,q)$-string, with $p$ $q$ coprime. The $(1,0)$ string is
the fundamental string, and $(0,1)$ is the D-string. A $SL(2,Z)$
map can bring, say, a $(1,0)$ string to a $(p,q)$-string. The
string tension formula is
\eqn\pqt{T_{p,q}=T\sqrt{p^2+q^2/g^2},}
where we assumed that vev of the R-R scalar $C^{(0)}$ vanishes.

Witten argued for the existence of such a string as a bound state
of $q>1$ D-strings, although a more rigorous argument is still
being awaited. However, when $q=1$, the bound state $(p,1)$ can be 
realized by a D-string with a constant electric field $2\pi\ap
F_{01}=pg$, if $g$ is small enough. This can be seen by examining 
the $B$ field induced by $F$, starting from the B-I action \bia, or 
simply by examining the energy of this configuration using the B-I 
action. Ignoring other fields, the B-I action of an electric field 
on the D-string is
\eqn\delec{S_{BI}=-{T\over g}\int d^2x\sqrt{1-(2\pi\ap E)^2},}
where $E=\p_t A$, and we have set $A_0=0$, the temporal gauge.
The conjugate momentum of $A$ is
$$P_A={2\pi\ap E\over g}(1-(2\pi\ap E)^2)^{-1/2},$$
where we used $T2\pi\ap =1$. The energy of per unit length is
then
\eqn\pone{E={T\over g\sqrt{1-(2\pi\ap E)^2}},}
which is equal to the desired tension formula if $2\pi\ap E=pg$
is small enough. To yield the exact formula we require
\eqn\crelec{2\pi\ap E={pg\over\sqrt{1+(pg)^2}},}
we see that there is a limit on the possible field strength which is
just $E_c=1/(2\pi\ap)=T$. This is related to the well-known
phenomenon that if the electric field is larger than the critical
value, the open string pair product rate diverges and such a state
is unstable.

From the Chern-Simons coupling of \nona, we also see that a R-R
scalar field is induced by the constant $E$ on a D-string.

\noindent {\it 3. Heterotic string as the D-string in type I theory}

Type I string theory contains a nonorientable open string sector. The 
realization of the Chan-Paton factor $SO(32)$ may be interpreted by
open string ends attached to different D9-branes. The are total $32$
D9-branes. The existence of branes breaks half of supersymmetry, the
16 supercharges are given by the combination $Q-\gamma^{0\dots 9}
\tilde{Q}$. In other words, the background induced by D9-branes puts 
constraint $Q=\gamma^{0\dots 9}\tilde{Q}$.

There is a rank two R-R tensor field $C^{(2)}$ in the theory, as we argued
in sect.2 that the solitonic string charged under this field is just
the heterotic string. Here we interpret this solitonic string as a
D-string. The introduction of a stretched d-string along, say $X^1$
introduces another constraint $Q=\gamma^{01}\tilde{Q}$. It is easy
to show that this is compatible with the constraint coming from D9-branes.
There are 8 unbroken supercharges satisfying both constraints.

The D-string introduces two open string sectors, call them the DD sector
and the DN sector respectively. The DD sector contains open strings with 
both ends attached to the D-string. As before, in the NS sector
there are possible vertex operators $A_\mu\p_tX^\mu$ and $\phi_i\p_n
X^i$, where $\mu=0, 1$ and $i=2, \dots , 9$. The exchange of the two ends
of an open string is realized by  $\sigma\rightarrow \pi
-\sigma$. So the gauge field is odd (since it is proportional to the first
oscillator) and the scalars are even. And the
gauge field is projected out for a nonorientable string. The fermions
in the R sector are subject to the same constraint on the supercharges,
therefore they are left-movers. 

The DN sector consists of open strings with one end attached to the D-string,
and another end to one of $32$ D9-branes. detailed analysis shows that
only the R sector contains massless states, which comes from quantization of 
the world-sheet fermions $\psi^\mu$. There are two states, again subject
to the constraint $\lambda=-\gamma^{01}\lambda$. Therefore there is only one
right-moving fermion from each D9-brane Chan-Paton factor. In all, there
are $32$ right-moving fermions. To summarize, the massless states from both
the DD sector and the DN sector are exactly those expected of the
``matter '' content on the world-sheet of a heterotic string.

\noindent {\it 4. D4-D0 bound states}

As the final example, let us consider the realization of bound states
of D0-branes and D4-branes. The existence of these bound states are
also required by string duality. The simplest way to see this is to
lift the IIA theory to 11 dimensions, and D4-branes are fivebranes
wrapped along $X^{11}$, and D0-branes are just momentum modes along
$X^{11}$, the state can be obtained by boosting fivebranes along a 
uncompactified $X^{11}$ and then periodically identifying $X^{11}$.
The total energy must be the sum of energy of D4-branes and that of
D0-branes. This is just a threshold bound state.

Consider an instanton solution along the 4 spatial directions of the
world-volume of D4-branes \refs{\ewit, \mld}, and 
$\int \tr F\wedge F\sim k$, $k$ is the 
instanton number. This, according to the Chern-Simons coupling in 
\nona\ generates field $C^{(1)}$. A careful computation shows that
it has exactly $k$ D0-brane charge. It is then natural to interpret
this configuration as a bound state of D4-branes and $k$ D0-branes.
It remains to check whether the binding energy vanishes.

The total energy of the system is given by the sum of the energy of the
un-excited D4-branes and $\int d^4x\tr F^2$ up to a numerical coefficient.
It is the property of an instanton solution that $\int d^4x \tr F^2 =
{1\over 2}
\int\tr F\wedge F$. Indeed the excess of energy is proportional to the
energy of $k$ D0-branes. After collecting all the coefficients it is seen
that it is equal to the energy of $k$ D0-branes.

A quantum state is obtained by quantization over the moduli space of 
instanton number $k$ in the weak string coupling limit. If the string
coupling is not small, there is a finite probability for D0-branes to
escape away from D4-branes, then quantization over the so-called Coulomb
branch, the branch describing detached D0-branes, is necessary. This
problem has not been solved, although the classical action for this
system is available.

\newsec{ The matrix theory conjecture}

\subsec{Why matrix theory works}

We have seen that the only stable states in M theory when none of the
eleven spacetime dimensions is compactified are those of supergravitons,
and multi-particle states of supergravitons. Membranes and fivebranes
are not stable, unless they are stretched along an infinite hyper-plane.
Those states have infinite energy and therefore are not visible
in any dynamical process. That supergravitons should be the only particle
states is a consequence of the eleven dimensional super Poincare
algebra. To this author, there are two possibilities that may help to
avoid this hasty
conclusion. The first is the postulate that there are some constituents
which do not form a single particle representation of the super
Poincare algebra. Rather, a spin two state, for example, is a composite
of these constituents. In particular, this implies that there is
no supersymmetry operating directly on these constituents. However, we know
of no such example in any dimensions being proposed. The closest thing
coming in mind is some effective supersymmetry in certain nuclei,
although the underlying theory has no such symmetry. The second
possibility is similar to the first in spirit. Here instead of looking
for constituents with a different symmetry structure, one might
look for a theory in which super Poincare symmetry manifests
only when one specifies the Minkowski background. In such a scheme,
one will be forced to abandon Einstein's equivalence principle, since
this principle dictates that locally there is a Minkowski frame,
and therefore there is local super Poincare symmetry. Thus, even
local Poincare symmetry would have to be a consequence of emergency
of spacetime.

Matrix theory in the above regard is a rather conservative scheme
\refs{\matri, \marev}. Here
one takes seriously the conclusion that supergravitons are the most
fundamental. Furthermore, not all supergravitons are equally fundamental.
This comes about from the D0-brane picture. We learned that D0-branes
are just supergravitons with a unit momentum on the M circle.
Higher KK modes are bound states of these D0-branes. Now, if one is to
assume that all stable objects upon compactification are composite
of D0-branes too, one is forced to focus on those states with
nonzero M momentum. Therefore, the infinite momentum frame interpretation
seems very natural with this scheme. In the IMF, every system carries
an infinite amount of longitudinal momentum. Here the longitudinal
direction is taken to be the M direction. We will always denote this
dimension by $X_{11}$.

Define the rest mass of a system through the relativistic relation
$p^2 =-M^2$, where $p$ is the eleven momentum of the system. If one of
the momentum component $p_{11}$ is much greater than others, then
$p_+=E+p_{11}\sim 2p_{11}$. The light-cone energy $E_{LC}=E-p_{11}$
is
\eqn\lcener{E_{LC}={p_i^2+M^2\over 2p_{11}},}
where $p_i$ is the transverse momentum, having 9 component.
This simple kinetic relation indicates that the system
in the IMF is a nonrelativistic system, and is the source of much
simplification of physics.

Upon compactifications, more stable states will be generated.
Due to the special kinetics in the IMF, the ability to describe
various states will impose strong constraints on the structure of
bound states. We will see that indeed the IMF kinetics is closely
related to the duality properties of the bound states. We will go
up in compact dimensions starting with a simple circle.

\noindent 1. M theory on $S^1$ \refs{\matri, \wt}

Compactifying M theory on a circle $X_9$ yields the IIA theory. This is the
first example in which we expect new states in the spectrum. As usual
the wrapped membrane around $X_9$ gives rise to a string. Such a string
state is not stable, however. The stable states are those KK modes
associated with $X_9$. These are new D0-branes. Since such a state
carries KK momentum in $X_9$, we may expect to obtain it by boosting
a D0-brane parton in $X_9$ direction. The IMF physics forces us to boost
a large collection of partons, when $R_9\rightarrow 0$, since for a
fixed $p_9$, the velocity $v_9$ would grow too large for a small mass.
We will see this picture will come out nicely in the 1+1 SYM description
of matrix theory \mtst.

As we shall see, a cut-off in $X_{11}$ would be essential for writing down
the dynamics of matrix theory. For a finite cut-off, a finite energy
state is obtained by wrapping a membrane around $X_{11}$ as well as around
$X_9$. Due to boost invariance along $X_{11}$, the light-cone
energy is expected to be independent of $p_{11}=N/R$. 
Now, if we are to hope that such a state can be regarded as a
composite state made of D0-brane partons, we must have $E_{LC}=
E-p_{11}$ to be independent of $p_{11}$ in the large $p_{11}$ limit.
Is this true? If we still regard $X_{11}$ as the M theory circle,
then we have a bound state of wrapped fundamental string with winding
$w_9$ and D0-branes. By doing T-duality along $X_9$, we have a bound
state of N D-strings and a fundamental string of momentum $p_9=w_9/
\tilde{R}_9$. This bound state is described by the Born-Infeld action
for a D-string. Physically, the momentum $p_9$ is realized by open
string modes moving along the $\tilde{X}_9$ direction on the D-strings. 
Since these
modes are massless, the net energy is proportional to $p_9$ and therefore
to $w_9$. This net energy is just
$E_{LC}=E-p_{11}$. We see that indeed $E_{LC}$ of the original state, the
wrapped longitudinal membrane, is independent of $p_{11}$.  Thus,
without running into the technical details, we already see that matrix
theory can work on a circle.

\noindent 2. M theory on $T^2$ \bs
 
Let $T^2=(X_8, X_9)$.
Various KK modes and longitudinally wrapped membranes are described
in the same way as  discussed before. Now we have a first nontrivial 
transverse state, the membrane wrapped
around $T^2$. The membrane is a transverse object and its mass should be
boost invariant. According to the general formula
$E_{LC}=E-p_{11}=M^2/2p_{11}$, where $M$ is the energy of the membrane 
in the rest frame, the light cone energy shall scale to zero in the large 
$p_{11}$ limit, and moreover it is proportional to $w^2$ where $w$ is 
the wrapping number of the 
membrane. Again this is a consequence of duality. Taking $X_{11}$ as the 
M circle, the membrane is interpreted as a D2-brane wrapped on $T^2$. 
We are studying the
bound state of this D2-brane with many D0-branes. Now the energy can be 
calculated
using the D2-brane Born-Infeld action. Again, there is an alternative and
more physical way to do this. Suppose we do T-duality along
$X_9$, we obtain a D-string of wrapping number $N$ from $N$ D0-branes. 
This D-string
is wrapped around $\tilde{X}_9$. Another D-string wrapped around $X_8$ is
obtained
from the D2-brane. The wrapping number is $w$. Naively, one would say
that the total energy is proportional to $\alpha N+\beta w$, where 
$\alpha$ and $\beta$ depends on the radii. This is wrong. We are looking 
for a ground state of a D-string with quantum numbers $N$ and $w$. 
Apparently the lowest energy state
is given by a straight D-string wrapped along a diagonal of the relevant 
torus. Thus the energy of this state is given by $\sqrt{(\alpha N)^2 
+(\beta w)^2}$.
In the large $N$ limit, we have $E-\alpha N=(\beta w)^2/(2\alpha N)$. 
Indeed $E_{LC}$ behaves as exactly what was expected. Here we see that 
the energy of the
membrane is completely soaked up by D0-branes in the large N limit. This 
is typical
of boosting a transverse object, and is the reason why interaction 
properties can be computed in the large N limit.

\noindent 3. M theory on $T^3$ \grts

We expect no new type of states. However, interpreting one of 
three circles of $T^3$
as the M circle enables one to perform T-duality along the other two 
circles to get to a new IIA theory. This T-duality will have a surprising 
manifestation in matrix theory.

\noindent 4. M theory on $T^4$ \tfour

Two classes of new states arise. Both are related to M theory fivebranes.
Wrapping a fivebrane around $T^4$ as well as the longitudinal direction 
gives rise to the longitudinal fivebrane. A string is obtained by wrapping 
a fivebrane around $T^4$ only.
The latter is not a stable state, although strings are always special 
since they can be quantized. As in the longitudinal membrane case, we 
expect that the light-cone energy of a boosted longitudinal fivebrane is 
independent of the boosting.
How does this come about? Again the reason is to be found using duality.
With $X_{11}$ interpreted as the M circle, we are studying the bound 
state of a D4-brane and many D0-branes. This bound state is a threshold 
bound state, namely the binding energy is zero. One way to see this is to 
look at the Higgs branch of the D4-brane, where D0-branes get interpreted 
as instantons in the world-volume theory.
This argument is suggestive, but not conclusive, since its nature is 
classical. A much more comforting picture is obtained by performing two 
steps of duality. T-dualing along a circle gives rise to a D3-brane and 
D-strings. S-dualing in this IIB picture we obtain a D3-brane with 
fundamental strings threading in the orthogonal
direction. This state has a vanishing binding energy according to the 
D-brane theory.

The mere existence of the above discussed threshold bound states with
the presence of many longitudinal fivebranes
indicates a hidden dimension in the matrix theory. Details are postponed.

\noindent 5. M theory on $T^5$ \seiberg

In addition to states already exist upon compactification on $T^4$, 
a new class of transverse states is derived from wrapping fivebranes 
around $T^5$. If we take $X_{11}$ as the M circle, we are talking about 
a bound state of a NS fivebrane and D0-branes. T-dualing along one circle 
of $T^5$, we obtain a bound state of D-strings
and a NS fivebrane. S-dualing in the IIB theory, we have a D5-brane and 
fundamental strings. T-dualing in the remaining four directions on $T^5$, 
we have a bound state of a D-string and parallel fundamental strings. 
Let the wrapping number of the
D-string be $w$, the total energy of this dyonic string is given by
$\sqrt{(\alpha N)^2+(\beta w)^2}$, a well-known formula. This is 
the correct answer for boosting a transverse fivebrane.

\noindent 6. M theory on $T^6$

Up to $T^5$, we have argued that the boosted longitudinal as well as 
transverse objects all have the right energy relation, using various 
duality transformations. These transformations, as will be seen, all 
have realization in matrix theory, therefore
the corresponding objects can be constructed as excitations in matrix 
theory.

Things become nasty on $T^6$. Taking one circle of $T^6$ as the M circle,
we can have a D6-brane wrapped around the remaining $T^5$. This gives 
rise to a new string, may be called KK monopole string, However, 
a KK monopole involves a nontrivial topology. 
Asymptotically, the topology of the compact space is $T^6$. At the 
core of the monopole, the M circle shrinks to a point. We do not have 
a nice matrix description of this.
We can also consider a D6-brane wrapped around $T^5$ of $T^6$ and the 
longitudinal direction. Again one wishes to be able to describe this 
longitudinal brane with many D0-branes. In discussions on tori of fewer 
dimensions, it has been always useful to interpret $X_{11}$ as the M 
circle. But now with the presence of a 6-brane we run
into trouble. Since one of the circles is not a standard one, D0-branes 
can not be T-dualized along that circle. If one T-dualizes along the 
other 5 circles, one obtains D5-branes with a transverse circle collapsing 
to a point at the core of the 6-brane. (The 6-brane is no longer a D6-brane, 
since $X_{11}$ is taken as the M circle.)
The winding number of the open string sector in the D5-brane theory is 
not conserved.

A description of the bound state of D0-branes and the 6-brane can be 
obtained only
when one performs T-duality along $X_{11}$. This is an operation that we 
have not used before. In any case, this allows one to show that indeed 
the desired energy relation for boosting a longitudinal object is valid. 
We do not know how to described a T-duality along the longitudinal 
direction in matrix theory, thus we do not know whether it is
possible to construct the 6-brane as an excitation in matrix theory. (The
recently discussed N-duality might be useful in this regard)

As we shall see later, there is a much more serious problem with matrix 
theory on $T^6$.

\noindent 7. M theory on $T^7$

To make the matter worse, let us consider compactification on $T^7$. 
Taking again one circle of $T^7$ as the M circle, and wrapping a D6-brane 
around the remaining $T^6$, this is a new transverse state. The microscopic 
picture of the bound state of this transverse 6-brane with D0-branes is 
hard to come about too. Again, the desired energy relation can be proved 
by invoking T-duality along the longitudinal direction as well as along 6 
directions in $T^7$. This combined operation yields a D-string and N 
fundamental strings wrapped around $X_{11}$, and is the same configuration 
we used to argue for the bound states of D0-branes
and a transverse fivebrane. Since a T-duality along the longitudinal 
direction is involved, this makes it difficult to construct this bound 
state in matrix theory.

\subsec{ The Hamiltonian}

The action of N D0-branes, extrapolated to large $R$ (thus strong IIA
coupling) regime, is an action of $9$ bosonic Hermitian matrices 
$X^i$ and $16$ fermionic Hermitian matrices $\theta_\alpha$, 
supplemented by a Hermitian gauge field $A_0$. The role of $A_0$ is 
to impose the $U(N)$ gauge invariance, and is also crucial for the 
existence of supersymmetry \refs{\dan, \dkps},
\eqn\mlag{S={1\over 2R}\int dt\tr\left((D_tX^i)^2+{R^2\over l_p^6}
[X^i,X^j]^2+i\theta D_t\theta -{R\over l_p^3}\theta\gamma_i[X^i,
\theta]\right),}
where $D_i=\p_t+i[A_0, $. The role of $\theta$ is to generate
a short representation of super Poincare algebra. In the $U(1)$ case,
there is a single D0-brane, $\theta$ upon quantization forms a 
$16$ dimensional Clifford algebra, therefore a spinor representation
of this algebra has dimension $2^8=256$, the one required of a 
graviton supermultiplet of 11D ${\cal N}=1$ super Poincare algebra.

The Hamiltonian of the system reads, in the gauge $A_0=0$,
\eqn\mham{H={R\over 2}\tr\left(P_i^2-{1\over l_p^6}[X^i,X^j]^2
+{1\over Rl_p^3}\theta\gamma_i [X^i\theta]\right).}
The $32$ supercharges have quite different representation in this
system. $16$ of them are ``dynamical", that is, the variables
transform nontrivially, such as $\delta X^i=\bar{\epsilon}\gamma^i
\theta$. Their anti-commutators yield the Hamiltonian. The other
$16$ supercharges are realized nonlinearly, $\delta\theta
=\eta$, $\delta X^i=0$. This is in accord with the fact that
these supercharges are broken by the presence of D0-branes, and
$\theta$'s are Goldstinos of this symmetry breaking. These
supercharges generate the short multiplet of the super Poincare
algebra. As such, their anti-commutators yield $P_{11}$, the
longitudinal momentum, since this quantity is preserved by
a given short multiplet. To summarize, we have the following
relations
\eqn\antic{\eqalign{\{Q_\alpha ,Q_\beta \}&=\delta_{\alpha\beta}H ,
\cr
\{q_\alpha,q_\beta\} &=\delta_{\alpha,\beta}P_{11},\cr
\{q_\alpha ,Q_\beta\} &=\gamma^i_{\alpha\beta}P_i.}}

The bosonic part of the super Poincare algebra consists of
the time evolution generator $H$, the boost generator $P_{11}$,
9D translation generators $P_i$, the 9D rotation generators
$J_{ij}$, and the transverse boosts $K_i$. $K_i$ and $P_{11}$
are those generators hidden in the quantum mechanical system.
The proof of symmetry generated by these operators will be
absolutely important for viability of matrix theory.

The first check that matrix theory indeed is a sensible theory of
gravity comes from the cluster decomposition property. For a fixed
N, decompose $N\times N$ matrices into $N_1\times N_1$ and
$N_2\times N_2$ blocks, where $N=N_1+N_2$. Each block describes
a subsystem. If the separation $1/ N_1\tr X_1^i-1/N_2\tr X^i_2$
is large enough, the off-diagonal blocks are heavy and must be
integrated out. This results in interaction between the two 
subsystem. The cluster decomposition requires that the interaction
tends to zero as the separation is increased. This property can be
checked by a direct calculation. It is important to have SUSY to
have this property. Note that if, say, there are only bosons,
at the one loop level the massive off-diagonal part is just a bunch
of harmonic oscillators. The interaction energy is just the
ground state energy of oscillators and it diverges lineally in
distance. It remains a deep mystery why spacetime properties
such as cluster decomposition is tied up with SUSY.

The interaction between two supergravitons with unit longitudinal
momentum is the dimensionally reduced form of Newton potential \dkps
\eqn\partp{V=cl_p^9{|v_1-v_2|^4\over R^3|r_1-r_2|^7}.}
Again a direct calculation shows indeed this is a result of the matrix
Hamiltonian. Here it is worthwhile to mention in the IIA string
context, the one-loop open string calculation gives the same result
for the term $v^4$ for both small separation and large separation.
This is basically due to some non-renormalization theorem \nonre.

If the one-loop calculation did not yield the desired potential
for two D0-branes, this would not fail matrix theory. Matrix theory
conjectures that the correct physics is to be reproduced only in
the large N limit. It is still an open problem whether this is 
true of the interaction between two large N threshold bound states,
since the wave functions are important for an actual calculation.

There is a curious generalized conformal symmetry in the D0-brane
dynamics \jy. It will be very interesting to explore consequences
of this symmetry. It is plausible that this symmetry together with
supersymmetry actually dictates all known results obtained in
the loop calculations, and implies much further results. Also, there
seems to be a link between this and 11 dimensional Lorentz boost
invariance.

There is evidence for holography, one of the main motivations for 
the matrix theory proposal, in a dynamic regime of a D0-brane gas.
We will postpone discussing this to the black hole section.

\subsec{Toroidal compactifications}

We argued in subsection 5.1 that matrix theory in principle works
for compactification on a torus, provided its dimension is not too large.
We are yet to see how the details work out. Indeed, proceeding in the
exact the same fashion as in the first subsection, we will be able 
to work out the matrix Hamiltonian on various tori.

\noindent 1. $S^1$

We argued before that the correct light-cone energy for a membrane
wrapped around $S^1$, parameterized by $X^9$, and around the 
longitudinal direction, is a result of formally doing T-duality 
along $X^9$.
D0-branes partons now become D-string partons. The length of an
individual D-string is the size of the dual circle, $\Sigma_9=
l_s^2/R_9=l_p^3/(RR_9)$. This is really a tiny size when $R$ is very
large. This is a prerequisite that these D-strings are partons 
and un-observable.

The quantum mechanics of these D-strings is given by the ${\cal N}
=8$, $1+1$ dimensional SYM on the tiny circle of size $\Sigma_9$.
There is a way to derive this $1+1$ dimensional theory from the
$0+1$ quantum mechanics formally. Imagine that compactification
of $X^9$ is effectively achieved by arranging infinitely many
images for each D0-brane, that is, start with the quantum mechanics
on the covering space of $S^1$. Thus, the rank of gauge group is
$N\times \infty$, where $\infty$ is the number of images. There
will be a new open string sector coming from open strings stretched
between a D0-brane and any image of another D0-brane, see
fig.4. Identifying images is achieved by the following periodic
conditions
\eqn\wati{\eqalign{UX^9U^{-1}&=X^9+2\pi R_9,\cr
UX^iU^{-1}&=X^i, i=1,\dots ,8,\cr
U\theta^\alpha U^{-1}&=\theta^\alpha.}}
where $U$ is a gauge transformation. A solution is achieved by the 
ansatz
\eqn\cova{X^9=i\p_\sigma I_N-A(\sigma), \quad X^i=X^i(\sigma),}
and $U=\exp (i2\pi R_9\sigma)$.  $A$, $X^i$ and $\theta^\alpha$
are $N\times N$ matrices. Note that the period of $\sigma$ is
$1/R_9$.
\bigskip
{\vbox{{\epsfxsize=2.5in
        \nobreak
    \centerline{\epsfbox{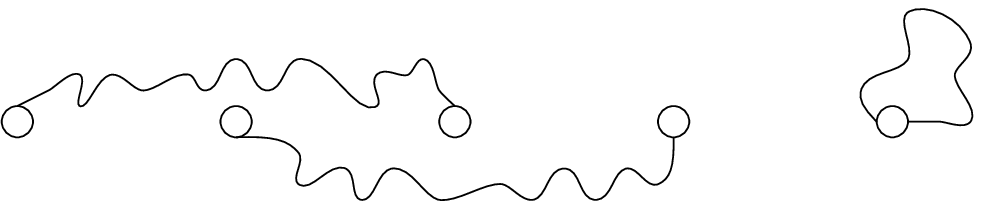}}
        \nobreak\bigskip
    {\raggedright\it \vbox{
{\bf Figure 4.}
{\it a periodic array of D0-branes and open strings stretched between
them .}
 }}}}}
\bigskip
The $1+1$ SYM thus obtained has a coupling constant $g^2_{YM}
=R^2/(R_9l_p^3)$. If we rescale the tiny circle of radius
$\Sigma_9$ to have the standard period $2\pi$, the theory is governed
by SYM with dimensionless coupling $g^2_{YM}\Sigma_9^2
=l_p^3/R_9^3$. Since the theory on a small $R_9$ is a weakly 
coupled IIA string theory, its string coupling is $g_s^2=R_9^3/l_p^3$,
we see that the effective YM coupling is proportional to
$1/g_s^2$. That is, for a weakly coupling IIA string, the YM
coupling is strong.

Now it is easy to see how a D0-brane arises as a momentum mode in $X^9$. 
$X^9$ gets replaced by the gauge field $A$, and
its conjugate momentum is just $E=\p_t A$. A constant $E$ 
configuration represents a D0-brane in the $X^9$ direction.

A IIA string is interpreted as a long string, composed of many
small D-string partons. For small $g_s$, $g_{YM}$ is large,
and the commutator $\tr [X^i,X^j]^2$ in the SYM action is weighted
by $g_{YM}^2$. To suppress this contribution, all $X^i$ must
be mutually commuting. The residual gauge symmetry is the Weyl
group $S_N$ of $U(N)$. The boundary condition for 8 scalars
$X^i(\sigma)$ can be twisted by this group. For instance,
$X^i_a(\sigma +2\pi)=X^i_{a+1}(\sigma)$. Thus many tiny D-strings
are sewed together to form a long string, fig.5. It is not hard to
see that the $1+1$ theory is capable of describing multi-string states.
Namely, we have a second quantized string theory in the
light-cone gauge.
\bigskip
{\vbox{{\epsfxsize=1in
        \nobreak
    \centerline{\epsfbox{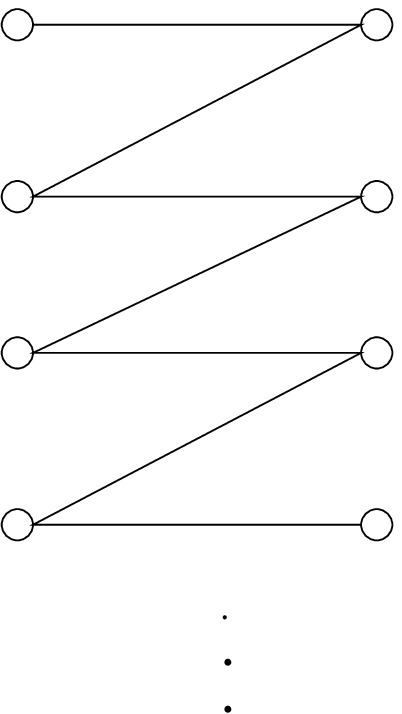}}
        \nobreak\bigskip
    {\raggedright\it \vbox{
{\bf Figure 5.}
{\it A long string in the twisted sector .}
 }}}}}
\bigskip

\noindent 2. $T^2$

The new transverse state is a wrapped membrane on $T^2$. Following the 
argument in subsection 5.1, we do T-duality along $X_9$ and obtain from
N D0-branes N D-strings. The transverse membrane becomes a D-string 
wrapped around $X_8$ $w$ times. The ground state of this system is a
long D-string wrapped around a diagonal direction. This configuration
can be obtained from N D-strings as follows. First we adjoin all N D-string
to form a single long D-string wrapped on $\tilde{X}_9$. This
is achieved by switching on a particular holonomy 
$$e^{i2\pi \tilde{R}_9A_9}=U$$
where $U$ is the t' Hooft circulation matrix $U_{ij}=\delta_{i,j-1}$. 
To wrap this long D-string
also in the $X_8$ direction, we need to switch on the expectation value
for the displacement $X_8$. Obviously, the answer is
$$X_8={wR_8\over N}\sigma 1_N +R_8w q$$
where the diagonal matrix $q$ has eigen-values $2\pi k/N$, $k=0, 1
\dots, N-1$.  

If we postulate that the matrix theory on $T^2$ is the low energy theory
of D2-branes on $\tilde{T}^2$, then both $X_8$ and $X_9$ are related to
gauge field components, $A_8$ and $A_9$. The above excitation get interpreted
in the $2+1$ SYM as a toron solution with $F\sim [X_8,X_9]\sim 1_N$.
It is easy to check that the $\tr [X_8,X_9]^2$ term gives the correct light
cone energy of the boosted membrane.

\noindent 3. $T^3$

Here an interesting phenomenon occurs. The U-duality group is
$SL(3,Z)\times SL(2,Z)$. The first factor is realized by the
geometric symmetry of $T^3$, and the second factor comes about 
a little more nontrivially. As before if we postulate that the
matrix model is the world-volume theory of tiny D3-branes, the
T-dual of D0-brane partons, then the natural explanation of
the second factor is the S-duality group of the $3+1$ ${\cal
N}=4$ SYM.

What is meaning of $SL(2,Z)$ group in M theory? Apparently it
has nothing to do with S-duality of string theory, as in the
case of compactification on $T^2$, this is contained in $SL(3,Z)$.
M theory on $T^3$ is a IIA theory on $T^2$, if we decompose
$T^3=S^1\times T^2$.  The T-duality along both directions of
$T^2$ yields another IIA theory, and this IIA theory has different
moduli. The whole T-duality group of $T^2$ is $O(2,2,Z)=
SL(2,Z)_T\times SL(2,Z)_U$. The first factor acts on the complex
structure moduli of $T^2$, therefore is part of the geometric 
symmetry $SL(3,Z)$. The second factor acts on a combination of the $B$ 
field and the Kahler moduli of $T^2$.

What we shall see here is that the "diagonal" of the S-generators
of the above two $SL(2,Z)$'s corresponds to the SYM S-duality 
generator. This diagonal is just the T-duality transformation along
both directions of $T^2$. Let the radii of the original $T^2$ be
$R_i$, $i=2,3$. The new radii of the T-dual $\tilde{T}^2$ are
$R'_i=l_p^3/(R_1R_i)$, where $R_1$ is the radius of the M circle $S^1$.

Initially, the matrix model is a large N SYM on the three
torus of radii $\Sigma_i=l_p^3/(RR_i)$, and the gauge coupling
$g^2_{YM}=l_p^3/(R_1R_2R_3)$. Assume that the new matrix model can be
obtained by starting with another M theory with Planck length
$L_p$, and infrared cut-off $R'$. The new matrix model is a matrix
model defined on a three torus of parameters 
$\Sigma_1$, $\Sigma'_2=L_p^3/(R'R'_2)=(L_p/l_p)^3 R_1R_2/R'$ 
and $\Sigma'_3=(L_p/l_p)^3R_1R_3/R'$. Since 
these are three scales in the theory, they must be equal to the 
original $\Sigma$'s up to a permutation. A little inspection shows 
that this is possible only when $\Sigma'_2=\Sigma_3$, $\Sigma'_3=
\Sigma_2$. This yields the condition 
\eqn\twom{({L_p^3\over R'})({R\over l_p^3})={l_p^3\over R_1R_2R_3}
=g^2_{YM}.}
We now show that $R'=R$. There are two more conditions: The
string scale is invariant under T-duality, so $L_p^3/R'_1
=l_p^3/R_1$; Further, $\Sigma_1=L_p^3/(R'R'_1)=l_p^3/(RR_1)$.
These two conditions are compatible if $R'=R$.
The new YM coupling is given by $(g'_{YM})^2=L_p^3/(R'_1R'_2R'_3)
=g^{-2}_{YM}$. We see that this is S-duality
of the SYM. Furthermore, the only one free parameter
$L_p$ is determined by relation which now becomes
\eqn\twomm{{L_p^3\over l_p^3}=g^2_{YM}.}

It can be further checked that the string couplings in the two IIA
theories are related by the usual T-duality relation. This is
quite nontrivial. The fact that the T-duality is valid
in matrix theory provides a check of the validity of this theory.

\noindent 4. $T^4$

Following the logic of compactification on $T^d$, $d\le 3$, we would
say that the matrix theory on $T^4$ is the $4+1$ SYM theory on 
the dual torus with radii $\Sigma_i=l_p^3/(RL_i)$. This is correct
in the low energy limit, when the energy scale is smaller than 
the light-cone energy of the longitudinal fivebrane, which is
$1/g^2_{YM}=1/\Sigma$, where
$$\Sigma ={l_p^6\over R\prod L_i}.$$

In the special dimension 4, the Yang-Mills coupling
has a length dimension. What does this length represent? In the D-brane
physics, this scale is nothing but the relevant string coupling constant
multiplied by the string length scale $l_s$, that is
$g^2_{YM}=g_sl_s$. This is nothing but the radius of the new M theory
circle. If we take $l_s$ as the one obtained from the original 
M theory, namely $l_s^2=l_p^3/R$, then $g_s$ is finite for the finite
cut-off $R$. The $4+1$ SYM is not renormalizable, therefore to 
regulate the theory we would have to take the whole string theory.
This is not in the spirit of matrix theory.

However, we want a well-defined theory which in the low energy limit
approaches the SYM with the fixed Yang-Mills coupling constant. This
is readily obtained by wrapping fivebranes around an M circle with 
radius $\Sigma$ in an M theory. Since only $\Sigma =R_{11}$ is fixed,
we can take, for example, 
$L_p\rightarrow 0$. The supergravity decouples in this limit.
This is the limit suggested by Berkooz, Rozali and Seiberg. In this
limit, they argued, the theory is a $(2,0)$ superconformal theory
on the wrapped 5-branes. The reason is that the separation between
fivebranes tends to zero as $L_p\rightarrow 0$. 

Longitudinal fivebranes in the original M theory have a simple 
explanation in this matrix model. They are just momentum modes in the
new direction $\Sigma$. The U-duality group of M theory on $T^4$
is $SL(5,Z)$, and is naturally interpreted as the geometric symmetry group
of the new 5 torus.

\noindent 5. $T^5$

Here we have a new transverse state again given by a boosted transverse
fivebrane. Treating $X_{11}$ as the M circle, there are a few ways to 
proceed to construct the boosted fivebrane. One of these we already 
described in 5.1. Another way is similar to that given by Seiberg. 
T-dualing over $T^5$, we obtain N D5-branes over $\tilde{T}^5$ from 
D0-branes. The transverse five-brane gets interpreted as a NS five-brane.
Its T-dual again is an NS fivebrane. Now we need to do S-duality in the 
IIB theory, we obtain a world-volume theory of N NS fivebranes. And
the transverse fivebrane gets mapped to a D5-brane. The boosted 
transverse fivebrane is then the bound state of N NS fivebrane and
a D5-brane. This picture as well as the one given in 5.1 gives the
correct formula for the light-cone energy, it is not the matrix theory
on $T^5$. The reason is that the NS fivebrane world-volume theory thus
obtained is not decoupled from the corresponding string theory. 

We must take the lesson learned with compactification on $T^4$ seriously.
When one of the five circles, say, $L_5$ is large, we should get back
to the theory on $T^4$. There the theory is a (0,2) superconformal
theory on a five torus with radii $\Sigma_i=l_p^3/(RL_i)$ and
$\Sigma = l_p^6/(R\prod L_i)$. Adding a new circle amounts to adding
a new transverse circle to fivebranes already wrapped on a 5 torus.
No we run into some trouble, if we simply take the limit $L_p
\rightarrow 0$. This is simply because the new dimension, if the 
previous decoupling argument is correct, is not felt by those fivebranes
in this limit. A resolution of this paradox is to add a vanishing 
circle of radius $R_{11}$, as proposed by Seiberg, with a fixed string 
length scale
$$l_s^2={l_p^9\over R^2\prod_{i=1}^5L_i}.$$
Since $l_s^2=L_p^3/R_{11}$, we see that $L_p\rightarrow 0$ as 
$R_{11}\rightarrow 0$. We now have IIA fivebranes wrapped on $\Sigma_i$
and $\Sigma$ with a vanishing string coupling constant. Seiberg
argued that this theory decouples from the bulk string theory.
There is a problem with this decoupling argument pointed out by Maldacena
and Strominger which we shall not run into here.

The length scale of the new circle $L_5$ is encoded in $l_s^2$.
The problem here is that we are treating $L_i$ $i=1,\dots, 4$ and
$L_5$ on different footing. Is there a symmetry among all five circles?
The answer is certainly yes. Note that there is still $SO(5,5,Z)$ 
T-duality that is inhered from the IIA string theory. A special
T-duality is the one along $\Sigma$. The dual radius in the IIB theory
is given by $\tilde{\Sigma}=\Sigma_5=l_s^2/\Sigma =l_p^3/(RL_5)$.
We thus see that indeed in this IIB description all five circles
are on the same footing. This is how the formula for $l_s^2$ was
postulated in the first place.

The gauge coupling constant on the IIB NS fivebranes is given by 
$l_s^2$. With $l_s^2$ as given above, this indeed agrees with the 
gauge coupling on the D5-branes on $\tilde{T}^5$. Our picture however 
is slightly different from the one obtained by the naive T-duality 
on $T^5$, since the string coupling in this naive IIB theory is 
always finite. After making S-duality to map the 
D5-branes to NS fivebranes, the string coupling is still finite.

In a IIB theory with string coupling $g_s$, the tension of the NS 
fivebrane is 
$$T_{NS}={1\over (2\pi g_s)^2(2\pi\alpha')^3},$$
and the tension of the D5-brane is 
$$g_sT_{NS}.$$
Let $V_5=\prod_{i=1}^5\Sigma_i$. The binding energy of the bound state 
of N NS fivebrane and a D5-brane is
$$V_5[(N^2T^2_{NS}+g_s^2T_{NS}^2)^{1/2}-NT_{NS}]=
{1\over 2N/R}({\prod_i L_i\over l_p^6})^2$$
indeed agrees with the light-cone of the boosted transverse fivebrane. 

The problem here is that unlike in the usual case with finite N, 
there is no correction to the light-cone energy. Also note that  $NT_{NS}
V_5$ 
has nothing to do with the longitudinal momentum of N partons. The 
longitudinal momentum is always finite for finite $N$. While the energy of 
$N$ NS fivebranes diverges in the limit $g_s\rightarrow 0$.

\noindent 6. $T^6$

As we discussed in subsection 5.1, $T^6$ is where the first obvious
decoupling problem appears. Again in the low energy limit, the theory
should be a $6+1$ dimensional SYM on $\tilde{T}^6$ of radii
$\Sigma_i$. The gauge coupling constant is given by
$$g^2_{YM}={l_p^{12}\over R^3\prod L_i}$$
Again we postulate that the D6-branes are the ones obtained from an M theory
with Planck length $L_p$ and M radius $R_{11}$. The gauge coupling is
given by $L_p^3$. Since only this constant and the size of $\tilde{T}^6$
are fixed, we are free to adjust $R_{11}$.

If we adopt the matrix theory proposal of Seiberg on $T^5$, then the free
parameter is fixed. Picking out $\Sigma_6$ from $\tilde{T}^6$, we
postulate that the matrix model reduces to the one on $\tilde{T}^5$
when $\Sigma_6$ shrinks to zero. In such a case we have D5-branes
on $\tilde{T}^5$. The string coupling constant diverges, since the string
coupling constant on the S-dual NS fivebranes vanishes. This implies
that $R_{11}$ diverges \sens. More precisely, the string coupling on D5-branes
is $g_s=R_{11}/\Sigma_6$.

\newsec{ Quantum properties of black holes realized in M theory}

Black holes present much enigma about issues in generalized 
thermodynamics including gravitation and quantum gravity. Many
of features are universal, regardless what is the underlying 
microscopic theory. For instance, with quite weak conditions,
it can be shown that a gravitationally collapsing system forming
an event horizon will 
eventually develop singularities in spacetime, thus physical laws
without gravity break down there. The strength of gravity 
becomes order 1 there, thus the usual semi-classical gravity
picture is not reliable and a quantum theory is a necessity.

The geometric entropy, first proposed by Bekenstein, is another
universal property of black holes \beken. Independent of spacetime 
dimensionality and the type of the black hole, the entropy is proportional
to the area of horizon. Classically, to a distant observer,
a black hole can carry a few conserved charges, in addition
to its mass and angular momentum, therefore the enormous amount
of entropy is inaccessible at the semiclassical level for most
of time. However, a black hole is not an absolutely stable state,
it radiates all kinds of particles, first discovered by Hawking.
This poses the well-known information problem, since most of
time the Hawking radiation can be treated semiclassically, and
no correlation between the particles radiated and the lump of
mass forming the black hole in the first place can be distangled
in Hawking's calculation.

For many particle physicists, in particular string theorists,
a perfect quantum evolution process must be involved in gravitational
collapse and black hole evaporation. It has been argued that
the theory of quantum gravity is so unusual that for an outside
observer, the semiclassical treatment breaks down near the horizon,
although to a geometer there is nothing unusual there for a 
large black hole. If string theory, and more recently M theory
or matrix theory is the correct theory of quantum gravity, such
a scenario must work out. Even when this is the case, some people
might argue that this is just a consistency check, not necessarily
implies that M theory is the only theory that is consistent with
quantum properties of black holes. But, as we shall see, what a
consistency check it is.

\subsec{D-brane black holes and matrix black holes}

D-brane technology is most powerful when dealing with a BPS black
hole. It can be proven that in order to construct a black hole with
a nonvanishing horizon area from a BPS state, there must be at least
three different charges carried by the black hole. By different we
mean that there exists no duality transformation to reduce them
into fewer charges. The first example was constructed by Strominger
and Vafa \dbh. This is a 5 dimensional black hole carrying electric and 
magnetic charge of a R-R field and a KK charge. Starting with IIB 
theory and compactifying it on $T^5$, there is
a abelian gauge field $C^{(2)}_{a\mu}$ resulting from $C^{(2)}$.
A wrapped D-string along $X^a$ carries its electric charge. 
Another abelian gauge field, $C^{(6)}_{1,\dots,5,\mu}$ results
from the dual of $C^{(2)}$. A D5-brane wrapped around $T^5$ carries
its electric charge. Apparently, the two SUSY conditions
$\epsilon=\gamma^{0a}\tilde{\epsilon}$ and $\epsilon=\gamma^{01\dots
5}\tilde{\epsilon}$ are compatible if $a$ is one of $1,\dots, 5$.
That is, the bound state of $N_5$ D5-branes wrapped around 
$T^5$ and $N_1$ D-strings wrapped around a circle of $T^5$ is a 
BPS state. Take $a=1$.

We need one more charge to construct a black hole. This is achieved
by adding momentum modes along $X^1$, namely along the D-string
direction. This introduces a further constraint on unbroken SUSY 
$\epsilon=\gamma^{01}\epsilon$, $\tilde{\epsilon}=\gamma^{01}
\tilde{\epsilon}$. This means that both $\epsilon$ and $\tilde{
\epsilon}$ are positive eigenstate of $\gamma^{01}$. Combined
with the D-string constraint, $\epsilon=\tilde{\epsilon}$.
Thus there are 8 unbroken super-charges. Finally the D5-brane
constraint eliminates half of them. The BPS black hole preserves
$4$ super-charges.

To see that this is a black hole, we need the metric.
\eqn\tend{\eqalign{ds^2&=(H_1H_5)^{-1/2}(-dt^2+dX_1^2+(H_p-1)(dt-dX_1)^2)+
H_1^{1/2}H_5^{-1/2}(dX_2^2+\dots +dX_5^2)\cr
&+(H_1H_5)^{1/2}(dr^2+
r^2d\Omega_3^2),\cr
e^{2\phi}&=g^2H_1H_5^{-1},}}
where $H_i$ are harmonic functions in 5 dimensions, 
$H_i=1+r_i^2/r^2$, where the parameter $r_i^2$ is proportional
to the corresponding charge. And the R-R fields
\eqn\rrf{C^{(2)}_{01}={1\over 2}(H_1^{-1}-1),\quad
F_{ijk}={1\over 2}\epsilon_{ijkl}\p_l H_5,}
where $i,j,k,l$ are indices tangent to the 4 open spatial dimensions. 
Let $(2\pi)^4V$ denote the volume of $T^4$ orthogonal to the D-strings,
and $R_1$ the radius of $X^1$. It is easy to see that
\eqn\param{r_1^2={gN_1\over V},\quad r_5^2=gN_5,\quad 
r_p^2={g^2N_p\over R_1^2V},}
where the momentum along $X^1$ is $N_p/R_1$. We have set $2\pi\ap=1$.
For fixed $V$ and $R_1$,
we see that all sizes $r_i$ become macroscopically large when
$gN_1\gg 1$, $gN_5\gg 1$, and $g^2N_p\gg 1$. We call this region of
the parameters the black hole phase.

When reduced to 5D, the Einstein metric reads
\eqn\fived{ds^2=-(H_1H_5H_p)^{-2/3}dt^2+(H_1H_5H_p)^{1/3}\left(
dr^2+r^2d\Omega_3^2\right).}
From the component $G_{00}$ we see that $r=0$ is the horizon, since
the red-shift factor becomes infinity at this point.
The Bekenstein entropy is easy to calculate, either by using the 8 
dimensional horizon if the hole is treated as living in 10 dimensions,
or by using the 3 dimensional horizon when it is treated as living in 5 
dimensions. It is relatively simpler to use the 5D metric. The horizon
area is given by $2\pi^2[r^2(H_1H_5H_p)^{1/3}]^{3/2}$ when the
limit $r\rightarrow 0$ is taken.  Thus $A_3=2\pi^2r_1r_5r_p$,
and $r=0$ is not a point. The 5D Newton constant is $G_5=
g^2/(4VR_1)$, so the entropy is
\eqn\entr{S={A_3\over 4G_5}=2\pi\sqrt{N_1N_5N_p},}
a nice formula. 

It can shown that all 5 dimensional black holes preserving $1/8$
of supersymmetry can be rotated into above black hole using U-duality,
here the U-duality group is $E_6$. If one can count the entropy
microscopically for one of them, then others must have a microscopic
origin too on the count of U-duality. For instance, a 5D black hole
in IIA theory is obtained by performing T-duality along $X^1$. The
hole is built with D4-branes, D0-branes bound to them, and string
winding modes around the dual of $X^1$. Now this has a simple M 
theory interpretation, the D4-branes get interpreted as fivebranes
wrapped around the M circle, winding strings get interpreted as
membranes wrapped around the M circle, and D0-branes are M momentum
modes. Thus, the hole is built using fivebranes intersecting
membranes along a circle with momentum modes running along this
circle.

Come back to the IIB 5D black hole. The simplest account of the
microscopic picture goes as follows. The D-strings are bound to 
D5-branes, and they live on the Higgs branch in the weak string
coupling limit, thus can oscillates only in the 4 directions along
D5-branes. If the size of $V$ is much smaller than $R_1$, the
oscillation is effectively described by a 1+1 conformal field theory.
The fluctuations correspond to wiggling of the open strings
stretched between D5-branes and D-strings, thus there are $4N_1N_5$
such bosons. Due to supersymmetry, there are also the same number of
fermions. The theory is therefore a conformal field theory with
central charge $6N_1N_5$. Since in a CFT a fluctuation is either
left-moving or right-moving, and we restrict our attention to BPS
states, there are only right-moving modes which contribute
to the total momentum $N_p/R_1$. Thus, $N_p$ is the oscillator
number. We are therefore interested in the coefficient of
$q^{N_p}$ in the expansion of the following partition function
\eqn\partiti{Z=\left(\prod_{n=1}{1+q^n\over 1-q^n}\right)^{4N_1N_5},}
and it is given, after a saddle point calculation, by 
$\exp(2\pi\sqrt{N_1N_2N_p})$, that is, the entropy agrees exactly
with \entr.

There is subtlety involved in the above calculation, makes it
invalid for large $N_1$ and $N_5$. A cure of this problem is
provided by the fractionation mechanism, whose details we will not 
run into here.

D-brane physics provides for the first time ever a microscopic
account of Bekenstein entropy. Even more surprisingly, further calculations
show that the usual string amplitudes associated to open strings 
colliding and combining into a closed string state reproduce
the Hawking radiation, and the greybody factor which takes the black
hole geometry into account \radiation.

As we already explained, the natural realization of the 5D black hole
in M theory is the intersection of fivebranes and membranes along the
M circle with momentum running along this circle. This in turn gets
interpreted in matrix theory. It is a 6 dimensional black string stretched
along the longitudinal direction. Here the matrix theory is described
by a $5+1$ dimensional SYM in low energy limit, the rank of gauge group
is just the number of D0-branes. Longitudinal fivebrane appears as
an instanton solution in a $4+1$ SYM theory, thus appears as an instanton
string in the $5+1$ SYM theory in question \mbh. Longitudinal membranes
are translated into momentum modes in SYM, which in turned can be carried
by the instanton string. For entropical reason, the string would like
to form a single long string. To break supersymmetry, one can even add
anti-fivebrane, or anti-instanton strings here. These are realized by
a long string that sometimes goes backward. One can also add anti-membranes,
or anti-momentum modes. Finally, the black hole is represented as
a single oscillating Hagedorn string, as indicated in the following
diagram
\bigskip
{\vbox{{\epsfxsize=3in
        \nobreak
    \centerline{\epsfbox{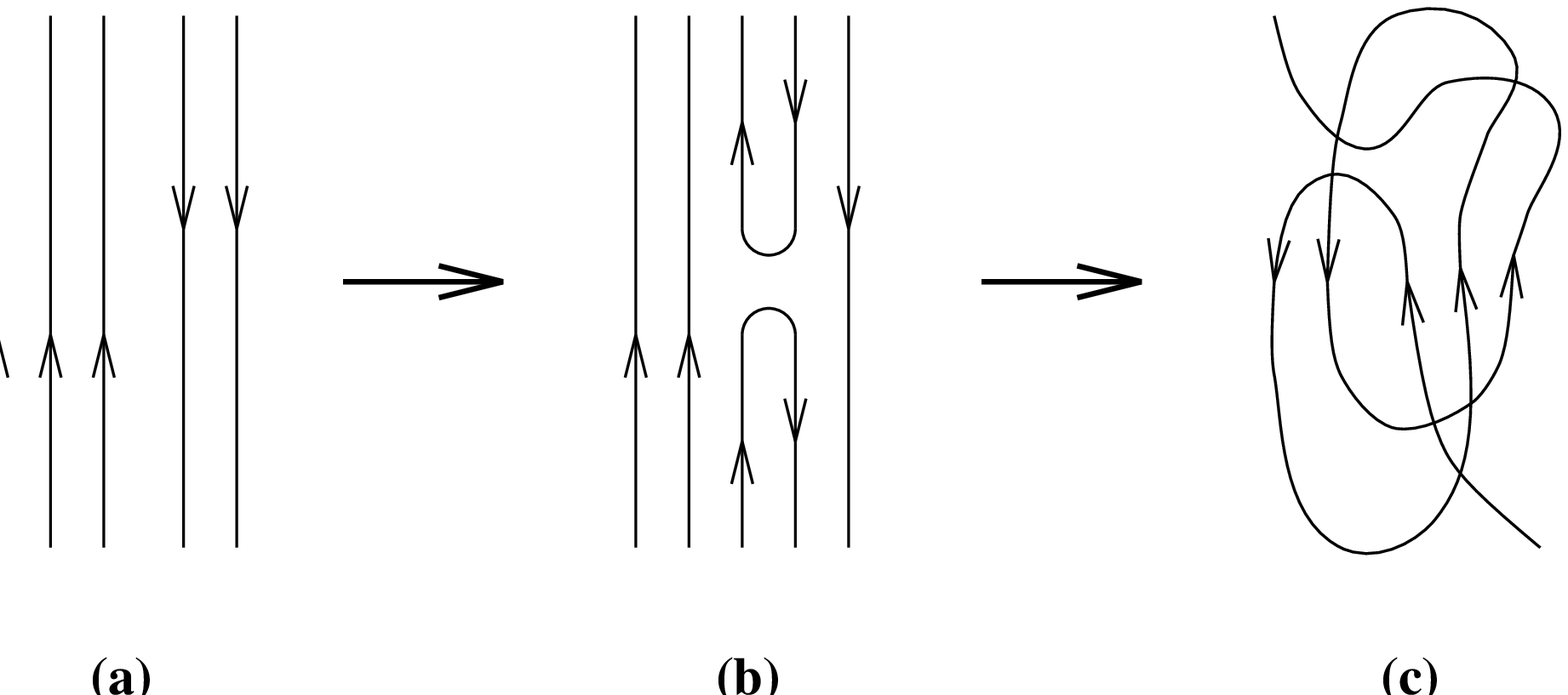}}
        \nobreak\bigskip
    {\raggedright\it \vbox{
{\bf Figure 6.}
{\it Maldacena's picture of the gas of `instanton strings'.
Through repeated joining/splitting interactions, the energy
is collected into the entropically preferred state --
one large string.}
 }}}}
    \bigskip}

The total ADM energy of the system is
\eqn\adm{
  l_p E_{ADM}=N{l_p\over R}
	+(N_2+N_{\bar 2}){RR_5\over l_p^2}
	+(N_5+N_{\bar 5}){RV\over l_p^5}\ .
}
The energy of the system not
carried by zero-branes is available to the string, 
since the IMF energy equals
\eqn\imfen{
  E_{LC}=p_+= E_{ADM}-{N\over R}.}
Note also that this energy is the Hamiltonian
of the 5+1 gauge theory.  The energy available to oscillators
of the instanton string is reduced by the constraint
that the black hole carry net two-brane and five-brane
charge, which are carried on the string as momentum
$l_p P=(N_2-N_{\bar2}){RR_5\over l_p^2}$ and winding
$l_p W=(N_5-N_{\bar5}){RV\over l_p^5}$.
Meanwhile, we treat the instanton string as noninteracting.
Then the left and right excitation numbers are 
\eqn\nlr{\eqalign{
  n_{L,R}=~&\ap_{eff}[E_{LC}^2 - (P\pm W)^2]\cr
	=~&{\ap_{eff} \over l_p^2}[{VRR_5\over 4l_p^8}
		 r_0^2]^2
		[(\cosh^2\sigma+\cosh^2\gamma)^2-(\sinh^2\sigma\pm\sinh^2
\gamma)^2]
\cr
=~&{\ap_{eff}  \over l_p^2}[{VRR_5\over 4l_p^8}
		 r_0^2 ]^2 4\cosh^2(\sigma\mp\gamma) .\cr
}}
The entropy is now evaluated as
\eqn\bhent{\eqalign{
  S=&2\pi [\sqrt{{1\over 6}{c_{\rm eff}}n_L} 
		+ \sqrt{{1\over 6} {c_{\rm eff}}n_R}]\cr
	=&2\pi [{ \ap_{eff}\over l_p^2}\cdot
		{VR_5R^2\over l_p^7}]^{1/2}
		(\sqrt{N_2}+\sqrt{N_{\bar{2}}})
		(\sqrt{N_5}+\sqrt{N_{\bar{5}}}).
}}
One must have $\ap_{eff}={N l_p^9\over VR_5R^2}$ to match the entropy.
Naturally, the energy per unit length of an instanton string
in $5+1$ gauge theory is 
\eqn\tension{
  T_{\rm eff}={4\pi^2\over g_{YM}^2 N}={VR_5R^2\over 2\pi Nl_p^9}
}
(the $1/N$ arises from the charge fractionalization mentioned above).
Then with the standard relation $T_{eff}=(2\pi\ap_{eff})^{-1}$,
the Hagedorn gas of instanton strings precisely accounts
for the Bekenstein-Hawking entropy.  
The combination $g_{YM}^2 N$ appearing in \tension\
suggests that conventional large-N techniques might be
useful for the study of the instanton string gas.
It is important to note that the factors 
in the entropy cannot be ascribed to
particular branes/antibranes; everything gets mixed up
in the `plasma' of light excitations, as we see from figure 1.
Another important feature is that the tension $\tension$
is finite in the limit $N,R\rightarrow\infty$,
$N/R^2$ fixed that characterizes the large $N$ limit
with fixed longitudinal momentum density and fixed
entropy per unit length.

\subsec{Matrix Schwarzschild black holes}

The most common black holes, the Schwarzschild black holes which
may exist in nature, have resisted understanding even in the
D-brane context. The difficulty in using D-brane technology to
deal with a neutral black hole stems from the fact that one need
both branes and anti-branes in order to keep the object neutral.
To have a macroscopic black hole, an appropriate combination of 
string coupling and the number of branes must be large. There
is no known world-volume theory which describes both branes and
anti-branes, leaving alone the strong coupling problem.

Matrix theory provides a unique opportunity to understand quantum
properties of Schwarzschild black holes \refs{\msbh, \lnmbh}. 
By boosting a hole with
an extremely large longitudinal momentum, one effectively puts
the hole against a background of a large number of D0-branes which
are BPS states. We already saw the advantage of this scheme in the
last subsection when we dealt with the 6D black string, there
one can include both fivebranes and anti-fivebranes, membranes and
anti-membranes. Here we shall show that matrix theory in principle
can be used to deal with neutral black holes in dimensions higher
than 4.

The first observation, due to Banks et al., is that for a finite
longitudinal cut-off $R$ and a black hole of radius $r_s>R$,
it is necessary to boost the hole in order to fit it into the
asymptotic box size $R$. Asymptotically, one can apply the Lorentz
contraction formula $r_se^{-\alpha}$, where $\alpha$ is the rapidity
parameter, roughly equal to $M/P_{11}$. The minimal boost is
determined by $r_se^{-\alpha}=R$, or $P_{11}=r_sM/R$. In matrix
theory, $P_{11}=N/R$, where $N$ is the number of partons. 
The above formula says that $N=r_sM\sim S$, where $S$ is the entropy
of the hole. This condition then says that the minimal number
of partons required to account for entropy $S$ is just $S$,
a physically appealing claim.

Geometrically, one might wonder how the Lorentz contraction could
happen to a horizon, since by definition horizon is a null surface
which is independent of the coordinates used. Indeed it can be shown
that in the boosted frame, the size of horizon remains the same.
What the boost does to the black hole is to change the relation
between the size of the horizon and the asymptotic radius of the
longitudinal direction, if the hole is put on a periodic circle.
It can be shown that for the horizon size to be $r_s$ while the
asymptotic box size to be $R$, the hole must carry a minimal
momentum as determined naively in the last paragraph.

Another point we want to emphasize here is that when the size of
the hole fits the box size, it looks more like a black string.
Indeed, a black string becomes instable at the special point
$N\sim S$. Since the horizon area of the black hole of the same
size and same momentum is greater than that of the black string
when one slightly increases the momentum, the black string will
collapse to a black hole. 

We will be able to explain the size of the hole and its entropy only
up to a numerical coefficient, thus whenever we write down a formula
that is valid only up to a numerical coefficient. In D dimensional
spacetime, the size of the Schwarzschild black hole and its entropy,
written in terms of the mass are given by
\eqn\sbh{r_s^{D-3}=G_DM, \quad S={r_s^{D-2}\over G_D}
=G_D^{{1\over D-3}}M^{{D-2\over D-3}}.}

At the special kinetic point $N\sim S$, we use the second relation 
in \sbh\ to solve $M$ in terms of $N$:
$$M=G_D^{-{1\over D-2}}N^{{D-3\over D-2}},$$
thus the light-cone energy 
\eqn\lce{E_{LC}=RG_D^{-{2\over D-2}}N^{{D-4\over D-2}},}
and the size of the hole
\eqn\sibh{r_s=(G_DN)^{{1\over D-2}}.}

As we argued before, the boosted black hole at the transition point 
$N\sim S$ can be either regarded as a black string, if the longitudinal
momentum is slightly smaller than the critical value, or a black hole
if the longitudinal momentum is slightly larger. In the former case,
one needs to excite longitudinal objects such as longitudinally 
wrapped membrane in matrix theory, thus the momentum modes in the
low energy nonabelian field theory are relevant. Actually the hole 
phase is easier to account for. Only the zero modes, in other words
the motion of D0-branes in the open space, are relevant.

When the Born-Oppenheimer approximation is valid, the one-loop, 
spin-independent potential between two D0-branes is given in \dkps. The
assumption that the Born-Oppenheimer approximation is valid for a
black hole implies that the dominant part of the black hole is a
gas of D0-branes, such that for dynamic purposes one can integrate
out off-diagonal variables. In D dimensional spacetime, when M
theory is compactified on a torus $T^{11-D}$, the analogous 
potential between two D0-branes can be obtained from that of \partp\ by 
summing over infinitely many images on the covering space of the 
torus:
\eqn\ddpot{L={1\over 2R}(v_1^2+v_2^2)+{c_DG_D\over R^3}{(v_1-v_2)^4
\over r^{D-4}},}
where $c_D$ is a positive constant, $G_D$ is the D dimensional 
Newton constant. Note that the above formula fails when $D\le 4$. 
The potential becomes logarithmic in $D=4$, where the transverse 
space is 2 dimensional. This potential is not well-defined without 
a cut-off.

We assume that the black hole at the transition point is a gas of
partons, and that the temperature of this gas is so low that the
kinetic energy is bound from zero only due to Heisenberg uncertainty
principle
\eqn\uncert{v\sim {R\over r_s},}
where demanding the nonrelativistic limit requires that $R\ll r_s$,
and indeed this is our starting point.
An individual parton feels the mean field caused by the rest of the
gas. The potential energy is roughly $N(G_D/R^3)(v^4/r_s^{D-4})$. 
Equating this to the kinetic energy by virtue of the virial theorem, 
we find
\eqn\virial{{1\over 2R}({R\over r_s})^2\sim N {G_D\over R^3r_s^{D-4}}
({R\over r_s})^4,}
this yields
$$r_s\sim (G_DN)^{{1\over D-2}},$$
the desired result.

The total light-cone energy is roughly
$$E_{LC}\sim {N\over R}({R\over r_s})^2\sim RG_D^{-{2\over D-2}}
N^{{D-4\over D-2}},$$
also the desired result. It remains to show that the entropy of the
system is given by $N$. This requires that D0-brane partons in the
gas are distinguishable particles, thus obey Boltzmann statistics.
This is possible when certain backgrounds such as a membrane
is switched on. This is a quite subtle point and we will skip it.

It turns out that the large N regime can also be understood at the
semi-quantitative level, and in this case the Boltzmann statistics
is easier to justify. Let us for the moment assume the relation
$S=TE_{LC}$ in the large N case, here $T$ is the temperature of the
system. We will justify this relation later. In the large N limit, 
the hole should behave as a transverse object, thus its light-cone
energy get smaller and smaller for larger and larger N. Thus the relation
$S=E_{LC}/T=M^2R/NT$ together with the Bekenstein formula results in
\eqn\largen{M=G_D^{{1\over D-4}}(NT/R)^{{D-3\over D-4}}.}
This implies
\eqn\radn{r_s=(NTG_D/R)^{{1\over D-4}},}
and
\eqn\entn{S=G_D^{{2\over D-4}}(NT/R)^{{D-2\over D-4}}.}
Note that these relations break down for $D\le 4$.

These two relations are not independent once we assume that 
the black hole is a gas of D0-branes, or a gas of clusters of 
D0-branes. To see this, use the virial theorem which says that the
kinetic energy is the same order of the total energy
\eqn\kinet{Nm\langle v^2\rangle\sim TS=T(NT/R)^{{D-2\over D-4}}
G_D^{{2\over D-4}},}
where $m=1/R$ for a D0-brane. If the time scale associated to a 
typical velocity is related to the temperature as $1/T$, then
the typical velocity scales as $v\sim Tr_s$. Substitute this into
the above relation we deduce
$$r_s\sim (NTG_D/R)^{{1\over D-4}},$$
the correct relation. Thus one has to determine either $r_s$ or
$S$.

In the first paper of refs.\lnmbh, it is suggested that some spin
dependent forces are responsible for the scaling laws concerning
the large N black holes. Another, more general, form of interaction
is proposed in the second paper of \lnmbh\ to account for these 
laws. There it is assumed that the black hole consists of clusters
of D0-branes. Each cluster has the size $N/S$, therefore there 
are roughly $S$ clusters. Assuming that the uncertainty relation
is saturated by a cluster, namely the typical velocity of a cluster
is $v\sim 1/(r_sm)=SR/(r_sN)$, then the total kinetic energy
scales as 
\eqn\kinett{E_T= Smv^2\sim {N\over R}({SR\over r_s N})^2
=({S\over r_s})^2{R\over N}.}

The potential energy between the two clusters assumes the form
$$G_D{m_av_a^2m_bv_b^2\over Rr_{ab}^{D-4}}$$
if the exchange of supergraviton producing the potential does not
cause longitudinal momentum transfer. The above form is certainly
appropriate for D0-branes, and for threshold bound states of D0-branes.
Here we need to take one step further, to assume that for processes
in which longitudinal momentum transfer occurs the interaction 
takes the more or less the same form, then the total interaction 
energy of the gas is 
\eqn\interac{\eqalign{E_{pot} &\sim G_D\sum_{\delta p_+=0}
^{N/(SR)}\sum_{a,b}{m_av^2m_bv^2\over Rr_s^{D-4}}\cr
&\sim G_D{N\over S}S^2{S^2R\over N^2r_s^D}\cr
&\sim E_T{G_DS\over r_s^{D-2}},}}
so the virial theorem implies that $S\sim r_s^{D-2}/G_D$, the
Bekenstein formula.

It can be shown that contribution to the potential energy from
other forms of interaction is the same order as \interac.
The fact that there are about $S$ clusters suggests that these
clusters obey Boltzmann statistics. This is easy to justify
for large N, since each cluster may have some fluctuation
in its longitudinal momentum. It is also possible that some
background whose kinetic energy is negligible is responsible
for the distinguishable clusters. In all, the relation $E\sim
TS$ we appealed to above must be valid.

Although matrix theory is successfully applied to account for 
scaling laws of a Schwarzschild black hole, much work remains
to be done. For one thing, we need to understand the exact 
numerical coefficients. We also need to understand the
detailed process of black hole collapsing and evaporation, in order to
resolve the information loss puzzle. Insights may be gained if
one can reconstruct the experience of an infalling probe.
All these cry for powerful large N techniques, or even some
conceptual leaps.

\newsec{M(aldacena) conjecture}

Much of the above materials had been written around February this year,
and since then the subject of Maldacena conjecture \mald\ has taken over
the community. This is a conjecture concerning duality between
string/M theory on an anti-de Sitter background and certain large
N field theory ``living at the boundary''. In the past, the brane theory
was employed to explain some of the black hole physics. Since
Maldacena made his conjecture, the course has been reversed. We are
now learning a lot about the large N strongly coupled gauge theories
using knowledge about supergravity in anti-de Sitter backgrounds.

The emergence of  Maldacena conjecture reinforces the believe
brought about by matrix theory, that quantum gravity is encoded in
the large N super Yang-Mills theory. In a certain sense Maldacena
conjecture implies matrix theory, although the precise relation
in all dynamic situations has not been clarified. Another important
reason for studying this conjecture intensively is the possibility
of solving the confinement problem of QCD in the large N limit,
by explicitly breaking supersymmetry in the SYM.

\subsec{The conjecture}

Instead of discussing the whole range this conjecture covers, we consider
one of the most interesting cases. This is the geometry induced by
a large stack of D3-branes. The ``near horizon'' geometry is obtained
from the D3-brane metric by throwing away the 1 in the harmonic function
that enters in the solution. The metric thus reads
\eqn\nhor{ds^2=\ap\left({U^2\over\sqrt{2\lambda}}(-dt^2+\sum dx_i^2)
+{\sqrt{2\lambda}\over U^2}dU^2+\sqrt{2\lambda}d\Omega_5^2\right),}
where the new radial coordinate $U=r/\ap$ has a mass dimension, and
$\lambda=2\pi g_sN=g_{YM}^2N$ is the 't Hooft parameter.
The five coordinates $(t, x_i, U)$ map out the five dimensional
anti-de Sitter space, the other five cover $S^5$ of radius $R=
(2\lambda )^{1/4}l_s$. The D3-brane near horizon geometry \nhor\
has a global symmetry $SO(4,2)\times SO(6)$, and this is precisely
the global symmetry of the ${\cal N}=4$, $D=4$ super Yang-Mills
theory.

Maldacena boldly conjectures that the full IIB string theory in the
background $AdS_5\times S^5$ is actually dual to the ``boundary
theory'' SYM, and this conjecture was formulated more precisely
in \gkpw. Given a field $\phi$ living on $AdS_5$ ($S^5$ reduced), one can
find a gauge invariant operator ${\cal O}$ in SYM, such that 
there is a coupling $\int d^4x \phi_0{\cal O}$, where 
$\phi$ approaches $\phi_0 U^\Delta$ at the boundary
$U=\infty$, $\Delta$ is the conformal dimension
of ${\cal O}$. For a scalar field of mass $m$ measured in
$m_s=l_s^{-1}$, there is the relation
\eqn\sdim{\Delta =2+(4+m^2\sqrt{2\lambda})^{1/2}.}
Now the exact correspondence relation is
\eqn\durl{\ln\langle e^{-\int \phi_0^I{\cal O}_I}\rangle_{SYM}
=S_{eff}(\phi^I),}
where on L.H.S. there is the generating functional for connected
correlation functions, and on the R.H.S. there is the effective
action of the whole string theory on the AdS space.

There are two interesting limits to consider. The first is the classical
supergravity limit, where one can ignore both the quantum gravity 
effects as well as massive string states. This requires the scale
of the AdS be much larger than $l_s$ and $l_p$. The first condition
leads to $\lambda\gg 1$, and the second leads to $N\gg 1$. In order
to suppress the string loop effects, $N\gg \lambda$. Thus the classical
supergravity is equivalent to the large N SYM in a strong 't Hooft
coupling limit. If one is willing to include stringy effects, but
ignore loop effects, then $\lambda$ can be arbitrary, and $N\gg
\lambda$. If Maldacena conjectures holds true in general, thus
the weak coupling regime of SYM can be approached only by understanding
the full classical string theory in a small AdS background.
Some observations similar to these were already made in the prescient
works \pres. Maldacena conjecture is so bold that for a while
since it was made in \mald, there had been confusion about the
question as to whether
the closed string sector and the open string sector are really
decoupled.

The AdS/CFT correspondence has been supported by several pieces of evidence. 
Most of evidence concerns quantities which are not corrected quantum
mechanically, for instance, the spectrum of chiral primary operators
is mapped to the KK modes in IIB supergravity, and the two point
and three point functions of these operators were computed \correl. 
Nontrivial predictions
such as the rectangular Wilson loops in the strong coupling limit \wils\
have not been verified.

Some massless fields and their corresponding operators deserve mentioning
explicitly. The massless graviton polarized in the longitudinal directions
$(x_\mu)=(t, x_i)$ is coupled to the stress tensor $T_{\mu\nu}$
in SYM. The dilaton field $\phi$ is coupled to $\tr F^2$, and the
massless R-R scalar $\chi$ is coupled to the topological term
$\tr F\wedge F$.

One interesting step towards proving the AdS/CFT correspondence is
the derivation of the ``anomalous'' conformal transformation of the
AdS space in SYM \jky. The special conformal transformation in
the 4D space is given by
$$\delta x^\mu = -2\epsilon\cdot x x^\mu +\epsilon^\mu x^2.$$
To have the metric \nhor\ invariant, both $x^\mu$ and the radial
coordinate $U$ must transform
\eqn\mtrans{\eqalign{\delta  x^\mu &= -2\epsilon\cdot x x^\mu +
\epsilon^\mu x^2 +\epsilon^\mu {2\lambda \over U^2},\cr
\delta U &= 2\epsilon\cdot x U,}}
where the last term in $\delta x^\mu$ is ``anomalous''. The transformation
of $U$ conforms with the fact that $U$ corresponds to the Higgs 
fields, thus has dimension 1. Without the last term in $\delta
x^\mu$, the metric $U^2dx^\mu dx_\mu$ is invariant. The additional
piece $U^{-2}dU^2$ necessitates the field-dependent term in $\delta x^\mu$.

The key observation of \jky\ is that the conformal transformation does
not commute with a gauge-fixing. In order to retain the same gauge,
a special conformal transformation must be accompanied by a field
dependent gauge transformation. Switching on background Higgs field
$\phi$, the additional transformation of the Higgs field is computed
at the one-loop level to be
\eqn\addg{\delta \phi = {2\lambda\over U^2}\epsilon\cdot \p\phi.}
And this is interpreted as introducing an additional piece
$2\lambda/U^2\epsilon^\mu$ into $\delta x^\mu$.

\subsec{The Wilson loops}

The original calculation \wils\ of the attractive force between a pair of
heavy quark and anti-quark was done with the metric \nhor. The idea
is the following. A heavy quark within D3-branes is represented
by an open string ending on D3-branes, with an infinite extension.
This configuration is realized in the near horizon geometry by an
open string extending from $U=0$ to $U=\infty$, with a constant
angle on $S^5$. The latter specifies the flavor of the heavy quark.
Note that as a BPS state, the heavy quark is charged not only with
respect to the gauge field, it is also charged with respect to
a scalar which is specified by the angle on $S^5$. Now given a pair
of heavy quark and an anti-quark of the same flavor, there are two
stretched open string with the opposite orientations. There will be
an attractive force between the two. According to the correspondence,
this interaction must be reflected by the AdS bulk physics. What is
more natural than the possibility of forming a single string out
the two open strings by joining them in the middle of the AdS space?

Many calculations become transparent if one switches from the coordinates
in \nhor\ to another coordinates system:
\eqn\newc{ds^2={R^2\over y^2}\left(dy^2+\sum dx_\mu^2\right),}
where $R^2=\sqrt{2\lambda}\ap$, and we have dropped the $S^5$ part.
Note that the role of $U$ now is played by $y$. As an exercise,
we would like to calculate the rectangular Wilson loop using this
coordinates system \foot{This was done together with Ruud Siebelink.}.
Unlike \wils\ where the Nambu-Goto action is used, we will use the
Polyakov action. Let the U-shaped string extend in two spatial directions,
$(y,x_1)$, where $y=y(x_1)$. At $y=0$, the boundary, the two ends of the
single string are separated by $L$. The minimal surface will be most
symmetric, so we assume that $y(-x_1)=y(x_1)$. The maximum $y$ is 
achieved at $x_1$, denote this value by $y_0$.

The action of the U-shaped string is
\eqn\uact{S={\sqrt{2\lambda}\over 4\pi}\int y^{-2}
\p_\alpha y^\mu\p_\beta y_\mu g^{\alpha\beta}\sqrt{g},}
where $(y_\mu )=(t, y, x_1)$ are functions of the world-sheet
coordinates $(\tau, \sigma )$. Going to the conformal gauge
yields the Virasoro constraints
\eqn\vira{\p_\tau y^\mu\p_\sigma y_\mu=0, \quad
\p_\tau y_\mu\p_\tau y^\mu+\p_\sigma y_\mu\p_\sigma y^\mu =0,}
which are satisfied by $t=\tau$, $\p_\sigma y=\sin\theta$,
$\p_\sigma x_1=\cos\theta$, where $\theta$ is a function of
only $\sigma$.

Now the action of the static string reads
$$S={\sqrt{2\lambda}\over 4\pi}\int y^{-2}\left(
1+(\p_\sigma y)^2+(\p_\sigma x_1)^2\right),$$
with the equations of motion 
\eqn\eomp{\eqalign{&\p_\sigma (y^{-2}\p_\sigma x_1)=0,\cr
&\p_\sigma (y^{-2}\p_\sigma y)+2y^{-3}=0.}}
The solution to the first equation, when combined with the
solution to the Virasoro constraints, is given by
$y=c^{-1}\sqrt{\cos\theta}$, $\p_\sigma\theta
=-2c\sqrt{\cos\theta}$. It is easy to check that this also
solves the second equation.

The maximum of $y$ is reached at $\theta =0$, so $y_0=
c^{-1}$. On the other hand, $y$ reaches the boundary at
$\theta =\pm \pi/2$. It follows from $\p_\sigma x_1=
\cos\theta$ and $\p_\sigma\theta =-2c\sqrt{\cos\theta}$
that
\eqn\diffe{\p_\theta x_1=-{1\over 2c}\sqrt{\cos\theta},}
and this gives rise to the condition
$cL=\int_0^{\pi/2}d\theta\sqrt{\cos\theta}$, or
\eqn\scon{c=(2\pi)^{3/2}\Gamma^{-2}(1/4)L^{-1}.}

Introducing a cut-off $T$ in time, the string action reads
\eqn\bare{\eqalign{S&=T{\sqrt{2\lambda}\over 2\pi} \int d\sigma y^{-2}\cr
&=T{c\sqrt{2\lambda}\over 2\pi}\int^{\pi/2}_{-\pi/2}d\theta
(\cos\theta )^{-3/2}.}}
The integral is divergent. This is not surprising, since we
expect that there is a contribution due to the infinite mass of
the two stretched open strings. The subtraction can be done effectively
by regulating the integral in \bare\ using the Euler beta function.
The result is 
\eqn\hpot{V=-{4\pi^2\sqrt{2\lambda}\over \Gamma^4({1\over 4})}L^{-1},}
the same as derived in \wils\ using the Nambu-Goto action.

\subsec{Large N QCD in the strong coupling limit}

Given a thermal state in SYM, there must be a corresponding state
in the AdS bulk theory. The natural candidate is the AdS black hole.
The AdS black hole is a vacuum solution to Einstein equations
with a negative cosmological constant. Thus the metric is a Einstein
metric. This leads Witten to propose that any Einstein metric
which asymptotes the AdS space represents a state in SYM \refs{
\gkpw, \hwit}. The earlier calculation of the black hole entropy \gkp\
supports this proposal, although there is a discrepancy in the
numerical coefficient between the AdS black hole entropy and that
of a free SYM.

The canonical ensemble of SYM is described by the QFT living on
$R^3\times S^1$, where the Euclidean time circle $S^1$ has a 
radius $\beta =1/T$. All fermions are anti-periodic, thus gain
a heavy mass in the reduced 3D theory when $T$ is large. As
standard in a thermal QFT, scalars as well as the time component
of the gauge field also gain a mass at the one loop level:
$m^2\sim \lambda T^2$. Thus for energies much below the scale
$T$, the theory is governed by an effective 3 dimensional
pure gauge theory, or the 3D QCD. If the quantum mechanically
generated masses persist to the strong coupling regime,
the effective theory is still the 3D QCD. Thus, Maldacena
conjecture leads to the exciting possibility, that the strongly
coupled 3D QCD can be understood in terms of the 5D AdS black
hole.

At a strong or intermediate coupling, the picture is not as
attractive as it first appears. The effective 3D gauge
coupling thus the 3D mass scale is $\lambda_3=\lambda T$.
This is no less than $T$, so the 3D interesting physics is
entangled with the KK modes. It is hard to tell whether what
the AdS physics teaches us is something about a 3D theory
or really a 4D theory.

Nevertheless, we still want to explore the physics of the AdS
black hole. In the Poincare coordinates, the metric is
\eqn\adsm{ds^2= {U^2\over R^2}[(1-{U_0^4\over U^4})
dt^2+\sum_{i=1}^3dx_i^2] +{R^2\over U^2}(1-{U_0^4\over U^4}
)^{-1}dU^2.}
The Hawking temperature is $T=U_0/(\pi R^2)$. This relation 
reflects the general physics called the UV/IR correspondence
\susswit. One can replace the temperature by an energy scale
$E$, and the corresponding radial distance $U$ that one explores
at this scale is $U=E\sqrt{\lambda}$. Now the calculation
of the maximal entropy a region inside a distance $U$ can
contain is similar to the calculation of the entropy of a
black hole of horizon size $U$:
\eqn\entrs{S\sim V_3(U/R)^3/G_5,}
where the five dimensional Newton constant $G_5=G_{10}/R^5$.
Using the UV/IR relation, we obtain
\eqn\fentr{S\sim N^2V_3E^3=N^2V_3/\delta^3,}
where we replaced $E$ by the UV cut-off in SYM. The entropy is
precisely what a QFT encodes with a UV cut-off $\delta$ and
volume $V_3$.

Next, to see the emergence of confinement, we need to prove that
the spectrum in the background of \adsm\ is discrete. This is
rather nontrivial in view that the manifold is not compact.
Witten showed for for a dilaton mode independent of time,
there is indeed a mass gap \hwit. For those KK modes in time,
the masses are even larger.

One can repeat the calculation of the Wilson loops as in the
zero temperature case. There are now a few possibilities.
First, one considers a single Wilson loop
wrapped in the time direction. This measures the effective mass of
a heavy quark in the 4D theory at a finite temperature. One does not
expect confinement, thus the mass correction to the infinite bare
mass is finite, and the expectation value $\langle W(C)\rangle$ must
be nonvanishing. Indeed this is the case, owing to the fact that
the world sheet with boundary $C$ can be extended in the AdS black
hole background. Note that the Euclidean time circle collapses to
a point at the horizon.

Next, one considers the correlation of two temporal Wilson loops.
In this case one expects the Debye screening. This is simply the
statement that the time component of the gauge field is massive,
so the interaction energy between a heavy quark and an anti-quark
must fall-off exponentially. The classical calculation shows that
beyond a certain separation of order $1/T$, the interaction
energy vanishes \reyith. To reproduces the Debye screening, one
need to take quantum fluctuations into account. So indeed the
electronic mass is nonvanishing in the strong coupling regime,
just as in the weak coupling regime.

The third case is a spatial Wilson loop. This can be interpreted
as a Wilson loop in the 3D pure gauge theory. It is argued in \hwit\
that there is an area law in this case. An explicit calculation shows
that the string tension is $\sqrt{\lambda} T^2$. Of course this
does not agree what one expects of the real 3D QCD, where the string
tension must be proportional to $\lambda_3^2\sim \lambda^2 T^2$.

One can choose another set of coordinates on $AdS_5$ such that
the topology of its boundary becomes $S^3\times R$. This is not
surprising, since this boundary is conformal to $R^4$, while
at the boundary of $AdS$ the metric blows up and only the conformal
structure is defined. Similarly, there is a black hole solution
\eqn\adsbh{ds^2= -(1+{r^2\over R^2}-{r_0^2\over r^2})dt^2
+(1+{r^2\over R^2}-{r_0^2\over r^2})^{-1}dr^2+r^2d\Omega_3^2.}
For a given Hawking temperature, there are two black holes.
The larger one has a size greater than $R$. It is this black hole
that corresponds to a thermal state in the boundary SYM living
on $S^3\times R$. It is easy to see that the specific heat is positive,
agrees with what one expects of a QFT thermal state. The smaller
black hole has a negative specific heat, thus corresponds to a
meta-stable state in SYM. 

The larger black hole goes over to the infinite volume limit.
There is an interesting twist in this case. As shown by Hawking and
Page \hawp, that there exist two manifolds for a given temperature.
The second one is obtained by periodically identify time in
the AdS metric. Comparing the two actions of the two manifold,
it was found in \hawp\ that there exists a phase transition at
the temperature $TR=3/(2\pi)$ (the original calculation
was done for $AdS_4$). For a higher temperature, the free
energy is dominated by the tree level effects thus is proportional
to $N^2$. This is called the high temperature phase. The tree level
free energy of the low temperature phase vanishes, thus the dominant
contribution comes from the one-loop effects, and the free energy
is independent of $N$. This low temperature phase is identified by
Witten as the finite volume confining phase \hwit. It can be
checked that the expectation value of a temporal Wilson $\langle
W(C)\rangle$ vanishes, so the $Z_N$ symmetry is unbroken, consistent
with the notion that the phase is a confining one. The reason
for vanishing the Wilson loop is rather simple: The Euclidean
time circle never collapses, so one can not find a smooth world
sheet with boundary $C$.

Indeed, the phase diagram of the 4D SYM is even more complicated
than the above discussion indicates. There exists another kind of
phase transition, the strong/weak coupling phase transition \phase.
This exists  at a finite temperature, for both the infinite
volume limit as well as a finite volume. For a larger coupling
$\lambda$, the phase may be termed as the supergravity phase, where
one trust the $\ap$ expansion in the background of the AdS
black hole. For a smaller coupling, one trust the perturbative
SYM, where the expansion parameter is $\lambda$, or $(\ap)^{-1}$.
The phase transition point is what one would call the correspondence
point \joeg. The order parameter of this strong/weak coupling
transition may be the Hawking-Page temperature at which the first
order Hawking-Page phase transition occurs. We expect that this
temperature drops to zero at the correspondence point \phase.

The 4D large N QCD at the strong coupling limit can be studied starting
with the near horizon geometry of D4-branes. All results obtained
using supergravity only again must be taken with a grain of salt, since
the large N strong/weak coupling phase transition is the thing we
have to live with. (Might it be the case that only with the presence
of such a phase transition, we can hope that the dimensional
transmutation in 4D QCD will emerge on the weak coupling side?)

\newsec{Conclusion}

We have learned a great deal in the past four years, and we are convinced
more than ever that string/M theory is the most promising approach
to unification of all forces in Nature, and to the elusive quantum
gravity theory. Admittedly, despite much has been learned about the
rich structure of M theory, and quantum properties of black holes,
we are still miles away from the goal of formulating a nonperturbative,
background independent M theory, and much still remains to be revealed
about quantum black holes, especially the prototypical of all,
the Schwarzschild black holes. In addition to the problem of uncovering
the principles, we have the eminent more ``technical'' problems
of relating M theory to the real world. How to break supersymmetry?
How has our own universe evolved to today's observed state,
thus it is accurately described by both the particle standard model
and the cosmological standard model?

One can spend hours speculating endlessly about the future of our efforts,
and about the ultimate formulation of M theory. The most important
thing the past experience teaches us is that we must always keep
an open mind, and many surprises are awaiting ahead for us. Undoubtedly
there have been many lines of thought as to where we should focus
our attention. To provide just one such thought, we recommend
ref.\liyo\ to the reader. 

\noindent{\bf Acknowledgments} 
We would like to thank the organizers of duality workshop at CCAST,
K. Wu and C.-J. Zhu for invitation to give lectures. The reading
of the manuscript by V. Sahakian is gratefully
acknowledged, his suggestions helped to improve the writing. 
This work was supported by DOE grant DE-FG02-90ER-40560 and NSF grant
PHY 91-23780.

\listrefs

\end